\documentclass[pdflatex,sn-nature]{sn-jnl}

\usepackage{graphicx}%
\usepackage{multirow}%
\usepackage{amsmath,amssymb,amsfonts}%
\usepackage{amsthm}%
\usepackage{mathrsfs}%
\usepackage[title]{appendix}%
\usepackage{xcolor}%
\usepackage{textcomp}%
\usepackage{manyfoot}%
\usepackage{booktabs}%
\usepackage{algorithm}%
\usepackage{algorithmicx}%
\usepackage{algpseudocode}%
\usepackage{listings}%

\theoremstyle{thmstyleone}%
\theoremstyle{thmstyletwo}%

\theoremstyle{thmstylethree}%

\begin{document}

\title[Article Title]{Numerical Simulation for General Relativistic Magnetohydrodynamics in Dynamic Spacetimes}


\author*[1]{\fnm{Beibei} \sur{Li}}\email{iauthor@gmail.com}



\affil*[1]{\orgname{Deep Space Exploration Lab}, \orgaddress{\country{China}}}




\abstract{We present a novel spectral solver for general relativistic magnetohydrodynamics on dynamical spacetimes.  By combining a high order discontinuous spectral method on mapped Chebyshev-Fourier grids, our scheme attains exponential convergence.  Implemented within a unified BSSN-Valencia framework, the code evolves both Einstein and MHD fields self‐consistently, enabling fully coupled simulations of black hole accretion jet systems.  We demonstrate spectral accuracy and entropy stability through convergence tests, and validate physical fidelity via equatorial embedding diagrams of horizon‐crossing GRMHD variables in Kerr-Schild coordinates.  Three‐dimensional scatter visualizations further highlight the solver’s capability to capture complex magnetized plasma structures around rotating black holes.  This approach paves the way for high‐order, low‐dissipation GRMHD simulations on exascale architectures, opening new avenues for precise modeling of strong‐field astrophysical phenomena.
}

\maketitle

\section{Introduction}\label{sec1}

Black holes are solutions to Einstein’s field equations in which spacetime curvature becomes so extreme that not even light can escape from within the event horizon.  The central singularity represents a point where curvature invariants diverge and classical general relativity ceases to provide a regular description of spacetime~\cite{Carroll1997,Alcubierre2006}.

The observational evidence for supermassive black holes comes from stellar and gas dynamics in galactic nuclei—most notably Sgr~A* at the center of our Galaxy—and from direct imaging. 
Multi‐frequency VLBI reveals that the M87 jet follows a single parabolic streamline over five orders of magnitude, consistent with a Poynting‐flux–dominated MHD nozzle~\cite{Nakamura2013}.  
The observed shift of M87’s jet boundary from parabolic to conical constrains the black hole’s mass and spin by linking magnetically dominated and equipartition flows~\cite{Nokhrina2019}.  
EHT imaging at 230\,GHz captures an asymmetric crescent shadow of $\simeq 42\,\mu$as, confirming the black hole’s mass in agreement with general relativistic predictions~\cite{EHT2019Shadow}.  
First horizon‐scale images of Centaurus~A reach down to $\sim200\,r_g$, revealing jet launching and collimation close to the black hole~\cite{Janssen2021}.  
Deep Horizon employs a deep‐learning network to infer black hole accretion parameters from synthetic horizon‐scale images~\cite{vanDerGucht2020}.  
Comparisons of multi‐frequency observations of M87 with GRMHD simulations constrain jet plasma composition and emission physics via polarized morphology~\cite{Anantua2023}.

GRMHD simulations solve the coupled Einstein–MHD equations in a 3+1 split of spacetime to probe the complex plasma dynamics near black holes.  High-resolution studies of magnetized accretion disks, jet launching, and feedback employ conservative, shock-capturing schemes such as HARM~\cite{Gammie2003}, vectorized implementations like \texttt{iharm3D}~\cite{Prather2021}, and adaptive mesh refinement codes such as WhiskyMHD~\cite{Giacomazzo2007}.  Convergence analyses have shown that diagnostics and spectra stabilize only above a critical grid resolution~\cite{Shiokawa2011}, while reviews summarize key numerical approaches and physical mechanisms in AGN jet simulations~\cite{Mizuno2022}.  Recent efforts include baryon inpainting via KL-loss–augmented neural networks~\cite{Dai2023}, Monte Carlo neutrino transport in $\nu$bhlight~\cite{Miller2019}, thin-slab binary merger setups~\cite{Fedrigo2024}, and multizone frameworks that bridge vast spatial and temporal scales~\cite{Cho2024}.  Additional studies incorporate radiative transfer comparisons against mm‐VLBI data~\cite{Dexter2012}, wind‐fed Sgr~A* models with dynamic electron heating and rotation‐measure analysis~\cite{Ressler2023}, analyses of radiative cooling thresholds~\cite{Dibi2012}, and investigations of how varying angular‐momentum profiles affect flow morphology~\cite{Bardiev2024}, collectively revealing how grid resolution, microphysics, spin, and radiation processes shape synthetic observables.

The GRMHD theory and computational frameworks provide the foundation for these studies. 
The HARM scheme introduces a conservative shock‐capturing GRMHD algorithm based on approximate Riemann solvers~\cite{Gammie2003}.  
WhiskyMHD provides a fully general‐relativistic MHD code with adaptive mesh refinement capabilities~\cite{Giacomazzo2007}.  
Generalized 3+1 GRMHD formalisms extend the Valencia framework to rotating black holes~\cite{Koide2009}.  
Foundational lecture notes offer comprehensive mathematical groundwork for general relativity~\cite{Carroll1997}.  
ADM and BSSN formalisms underpin 3+1 numerical relativity simulations~\cite{Alcubierre2006}.  
Performance‐portable implementations like KHARMA leverage Kokkos for efficient GRMHD on CPUs and GPUs~\cite{Prather2024}.  
GPU‐accelerated dynamical‐spacetime packages such as GRaM‐X facilitate exascale GRMHD research within the Einstein Toolkit~\cite{Shankar2023}.  
The Valencia flux‐conservative formulation ensures consistency across GRMHD codebases~\cite{Mewes2021}.

Numerical solutions of the Einstein–MHD system have become the cornerstone of modern strong‐field astrophysics.  For example, the first direct detection of gravitational waves by LIGO relied critically on banks of waveform templates generated by numerical relativity simulations of inspiraling and merging compact binaries. Likewise, the first image of a black hole shadow in M87 by the Event Horizon Telescope combined VLBI observations with GRMHD‐informed ray tracing through simulated accretion flows to confirm that the observed crescent and dark central depression indeed arise from light bending around a relativistic magnetized disk~\cite{EHT2019Shadow}.  Beyond these landmark discoveries, GRMHD and numerical relativity simulations continue to guide the design of next‐generation observatories by predicting how variations in spin, magnetic field strength, accretion physics, and equation of state will manifest in both gravitational‐wave and electromagnetic signatures, thereby shaping our strategies for probing gravity in its most extreme regime.

While existing GRMHD schemes have enabled many breakthrough simulations, they remain formally limited to accuracy in regions, and their numerical dissipation can obscure small‐scale features. However, existing computational frameworks for solving the coupled Einstein–MHD system remain relatively incomplete, with only a handful of codes supporting fully dynamical spacetime evolution coupled to GRMHD, limiting our ability to study feedback between strong‐field gravity and plasma flows in a unified manner.

We presents a high-accuracy spectral solver for fully coupled general-relativistic magnetohydrodynamics in dynamic spacetimes. Starting from the covariant Einstein–Maxwell–fluid system under the ideal-MHD approximation, we adopt horizon-penetrating Kerr–Schild coordinates and perform a 3+1 Valencia decomposition to cast the GRMHD equations in conservative form—identifying conserved variables, fluxes, geometric source terms, and the divergence-free magnetic constraint. Spatial derivatives are handled spectrally, using a Chebyshev–Gauss–Lobatto grid radially and Fourier pseudospectral methods angularly, with 2/3 dealiasing and exponential filtering to control nonlinear aliasing. Time integration employs a fourth-order Runge–Kutta scheme, and primitive variables are recovered via a damped Newton–Raphson iteration. The solver supports both fixed-background and fully coupled BSSN evolution incorporating XCTS initial data, 1+log slicing, and Gamma-driver shift within a single framework. A suite of numerical experiments, including equatorial embedding diagrams, 3D field visualizations, and discrete-entropy monitoring, verifies spectral accuracy, entropy stability, the monotonic entropy decay, and convergence per the Lax–Wendroff and Lax equivalence theorems.

\section{Results}


\subsection{General Relativistic Magnetohydrodynamics}
General Relativistic Magnetohydrodynamics describes the dynamics of a conducting fluid in curved spacetime, combining Einstein’s field equations with Maxwell’s equations and fluid conservation laws.

\subsubsection*{The Einstein Field Equations}
Einstein’s equations relate spacetime curvature to the total stress–energy tensor:
\begin{equation}
R_{\mu\nu} - \tfrac12 g_{\mu\nu} R = 8\pi T_{\mu\nu},
\end{equation}
where
\begin{equation}
T_{\mu\nu} = T^{\rm fluid}_{\mu\nu} + T^{\rm EM}_{\mu\nu}.
\end{equation}

\subsubsection*{The Energy of the Fluid and Electromagnetic Stress }
The fluid stress–energy tensor:
\[
T^{\mu\nu}_{\rm fluid} = \rho h\,u^{\mu}u^{\nu} + p\,g^{\mu\nu},
\]
and the electromagnetic contribution via the magnetic four–vector $b^{\mu}$:
\[
b^{\mu} = {}^*F^{\mu\nu}u_{\nu}, \quad b^{2}=b_{\mu}b^{\mu},
\]
\[
T^{\mu\nu}_{\rm EM} = b^{2}u^{\mu}u^{\nu} + \tfrac12 b^{2}g^{\mu\nu} - b^{\mu}b^{\nu}.
\]

\subsubsection*{Electromagnetic Fields in Curved Spacetime}
The electromagnetic field tensor and its dual:  
\begin{align}
F_{\mu\nu} &= \partial_{\mu}A_{\nu} - \partial_{\nu}A_{\mu}, \\
{}^*F^{\mu\nu} &= \tfrac12\epsilon^{\mu\nu\alpha\beta}F_{\alpha\beta},
\end{align}
with $\epsilon^{0123}=1/\sqrt{-g}$. Maxwell’s equations are:
\begin{equation}
\nabla_{\nu}F^{\mu\nu} = J^{\mu}, \quad \nabla_{\nu}{}^*F^{\mu\nu} = 0, \quad \nabla_{\mu}J^{\mu} = 0.
\end{equation}

\subsubsection*{The Covariant GRMHD Equations}
Mass conservation, energy–momentum conservation, and Maxwell’s equations in the ideal MHD limit $F^{\mu\nu}u_{\nu}=0$:
\begin{align}
&\nabla_{\mu}(\rho u^{\mu}) = 0, \\
&\nabla_{\mu}T^{\mu\nu} = 0, \\
&\nabla_{\nu}F^{\mu\nu} = 0, \quad \nabla_{\nu}{}^*F^{\mu\nu} = 0.
\end{align}
where the flux freezing implies
\[
{}^*F^{\mu\nu} = u^{\mu}b^{\nu} - u^{\nu}b^{\mu}.
\]

\subsection{GRMHD in Kerr-Schild Coordinates}
In this section we present all of the key formulas needed to evolve GRMHD in a Kerr–Schild background. We first give the horizon-penetrating Kerr–Schild metric in closed form and show how it decomposes into lapse, shift, and 3-metric for a 3+1 evolution. We then write the Valencia-form GRMHD system listing the conserved variables, their fluxes, the source terms built from the Christoffel symbols, and the divergence-free constraint on the magnetic field specialized to these coordinates.

\subsubsection*{The Kerr-Schild Metric and 3+1 Decomposition}
Kerr\textendash{}Schild coordinates are horizon-penetrating, regular at the event horizon, and suited for black hole simulations.

\paragraph{Metric:}
\begin{align}
ds^2 =\;& -\left(1 - \frac{2Mr}{\Sigma}\right)dt^2 + \frac{4Mr}{\Sigma}dtdr + \left(1 + \frac{2Mr}{\Sigma}\right)dr^2 + \Sigma d\theta^2 \nonumber\\
& + \sin^2\theta\left[\Sigma + a^2\sin^2\theta\left(1 + \frac{2Mr}{\Sigma}\right)\right]d\phi^2 \nonumber\\
& - \frac{4aMr\sin^2\theta}{\Sigma}dtd\phi - 2a\sin^2\theta\left(1 + \frac{2Mr}{\Sigma}\right)drd\phi
\end{align}

\begin{equation}
\Sigma = r^2 + a^2\cos^2\theta
\end{equation}

\paragraph{The Kerr-Schild Structure:}

The metric is decomposed as
\[
g_{\mu\nu} = \eta_{\mu\nu} + 2H\,l_{\mu}l_{\nu},
\]
with
\[
H = \frac{Mr}{\Sigma}, \quad \Sigma = r^{2} + a^{2}\cos^{2}\theta, \quad l_{\mu} = \bigl(1,\,1,\,0,\,\tfrac{a\sin^{2}\theta}{\Sigma}\bigr).
\]

\paragraph{The 3+1 Decomposition:}
In Kerr\textendash{}Schild form, lapse, shift, and spatial metric are:
\begin{align}
\alpha &= (1 + 2H)^{-1/2}, \\
\beta^{i} &= \frac{2H}{1+2H}\,l^{i}, \\
\gamma_{ij} &= \delta_{ij} + 2H\,l_{i}l_{j}.
\end{align}

\vspace{1em}
\subsubsection*{GRMHD Conservative Form}
These enter the Valencia formulation of GRMHD:
\begin{equation}
\partial_t(\sqrt{\gamma} U) + \partial_i(\sqrt{\gamma} F^i) = \sqrt{\gamma} S
\end{equation}
where $\gamma = \det(\gamma_{ij})$.

\paragraph{Conserved Variables:}
\begin{equation}
U = \begin{pmatrix}
D \\
S_j \\
\tau \\
B^k
\end{pmatrix}
\end{equation}

\begin{align}
D &= \rho W \\
S_j &= (\rho h^* + b^2) W^2 v_j - \alpha b^0 b_j \\
\tau &= (\rho h^* + b^2) W^2 - (p^* + (b^0)^2) - D
\end{align}

with 
\begin{align}
h^* &= 1 + \epsilon + \frac{p}{\rho} + \frac{b^2}{\rho} \\
p^* &= p + \frac{b^2}{2}
\end{align}

\paragraph{Fluxes:}
\begin{equation}
F^i = \begin{pmatrix}
D v^i \\
S_j v^i + p^* \delta^i_j - b_j B^i / W \\
\tau v^i + p^* v^i - b^0 B^i / W \\
v^i B^k - v^k B^i
\end{pmatrix}
\end{equation}

\paragraph{Lorentz factor:}
\begin{equation}
W = \frac{1}{\sqrt{1 - \gamma_{ij} v^i v^j}}
\end{equation}

\paragraph{Magnetic field four-vector:}
\begin{align}
b^0 &= \frac{W}{\alpha} (v_i B^i) \\
b^i &= \frac{B^i + \alpha b^0 u^i}{W} \\
b^2 &= b^\mu b_\mu
\end{align}

\vspace{1em}
\paragraph{Source term:}
\begin{equation}
S = \sqrt{\gamma} T^{\mu\nu} \left( \partial_\mu g_{\nu\alpha} - \Gamma^\lambda_{\mu\nu} g_{\lambda\alpha} \right)
\end{equation}
where $T^{\mu\nu}$ is the total stress-energy tensor.

\vspace{1em}
\subsubsection*{Magnetic Constraint}
\begin{equation}
\partial_i (\sqrt{\gamma} B^i) = 0
\end{equation}

\vspace{1em}
\subsubsection*{Christoffel Symbols and Metric Derivatives}

\paragraph{General formula:}
\begin{equation}
\Gamma^\lambda_{\mu\nu} = \frac{1}{2} g^{\lambda\sigma} \left( \partial_\mu g_{\sigma\nu} + \partial_\nu g_{\sigma\mu} - \partial_\sigma g_{\mu\nu} \right)
\end{equation}

\subsection{Equatorial Embedding Diagram of GRMHD Variables on the Black Hole Horizon}

\bigskip

\subsubsection*{Coordinate System}

\begin{itemize}
  \item We work in the equatorial plane $\theta = \pi/2$ of a Kerr--Schild black hole background with mass \(M\) and spin parameter \(a\).
  \item The \emph{areal radius} \(r\) is sampled from the event horizon radius
    \[
      r_{\rm h} \;=\; M + \sqrt{M^2 - a^2}
    \]
    out to an outer radius \(r_{\rm max}\).
  \item The angular coordinate \(\varphi\in[0,2\pi)\) parameterizes the equatorial circle.
  \item We embed the 2D spatial slice \(r\ge r_{\rm h}\), \(\theta=\pi/2\) into \(\mathbb{R}^3\) as a Flamm paraboloid
    \[
      Z(r)\;=\;\int_{r_{\rm h}}^r \sqrt{g_{rr}(r')-1}\,\mathrm{d}r',
      \quad
      g_{rr}(r)=1+2 H(r,\tfrac{\pi}{2}),
    \]
    where \(H(r,\theta)=M\,r / \Sigma(r,\theta)\) and \(\Sigma=r^2 + a^2\cos^2\theta\).
  \item The embedding coordinates are
    \[
      X = r\cos\varphi,\quad
      Y = r\sin\varphi,\quad
      Z = \pm Z(r),
    \]
    giving the familiar “upper” and “lower” Flamm surfaces.
\end{itemize}

\subsubsection*{GRMHD Conservative Variable Visualization}

\begin{figure}[h]
    \centering
    \includegraphics[width=0.3\textwidth]{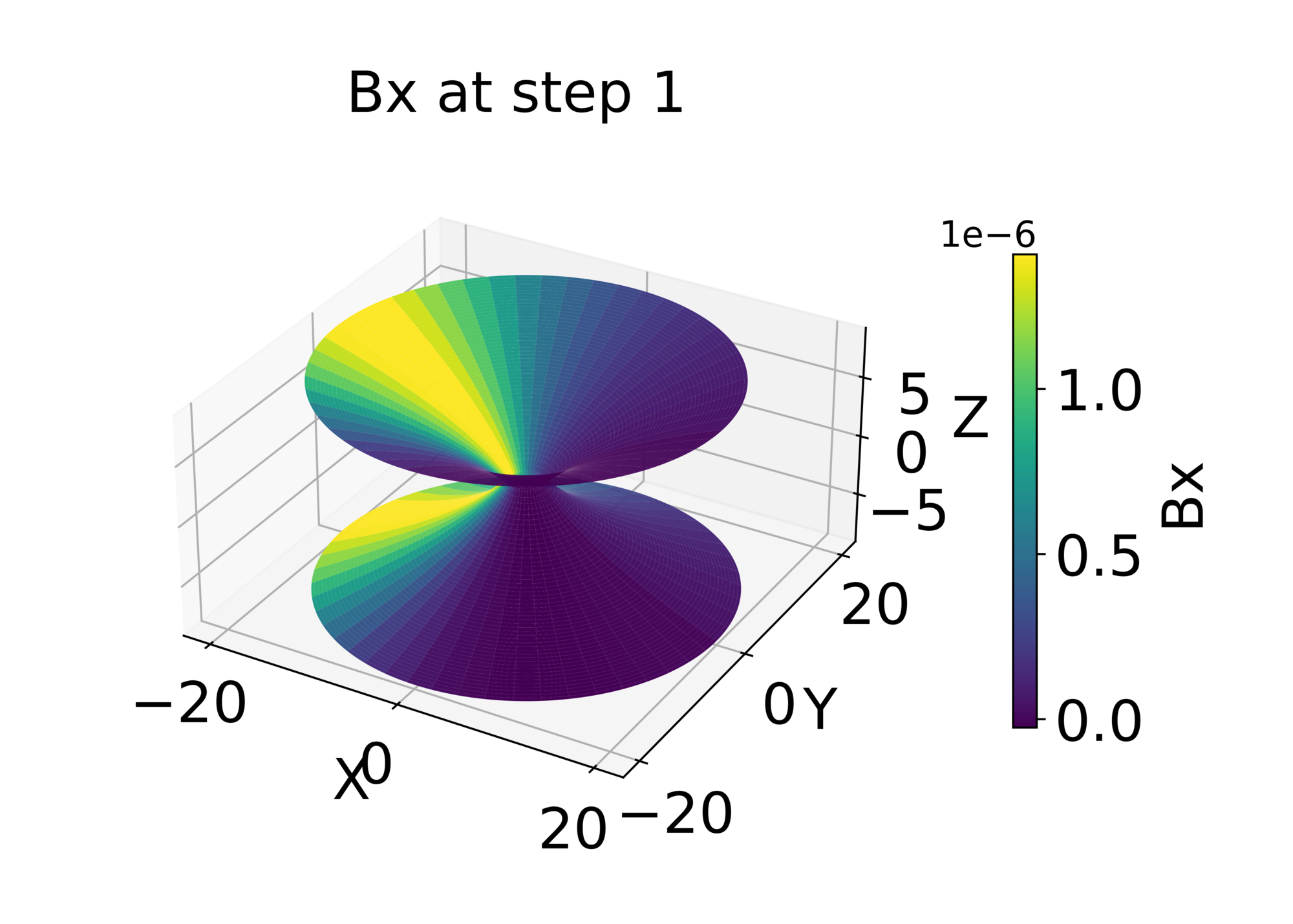}
    \includegraphics[width=0.3\textwidth]{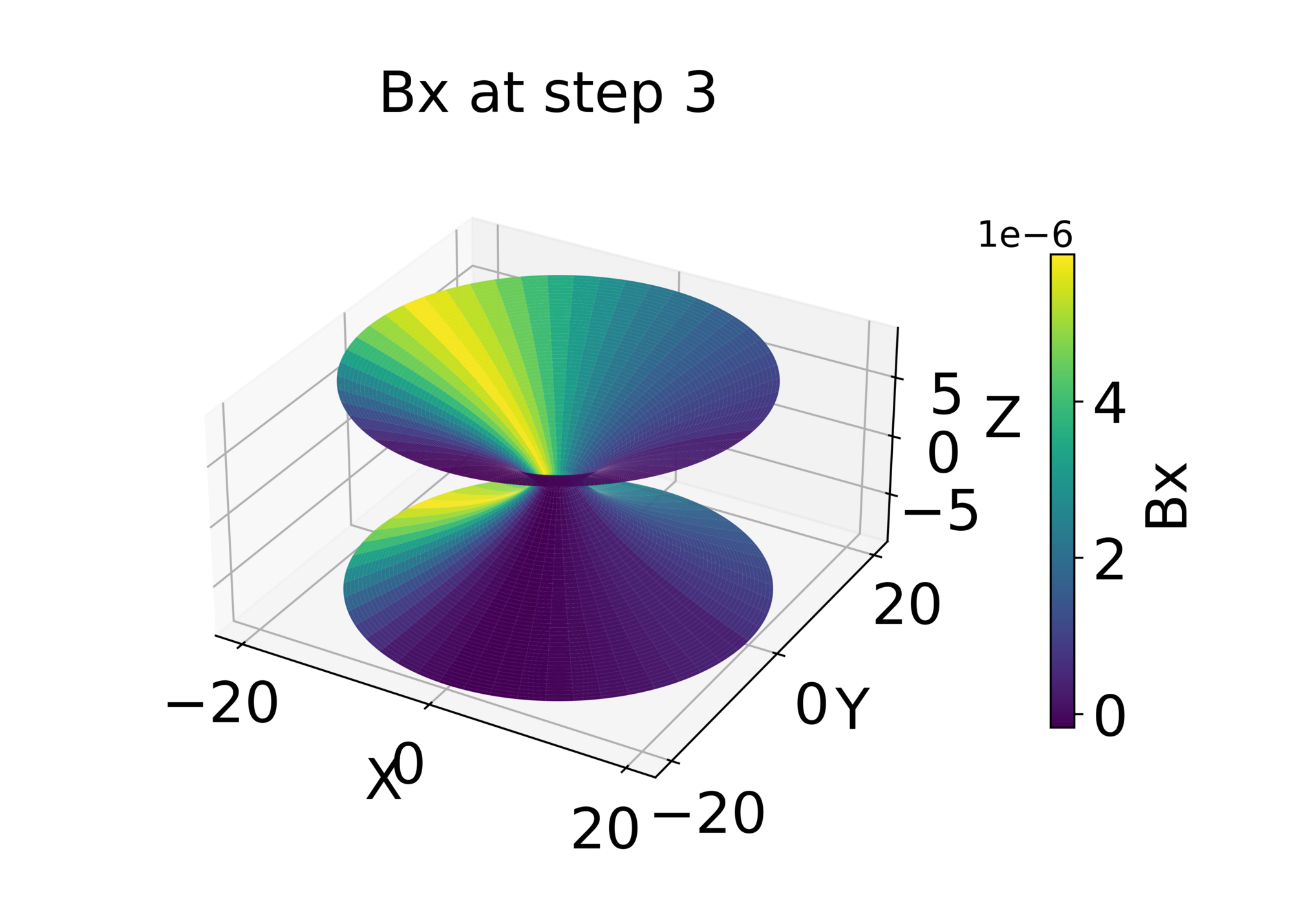}
    \includegraphics[width=0.3\textwidth]{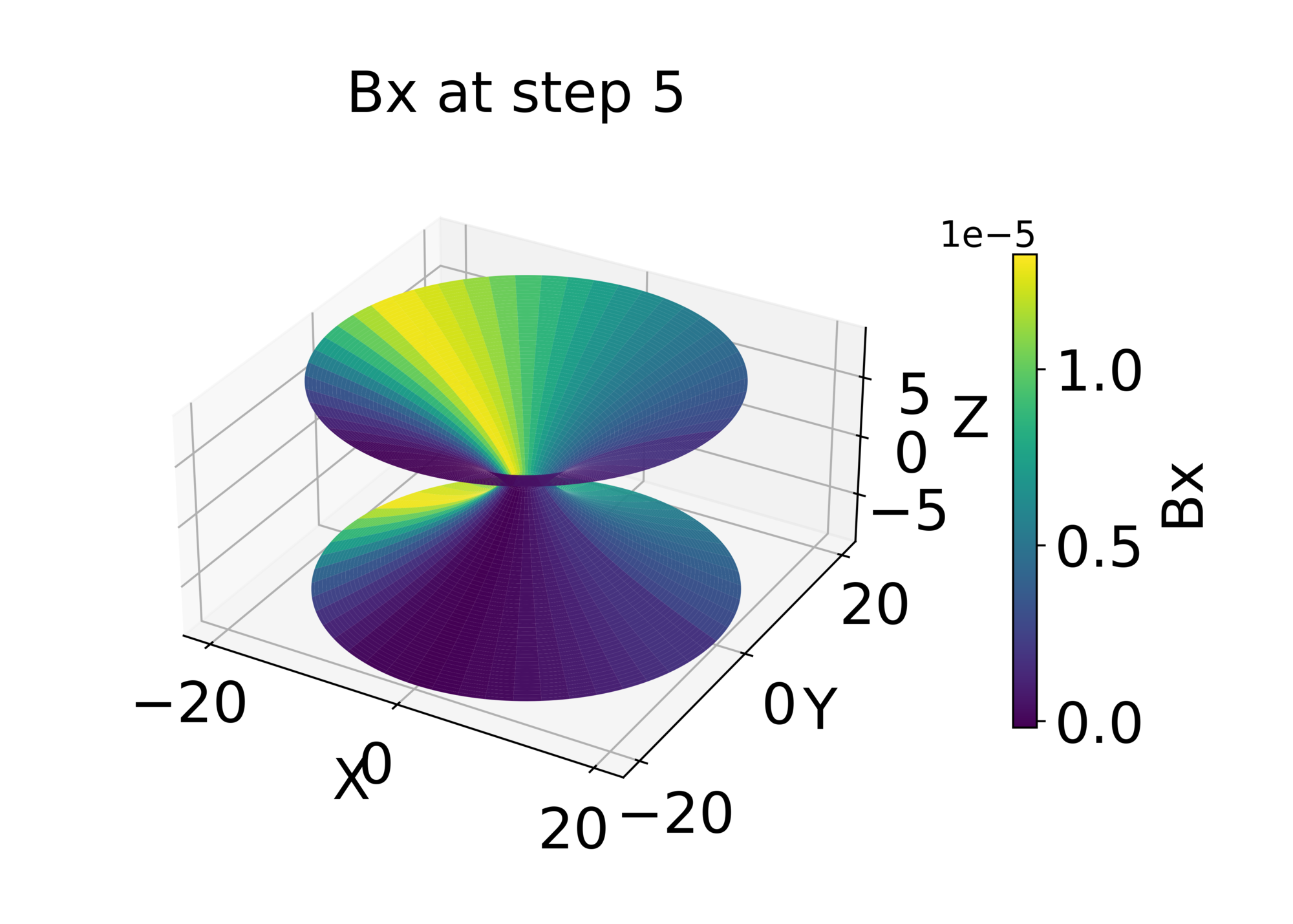}
    \includegraphics[width=0.3\textwidth]{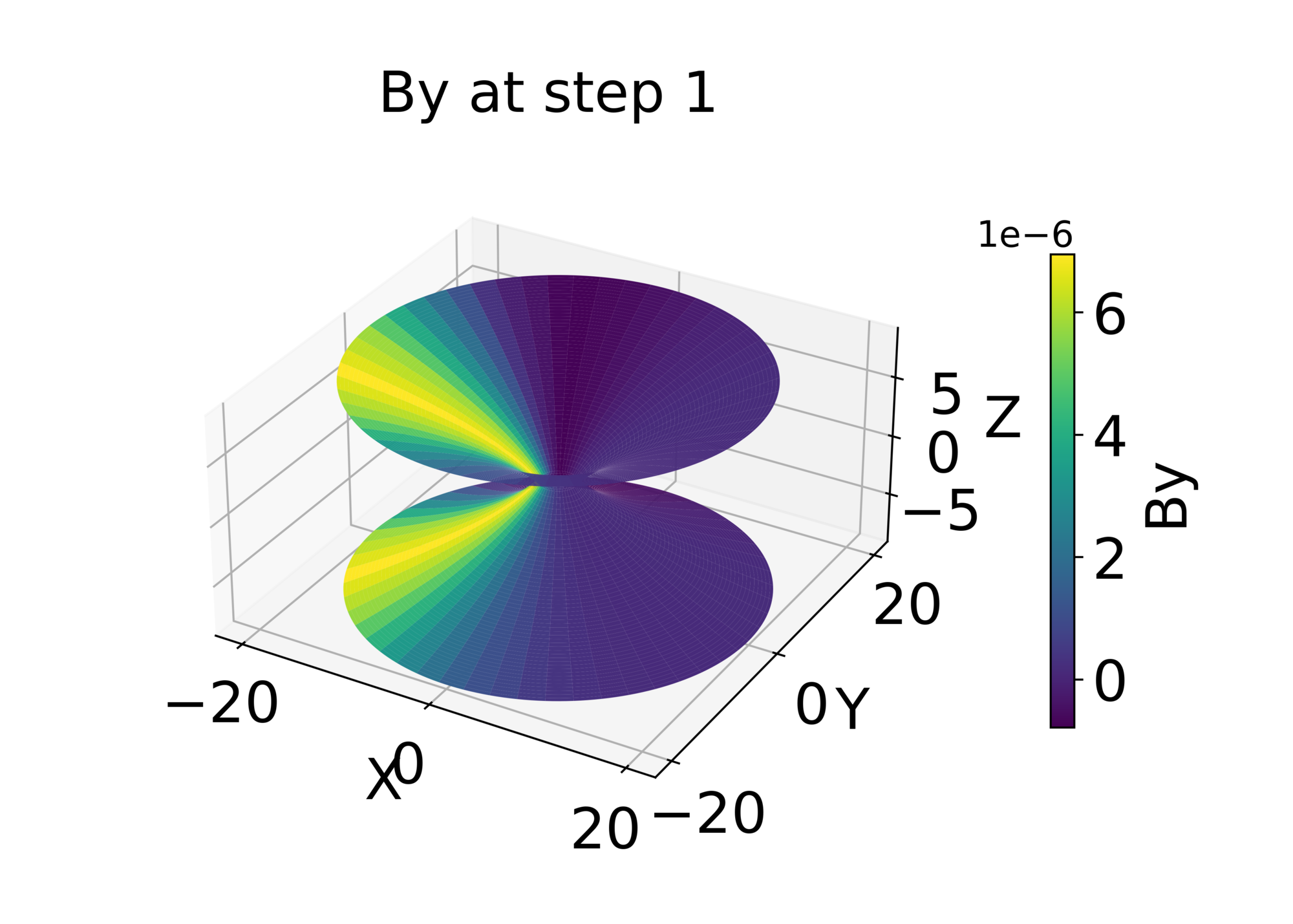}
    \includegraphics[width=0.3\textwidth]{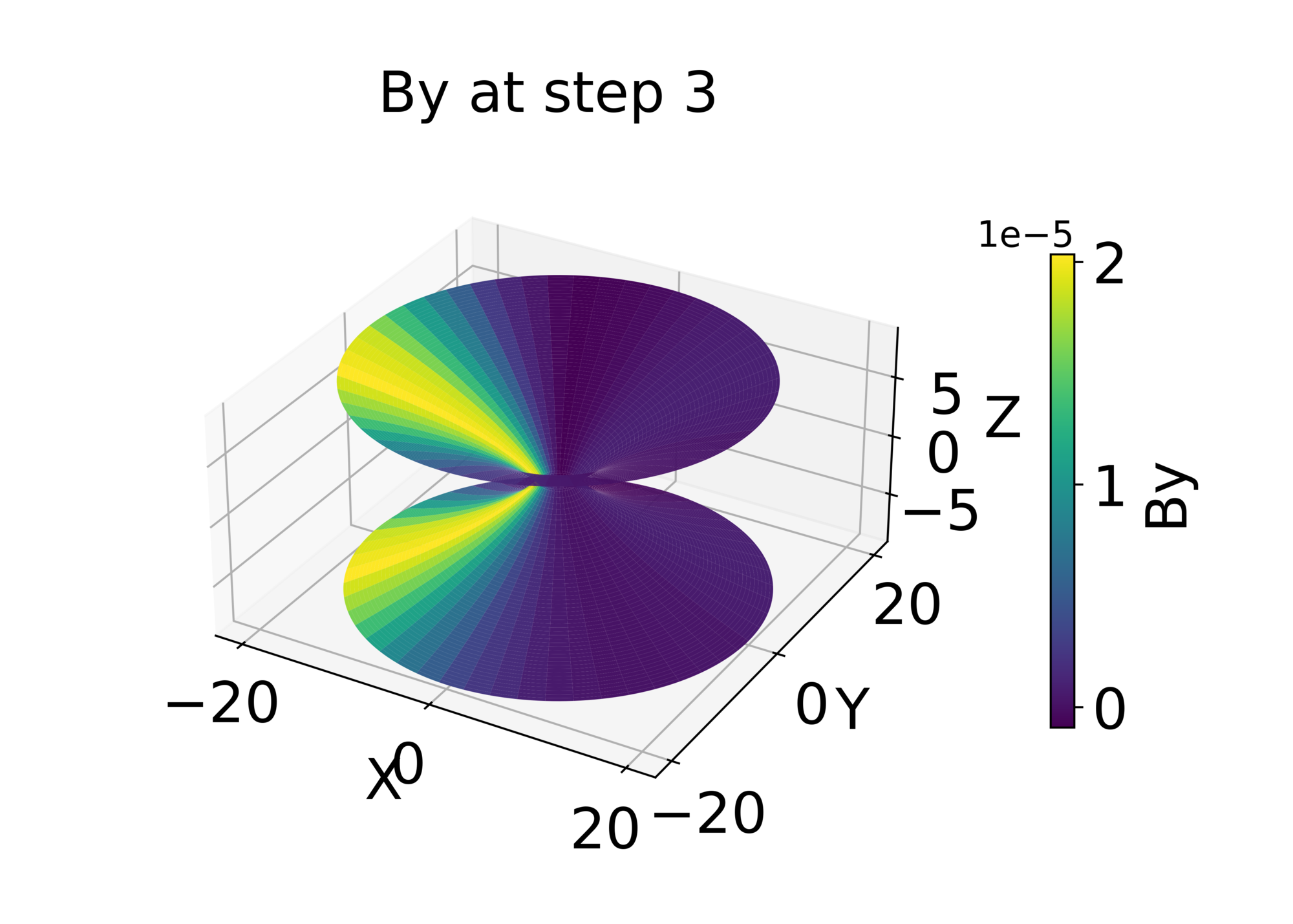}
    \includegraphics[width=0.3\textwidth]{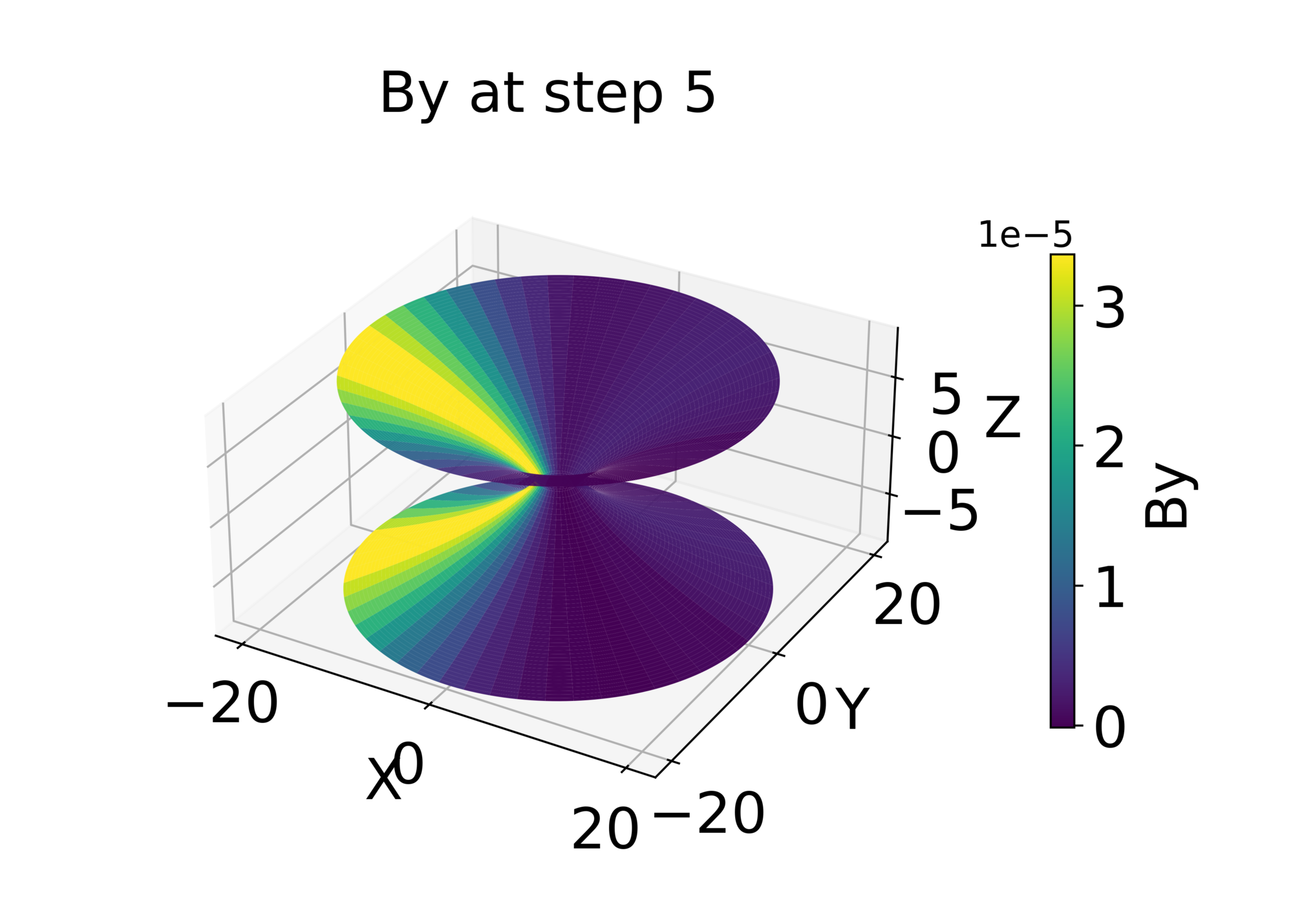}
    \includegraphics[width=0.3\textwidth]{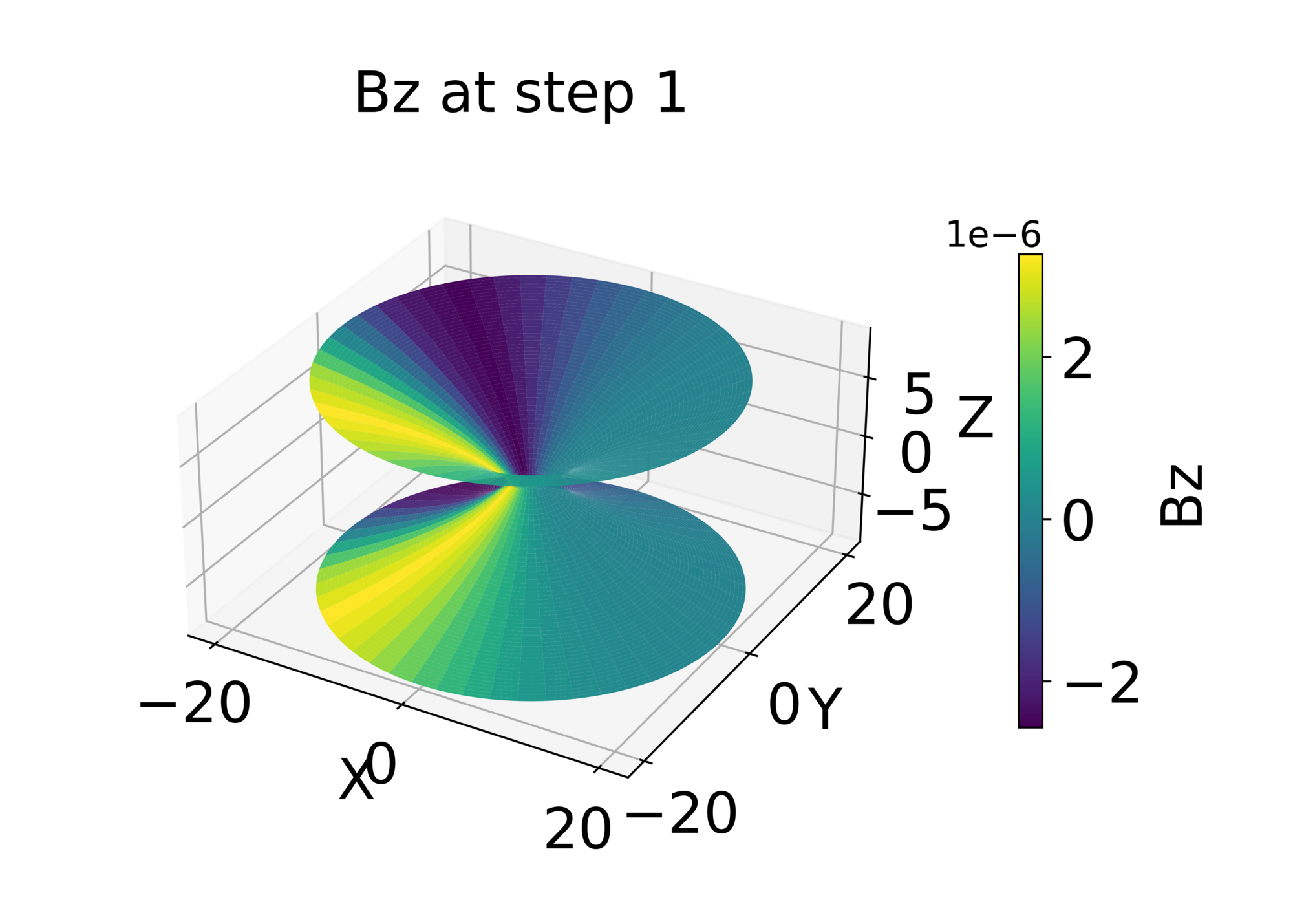}
    \includegraphics[width=0.3\textwidth]{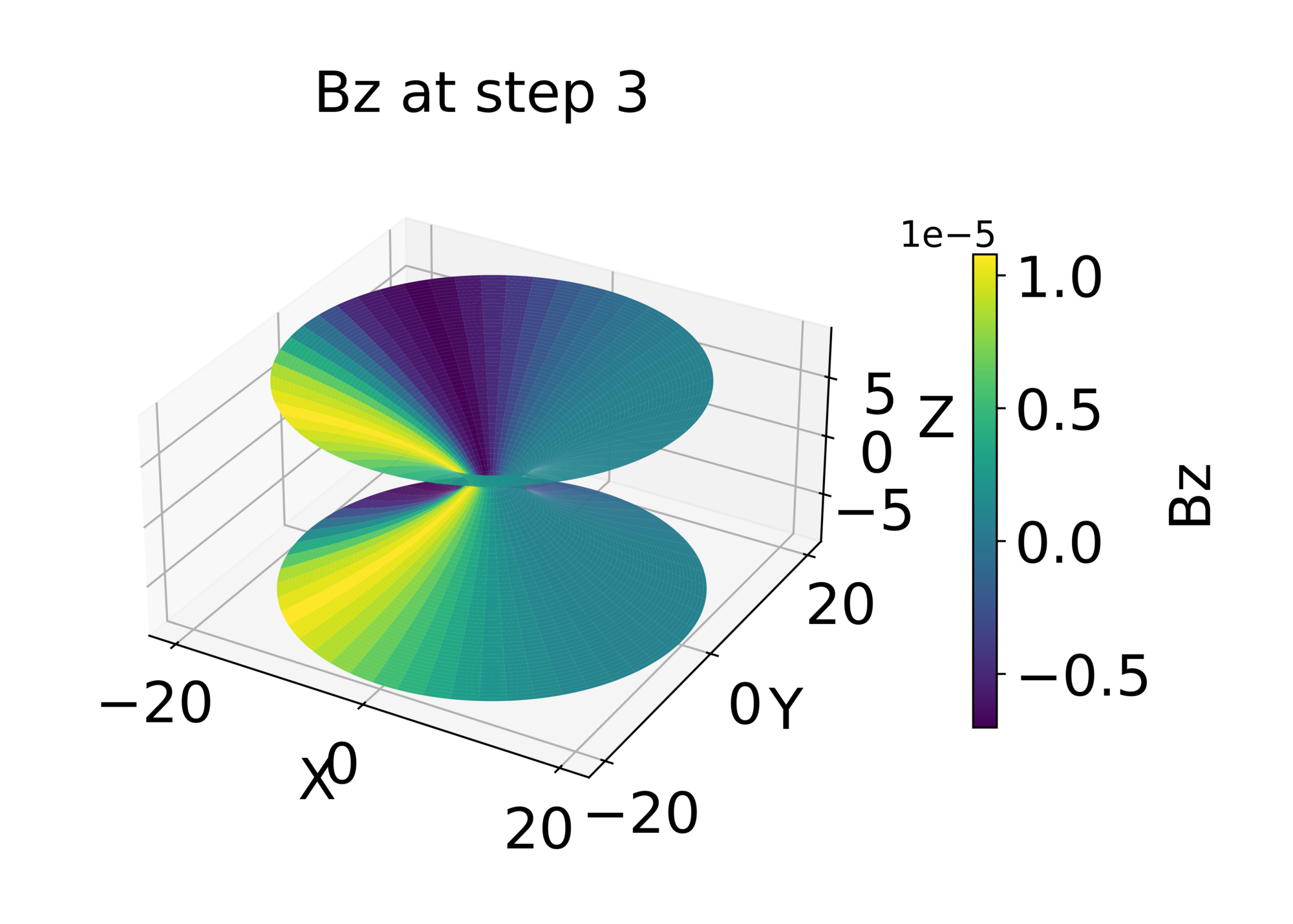}
    \includegraphics[width=0.3\textwidth]{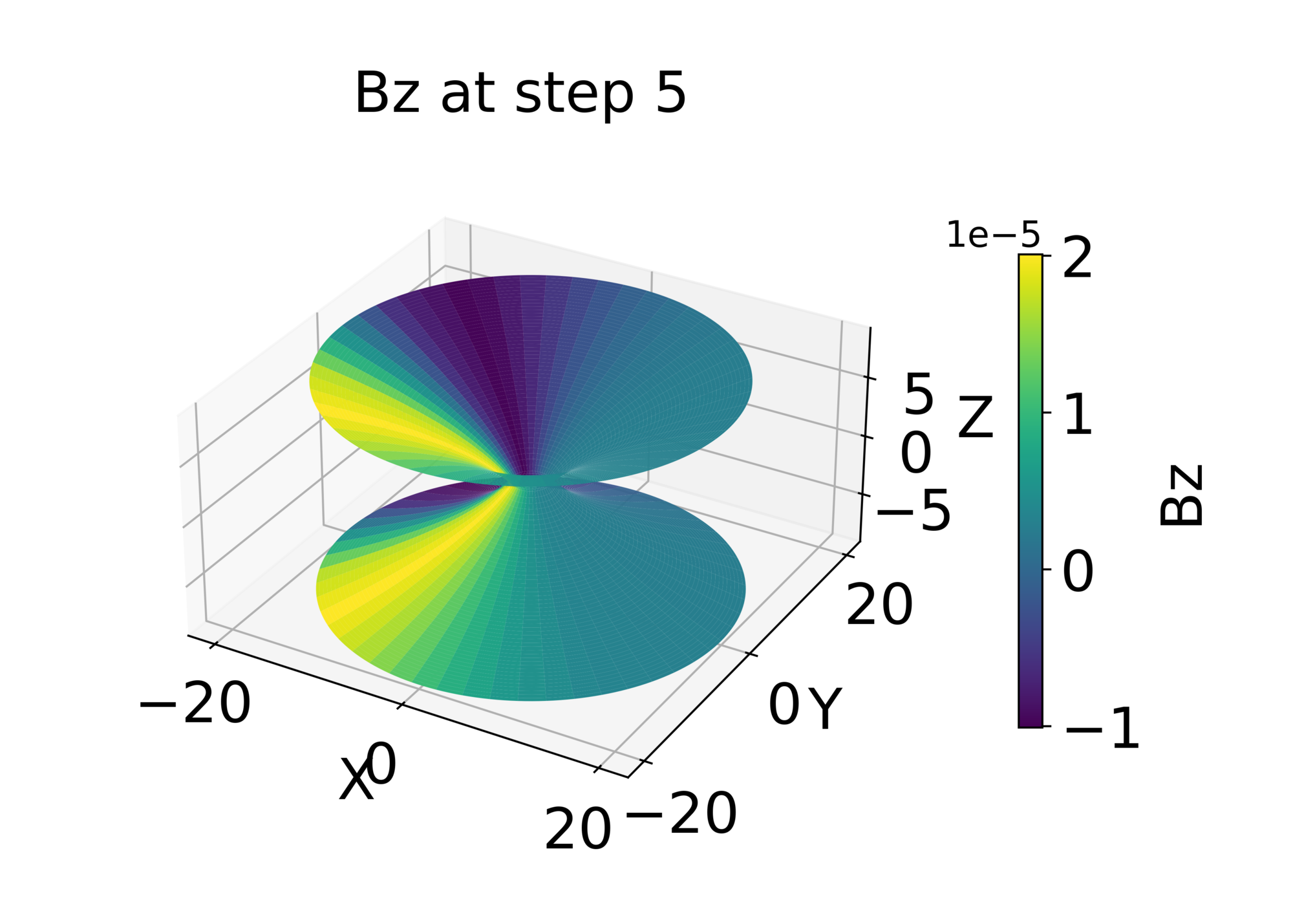}
    \includegraphics[width=0.3\textwidth]{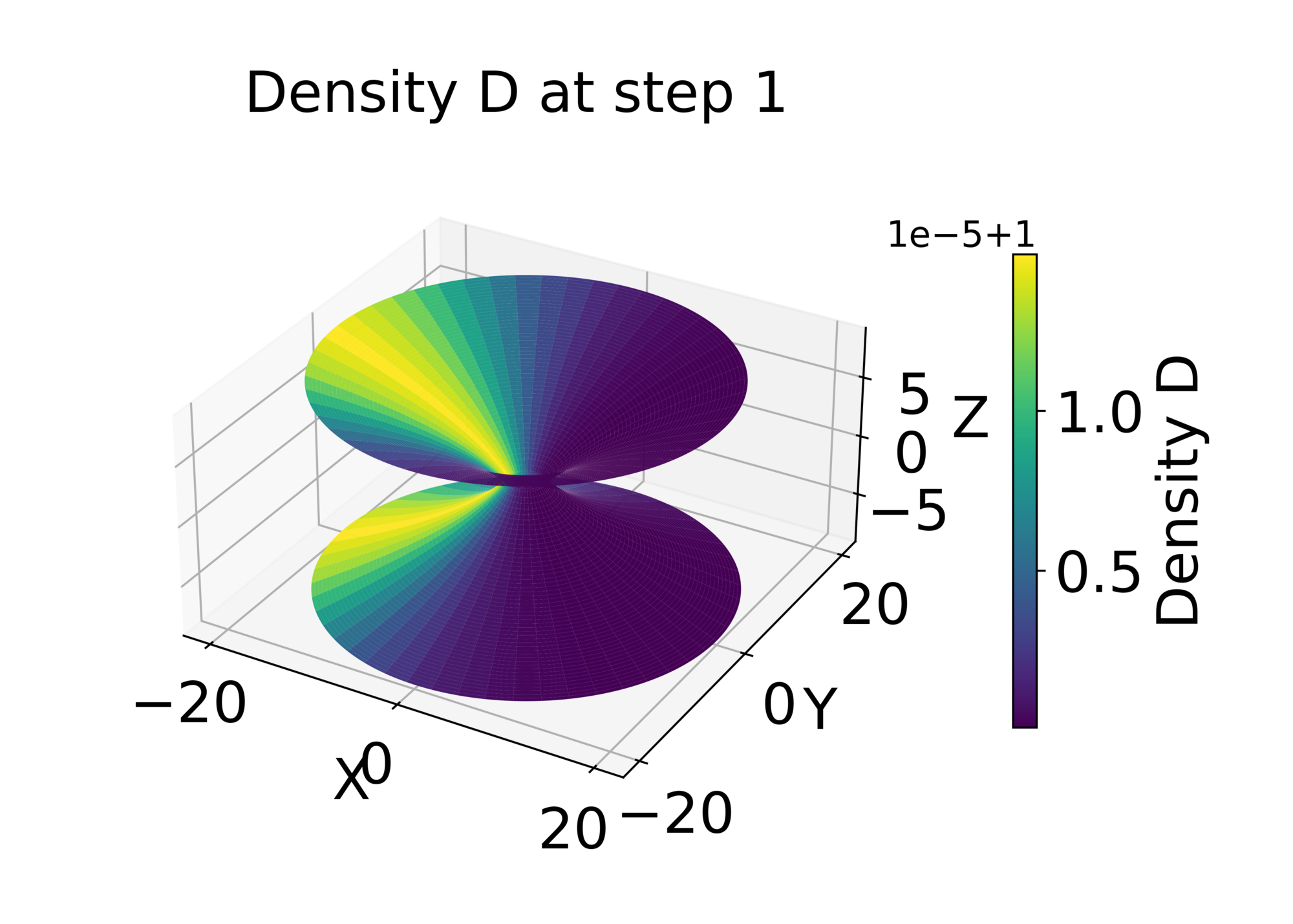}
    \includegraphics[width=0.3\textwidth]{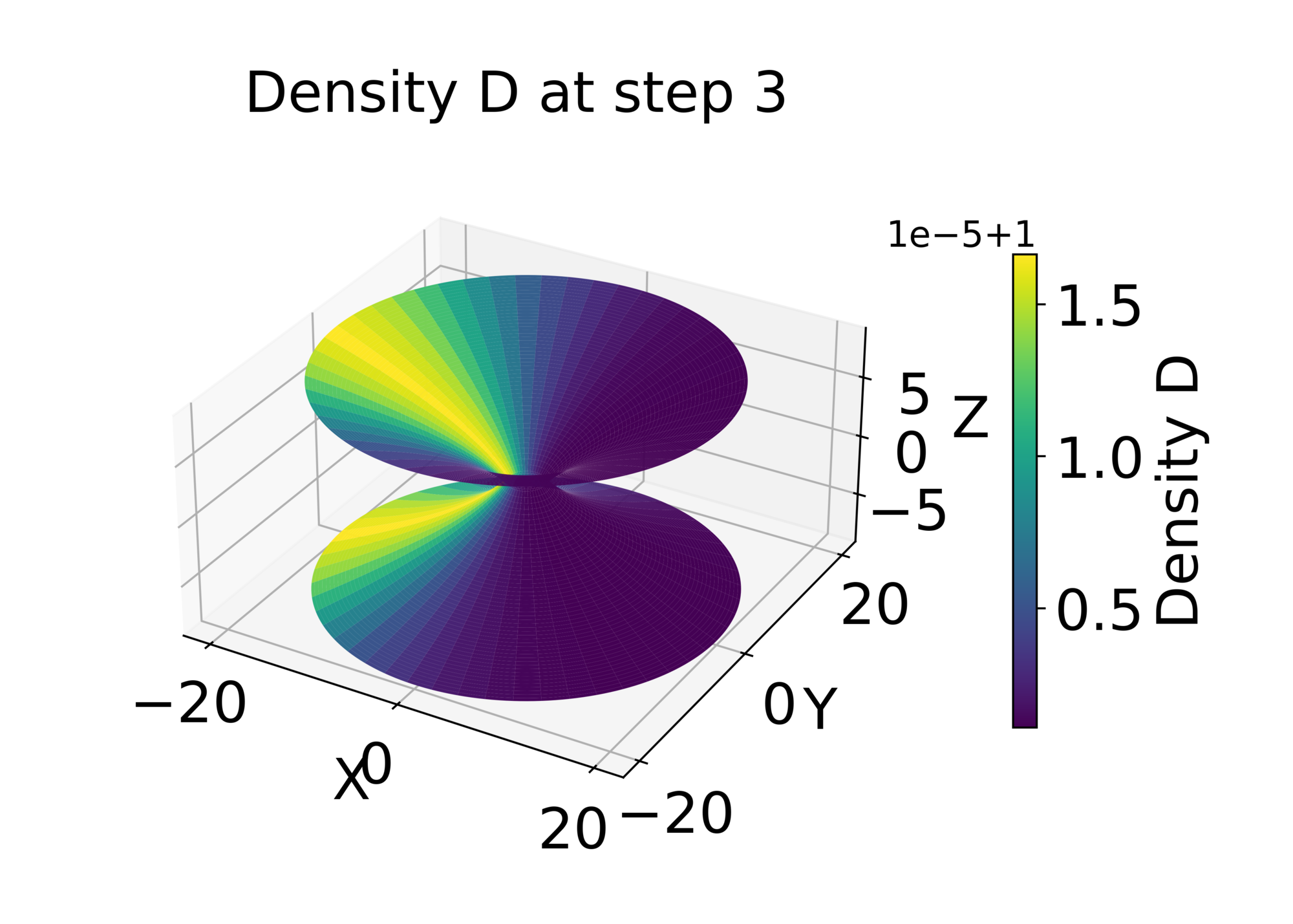}
    \includegraphics[width=0.3\textwidth]{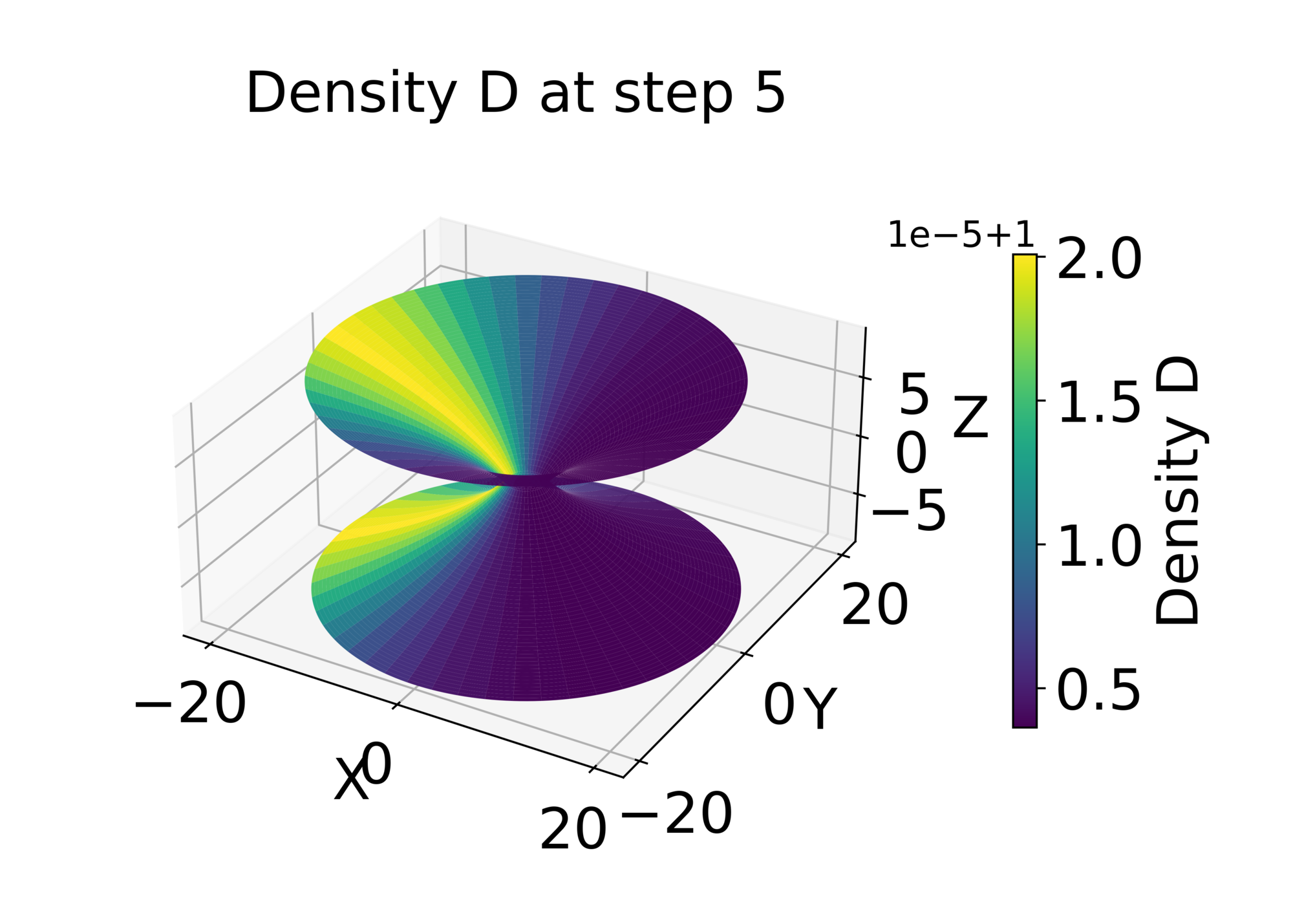}
    \includegraphics[width=0.3\textwidth]{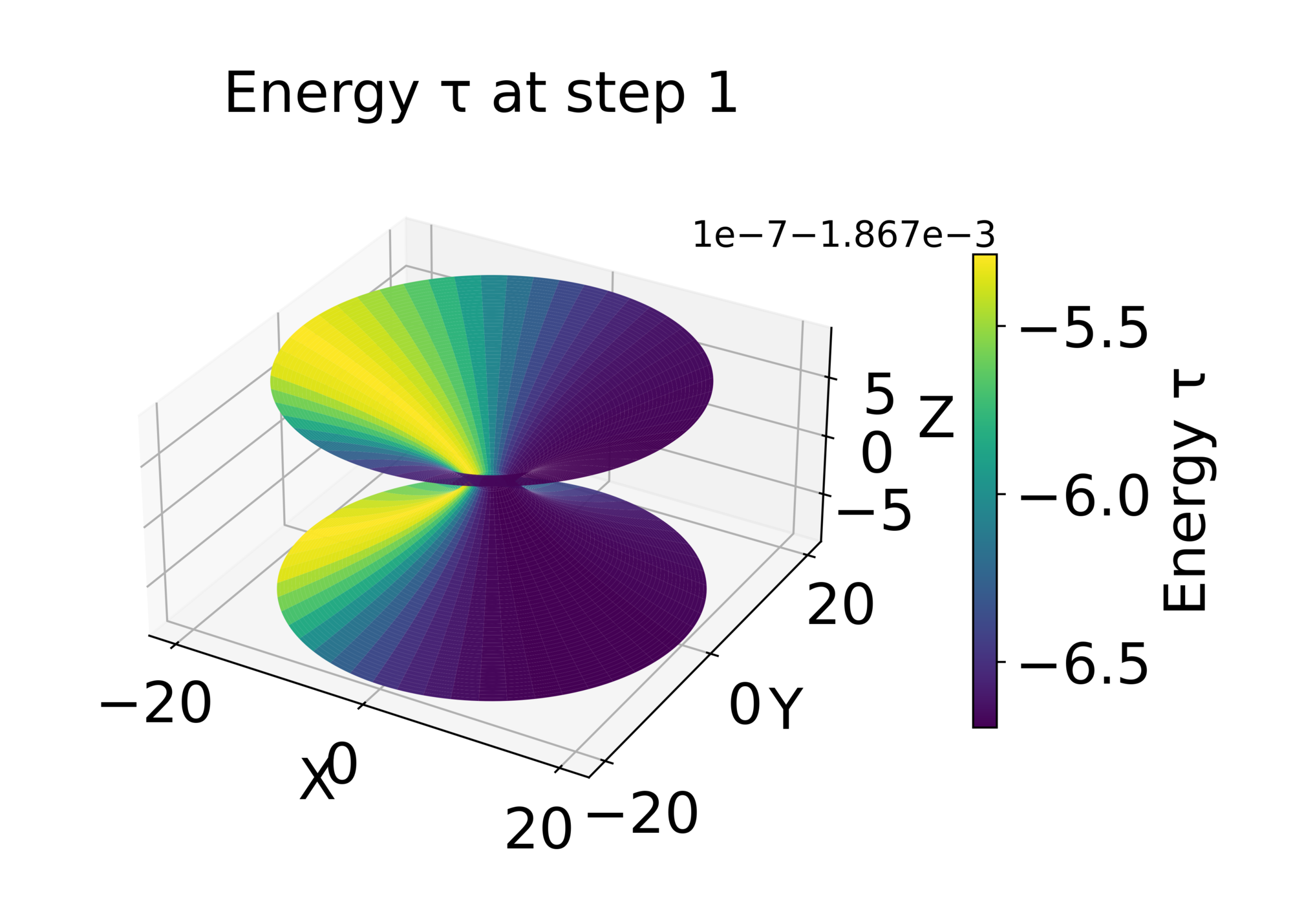}
    \includegraphics[width=0.3\textwidth]{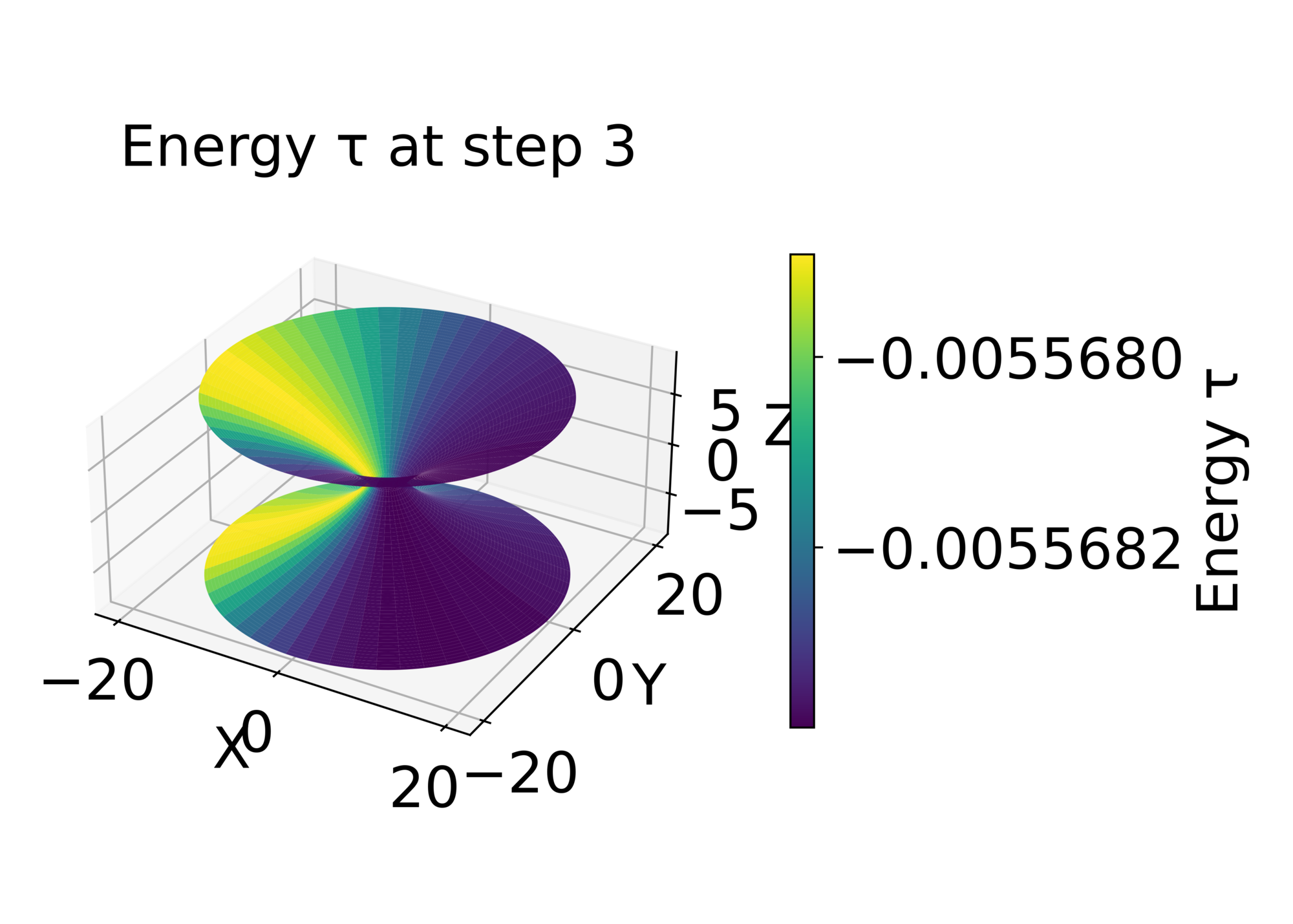}
    \includegraphics[width=0.3\textwidth]{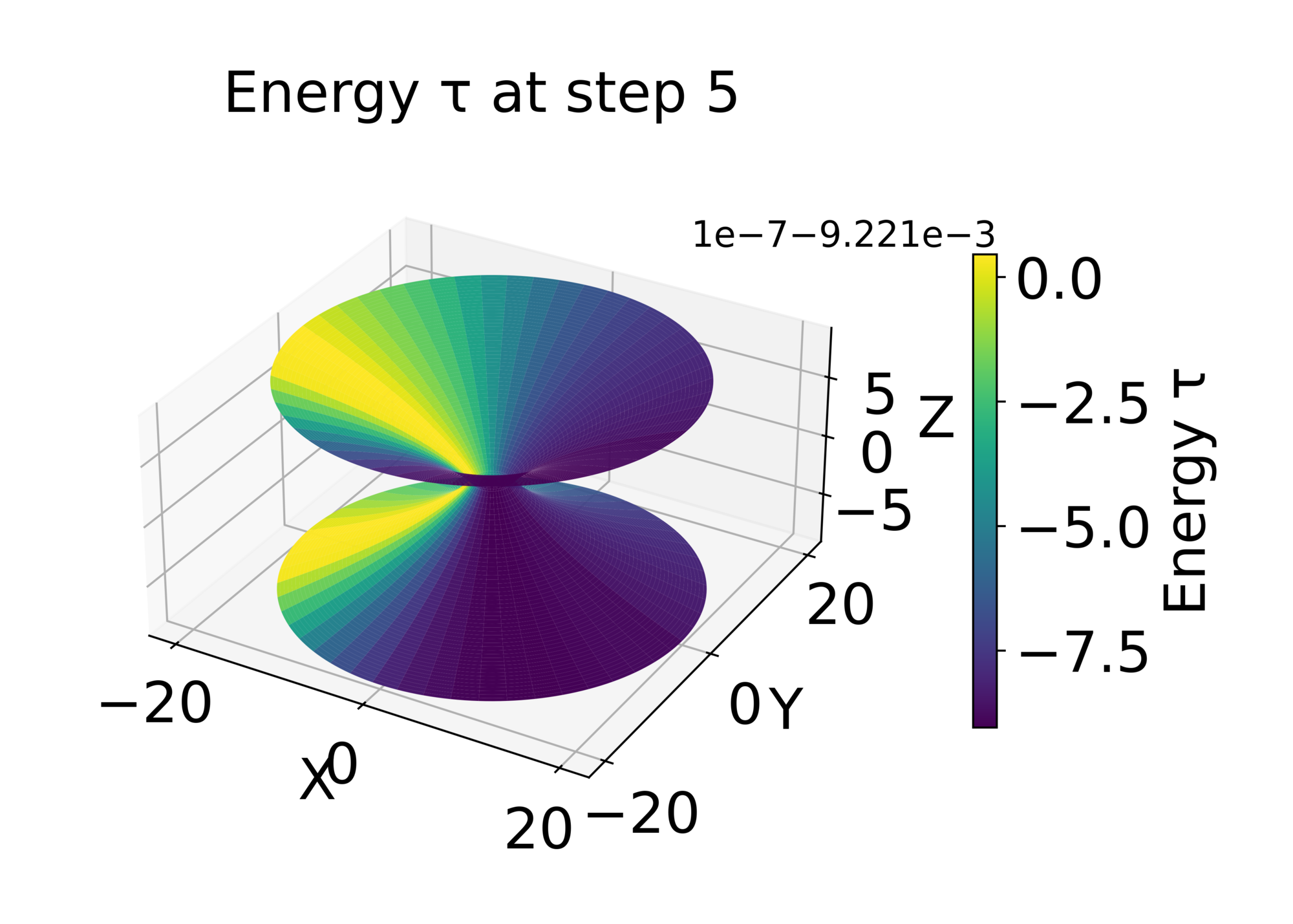}
    \caption{Equatorial embedding diagrams of the Kerr black hole event horizon \( r_h = M + \sqrt{M^2 - a^2} \), shown as a Flamm paraboloid in fictitious 3D Euclidean space. The surface is parametrized by \( X = r_h \cos\phi \), \( Y = r_h \sin\phi \), and the embedding height \( Z \) computed from the spatial metric. Colors indicate the value of each conserved variable of magnetic field components, density and energy along the equatorial ring \( \theta = \pi/2 \) at the event horizon. Both upper \( +Z \) and lower \( -Z \) surfaces are shown for symmetry.}
    \label{fig:DE}
\end{figure}

\begin{figure}[h]
    \centering
    \includegraphics[width=0.3\textwidth]{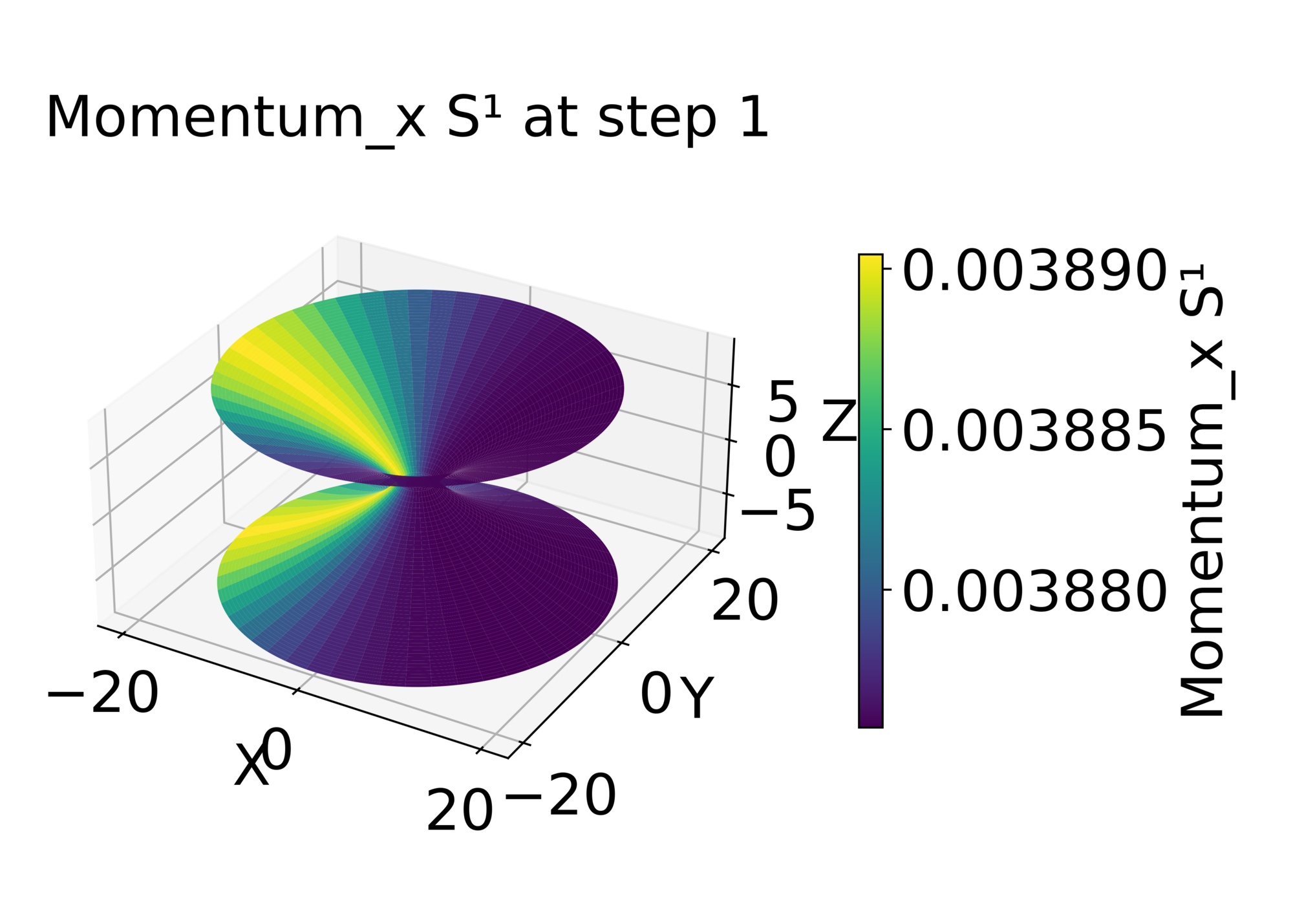}
    \includegraphics[width=0.3\textwidth]{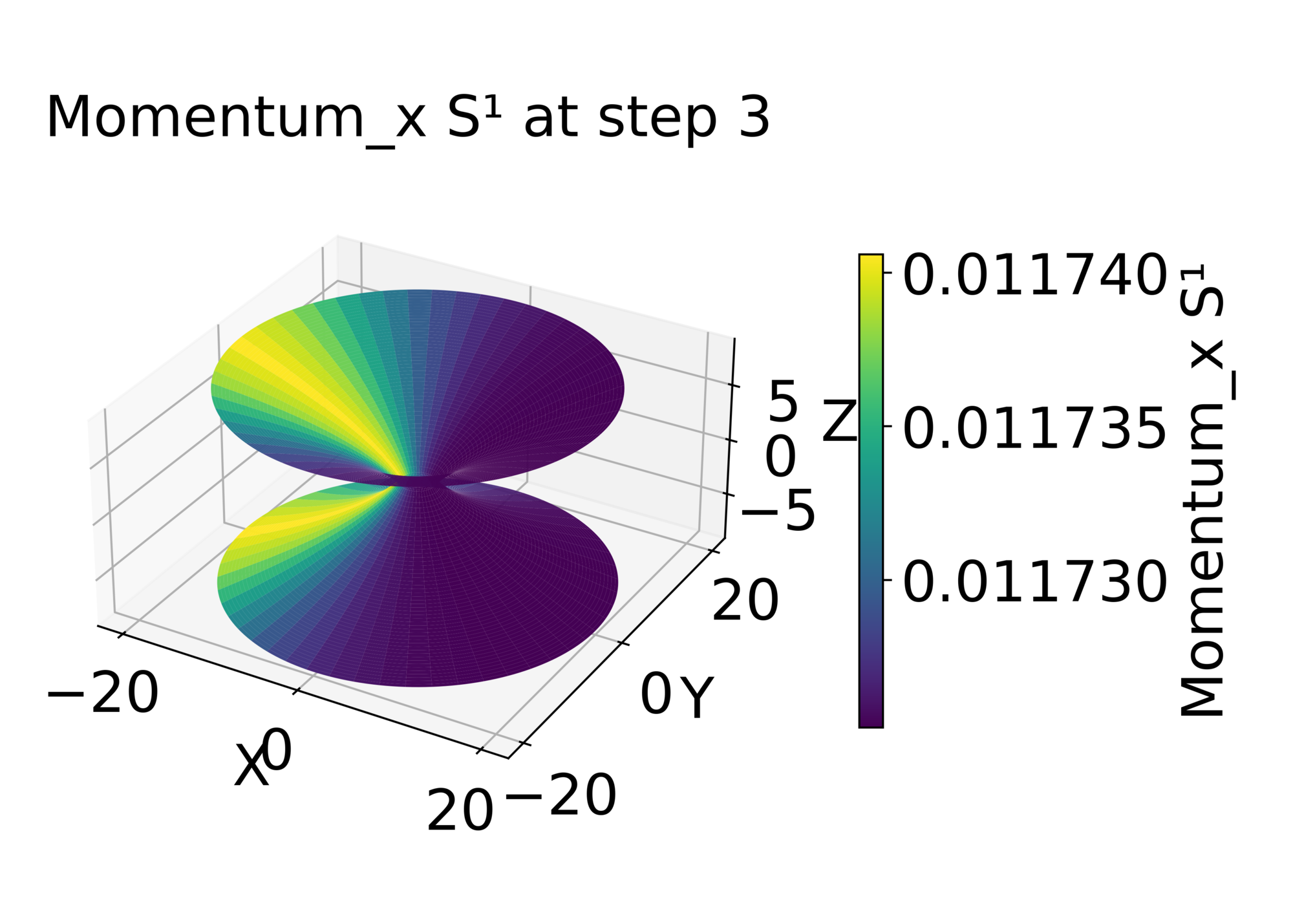}
    \includegraphics[width=0.3\textwidth]{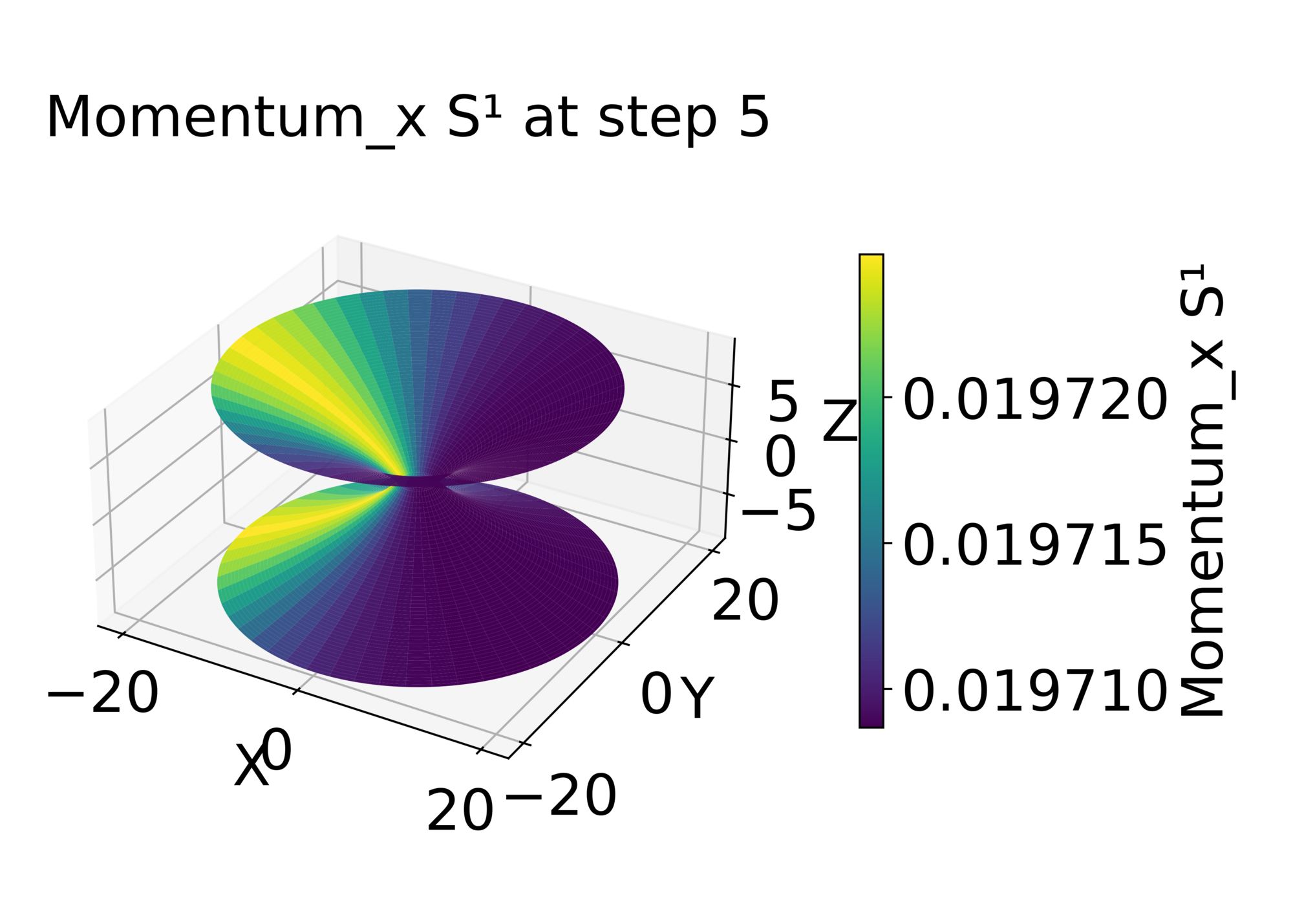}
    \includegraphics[width=0.3\textwidth]{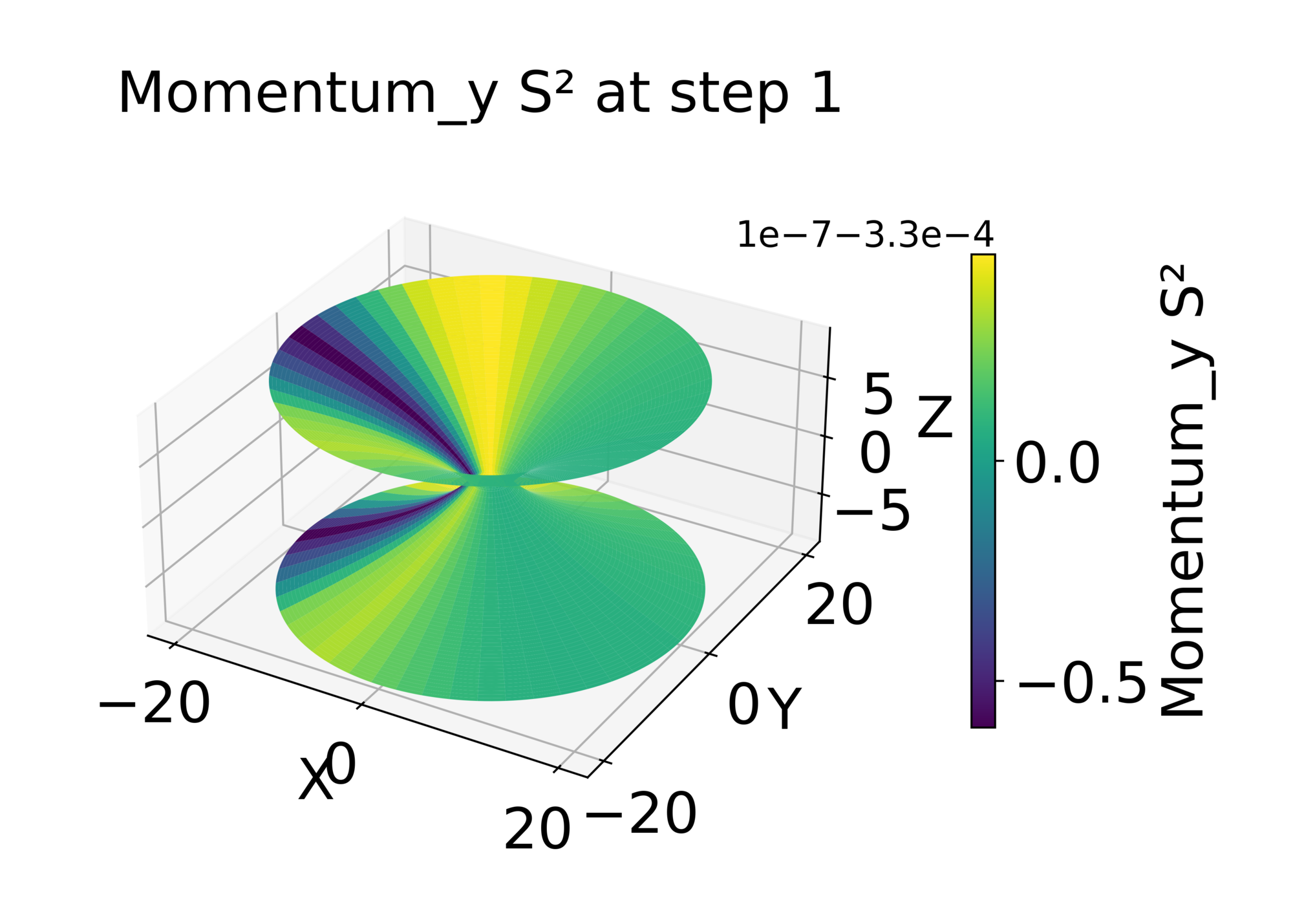}
    \includegraphics[width=0.3\textwidth]{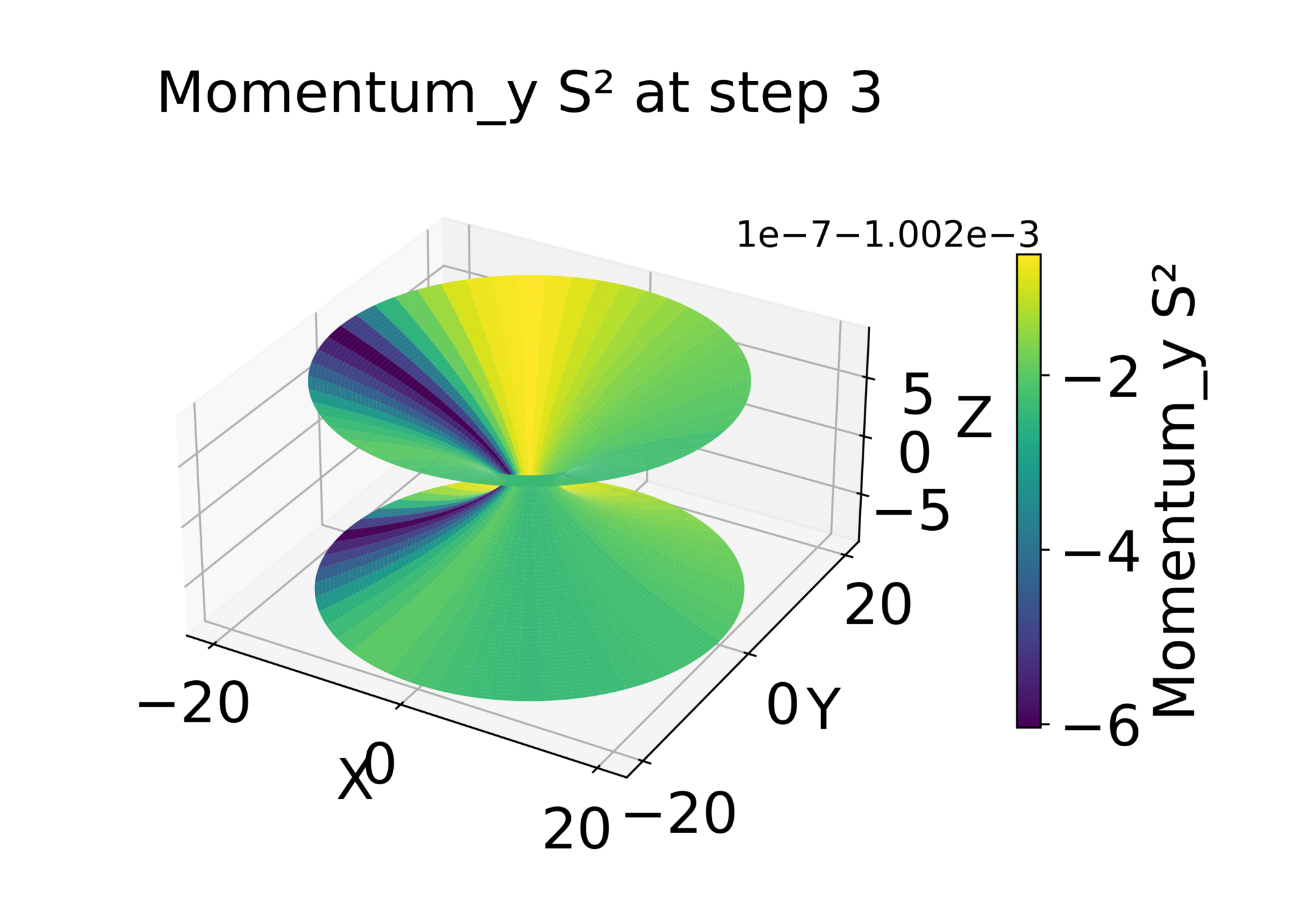}
    \includegraphics[width=0.3\textwidth]{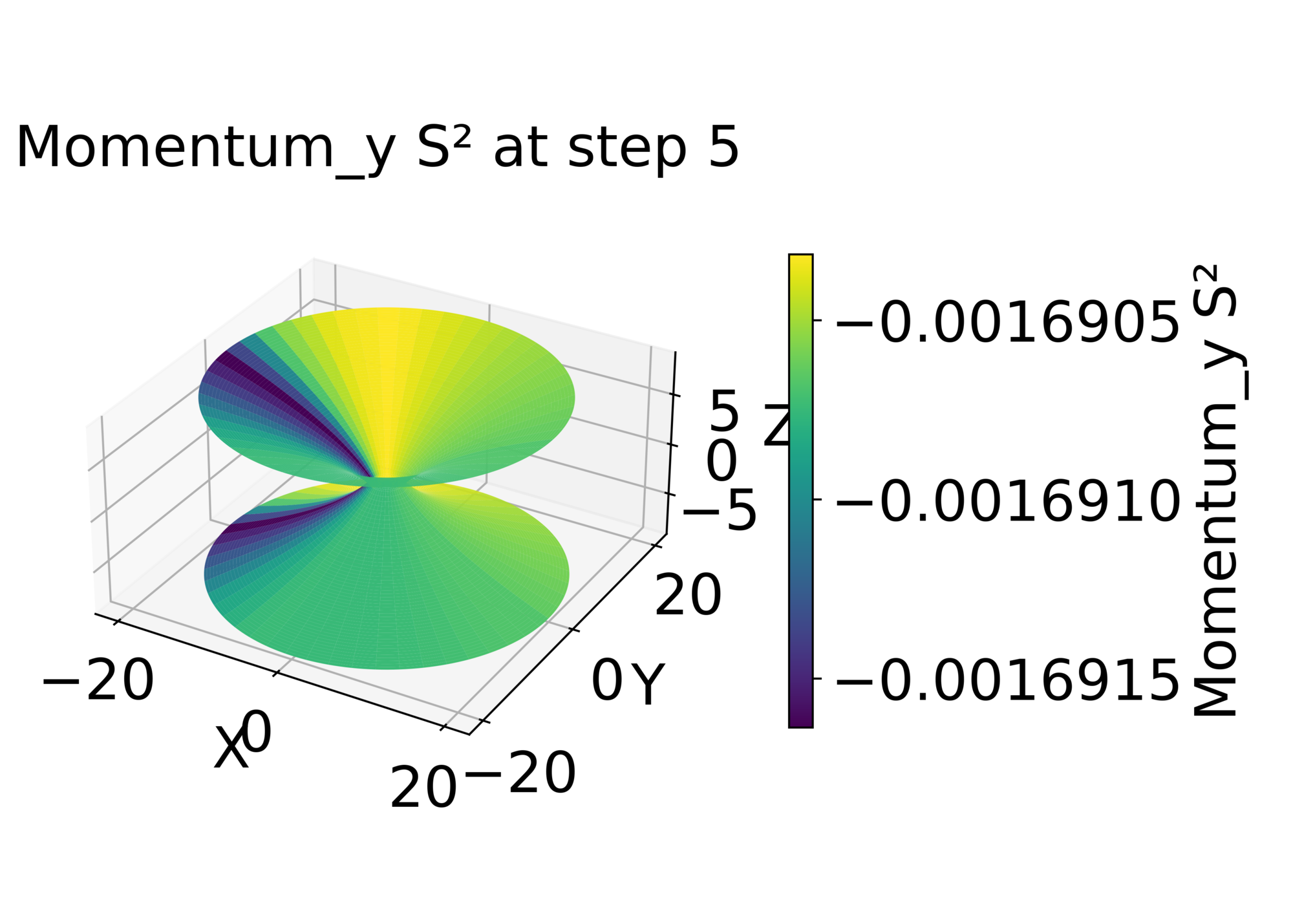}
    \includegraphics[width=0.3\textwidth]{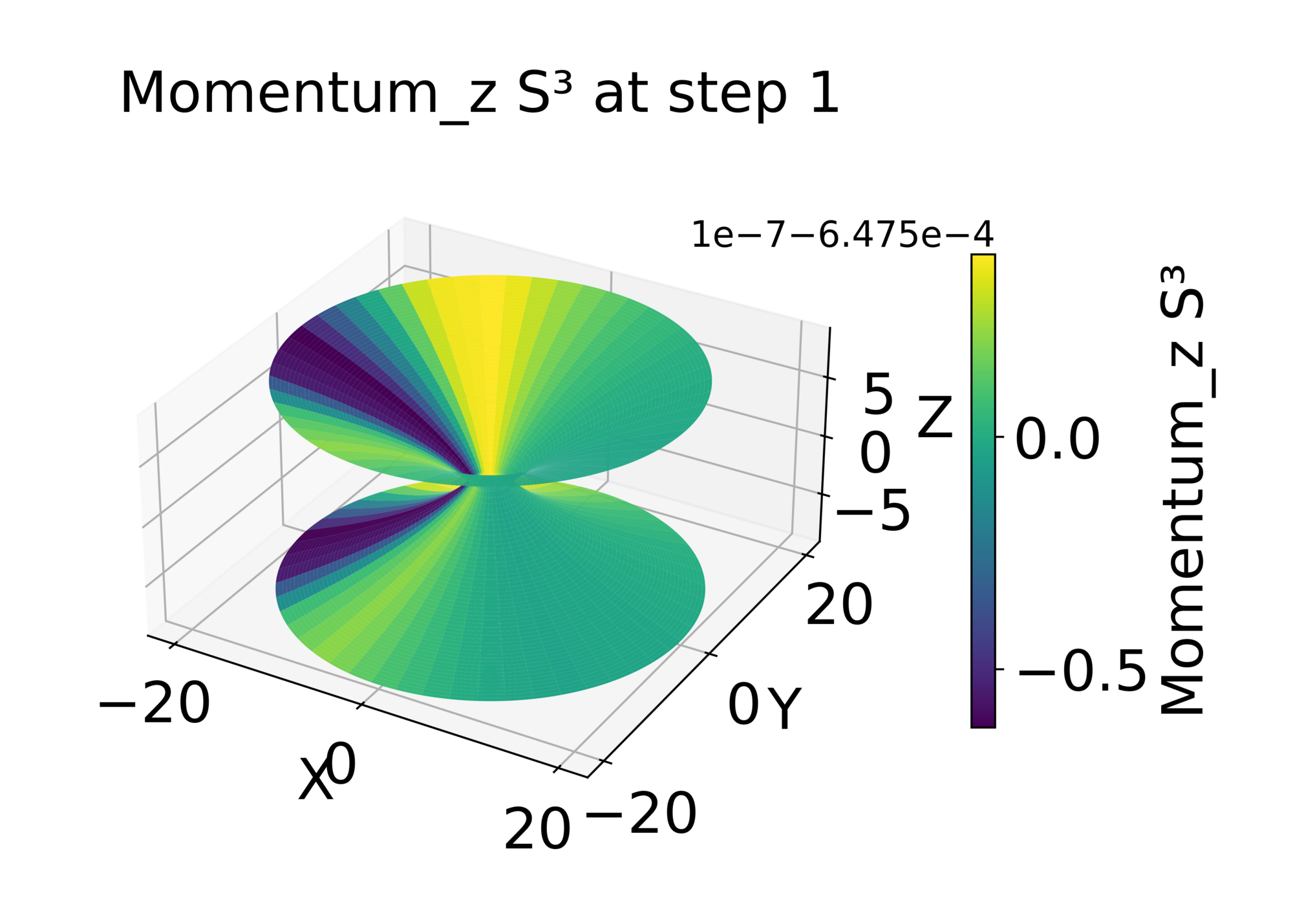}
    \includegraphics[width=0.3\textwidth]{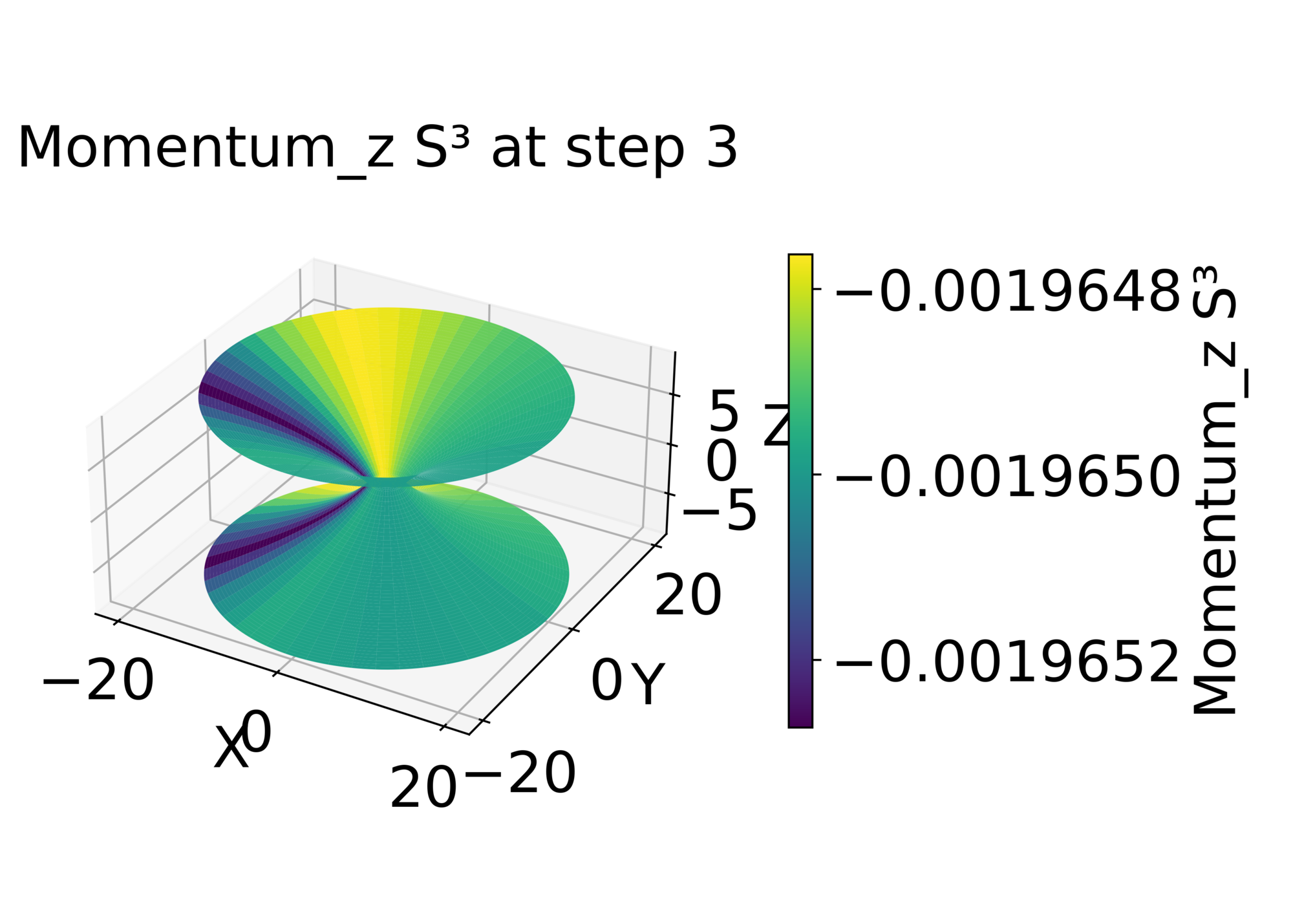}
    \includegraphics[width=0.3\textwidth]{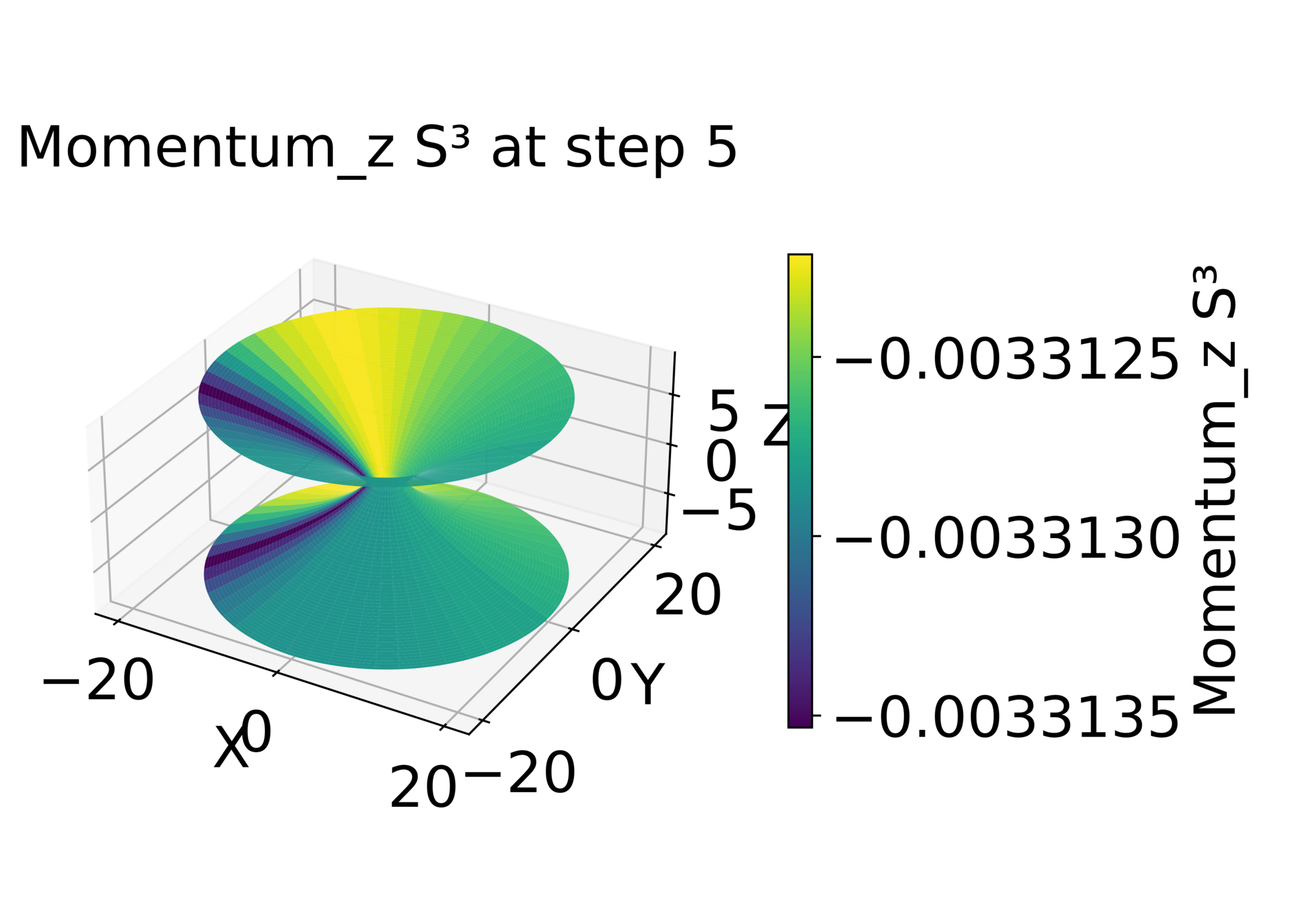}
    \caption{Equatorial embedding diagrams of the Kerr black hole event horizon \( r_h = M + \sqrt{M^2 - a^2} \), shown as a Flamm paraboloid in fictitious 3D Euclidean space. The surface is parametrized by \( X = r_h \cos\phi \), \( Y = r_h \sin\phi \), and the embedding height \( Z \) computed from the spatial metric. Colors indicate the value of each conserved variable of momentum along the equatorial ring \( \theta = \pi/2 \) at the event horizon. Both upper \( +Z \) and lower \( -Z \) surfaces are shown for symmetry.}
    \label{fig:TheM}
\end{figure}

We extract one of the conserved GRMHD variables in Figure~\ref{fig:DE} and ~\ref{fig:TheM} at different time step, 
  \[
    U = \bigl[D,\;S_x,\;S_y,\;S_z,\;\tau,\;B_x,\;B_y,\;B_z\bigr]
    \quad\text{at}\quad r=r_{\rm h},\ \theta=\tfrac\pi2,
  \]
  as a function of $\varphi$.
We interpolate these values onto a fine $\varphi$-grid and use them to color the embedding surface.
Thus each colored Flamm surface shows the angular variation of the chosen variable \emph{on the event horizon}.
Repeating over several time steps illustrates how the horizon‐crossing flow, momentum or magnetic field structure evolves.

Regions of high/low value in the color map correspond to hotspots or deficits of density, momentum, energy or magnetic field around the horizon.
The embedding surface makes the radial geometry explicit, while the color encodes the fluid/magnetic structure at the horizon.
By viewing both the “upper” and “lower” sheets, one visualizes the full 3D shape and field distribution of the equatorial slice.

The Figure ~\ref{fig:K} is 3D surface of the Kerr-Schild lapse, and Contour plots of the Kerr-Schild ADM split geometric quantities in Figures~\ref{fig:kerr-adm-contours} and Table~\ref{tab:kerr-adm-fields}.

\paragraph*{The Entropy Curve Implies Stability and Convergence}

We define the \emph{discrete total entropy} at step \(n\) as
\[
E^n \;=\;
\sum_{i,j,k}
\underbrace{\eta\bigl(u^n_{i,j,k}\bigr)}_{\rho\,s}
\;\underbrace{\sqrt{\gamma(r_i,\theta_j)}}_{\text{metric factor}}
\;\Delta V_{i,j,k},
\]
where
\[
\eta(u)=\rho\,s,\qquad
s=\frac{1}{\gamma-1}\ln\!\Bigl(\frac{p}{\rho^\gamma}\Bigr),
\]
\(\sqrt\gamma\) is the determinant factor of the Kerr–Schild spatial metric, 
and \(\Delta V_{i,j,k}=(\Delta r_i)(\Delta\theta_j)(\Delta\phi_k)\) is the cell volume.

\medskip

\begin{figure}[h]
    \centering
    \includegraphics[width=0.45\textwidth]{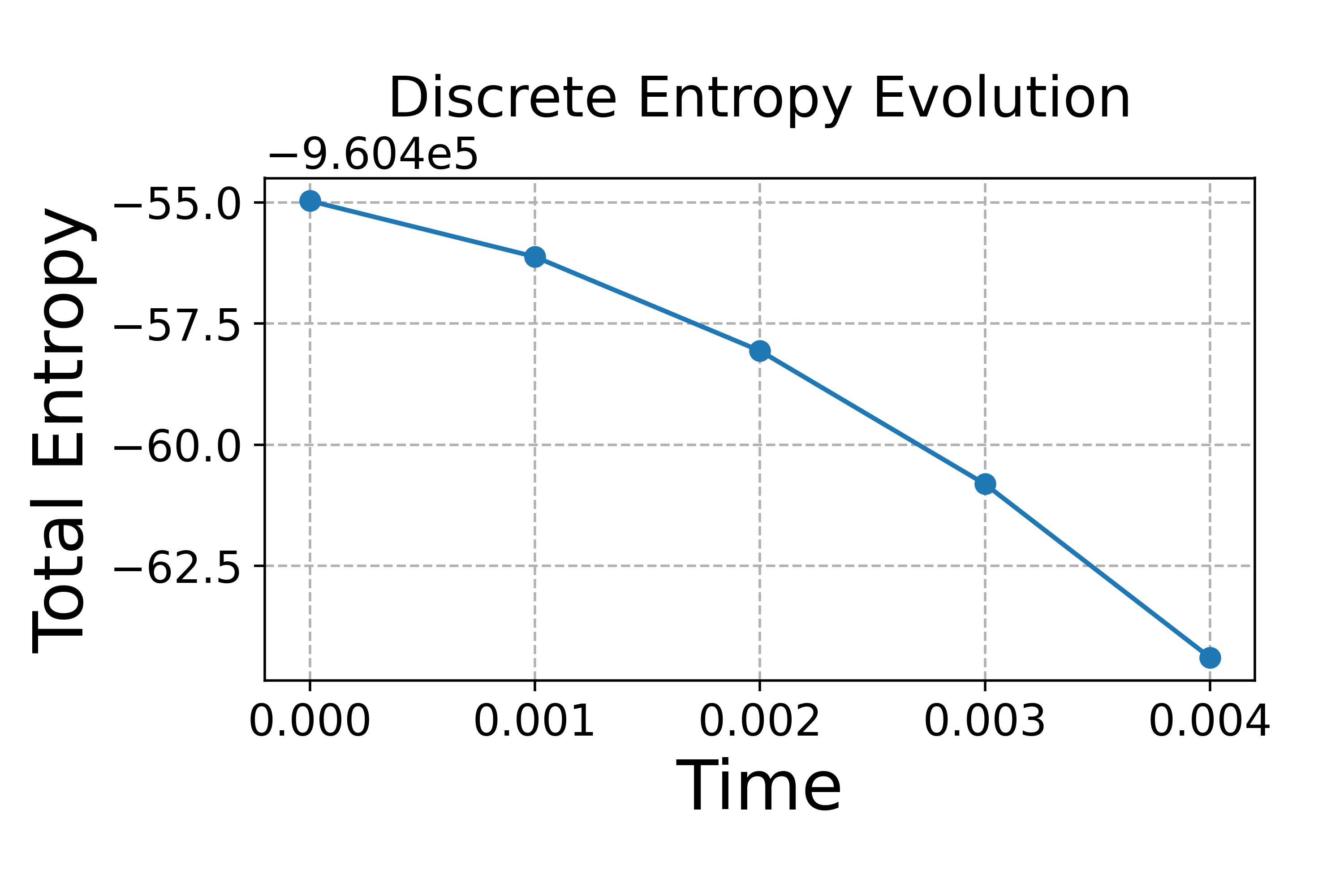}
    \caption{The Entropy Curve}
    \label{fig:Entrpy}
\end{figure}

\noindent\textbf{Entropy Stability.}
The plot ~\ref{fig:Entrpy} shows
\[
E^0 \;\ge\; E^1 \;\ge\; E^2 \;\ge\;\cdots,
\]
i.e.\ for every \(n\),
\[
E^{n+1} \;\le\; E^n.
\]
This \emph{discrete entropy inequality} means the scheme never produces
non-physical entropy growth, so the numerical solution is bounded
in the entropy norm and hence \emph{stable}.

\noindent\textbf{Consistency.}
\begin{itemize}
  \item The Chebyshev–Fourier spatial discretization has truncation error \(\to0\)
        exponentially fast for smooth solutions.
  \item The RK4 time integrator has global error \(O(\Delta t^4)\).
\end{itemize}

\medskip

\noindent\textbf{Convergence.}
\begin{itemize}
  \item By the \emph{Lax–Wendroff–Tadmor theorem} for nonlinear conservation laws:
        any scheme that is conservative, consistent, and entropy‐stable
        converges to the unique entropy solution.
  \item For linear problems, the \emph{Lax equivalence theorem} asserts:
        consistency plus stability implies convergence.
\end{itemize}

\medskip

\noindent\textbf{Conclusion.}
Monotonic decrease of \(E^n\) entropy stability together with spectral/RK4
consistency guarantees that the method is both \emph{stable} and
\emph{convergent} to the correct entropy‐satisfying solution.

\subsection{The 3D GRMHD Variable Visualization}

\bigskip

\subsubsection*{Coordinate System}

The code maps each simulation point from spherical Kerr–Schild coordinates $(r,\theta,\varphi)$ into standard Cartesian $\mathbb{R}^3$ via
\[
  \begin{aligned}
    X &= r \,\sin\theta\;\cos\varphi,\\
    Y &= r \,\sin\theta\;\sin\varphi,\\
    Z &= r \,\cos\theta.
  \end{aligned}
\]
Here:
\begin{itemize}
  \item $r$ is the areal radius (from just outside the horizon $r_h$ outwards),
  \item $\theta\in[0,\pi]$ is the polar angle measured from the rotation axis,
  \item $\varphi\in[0,2\pi)$ is the azimuthal angle.
\end{itemize}

\subsubsection*{The GRMHD Variable Visualization}

\begin{figure}[h]
  \centering
  \includegraphics[width=0.30\linewidth]{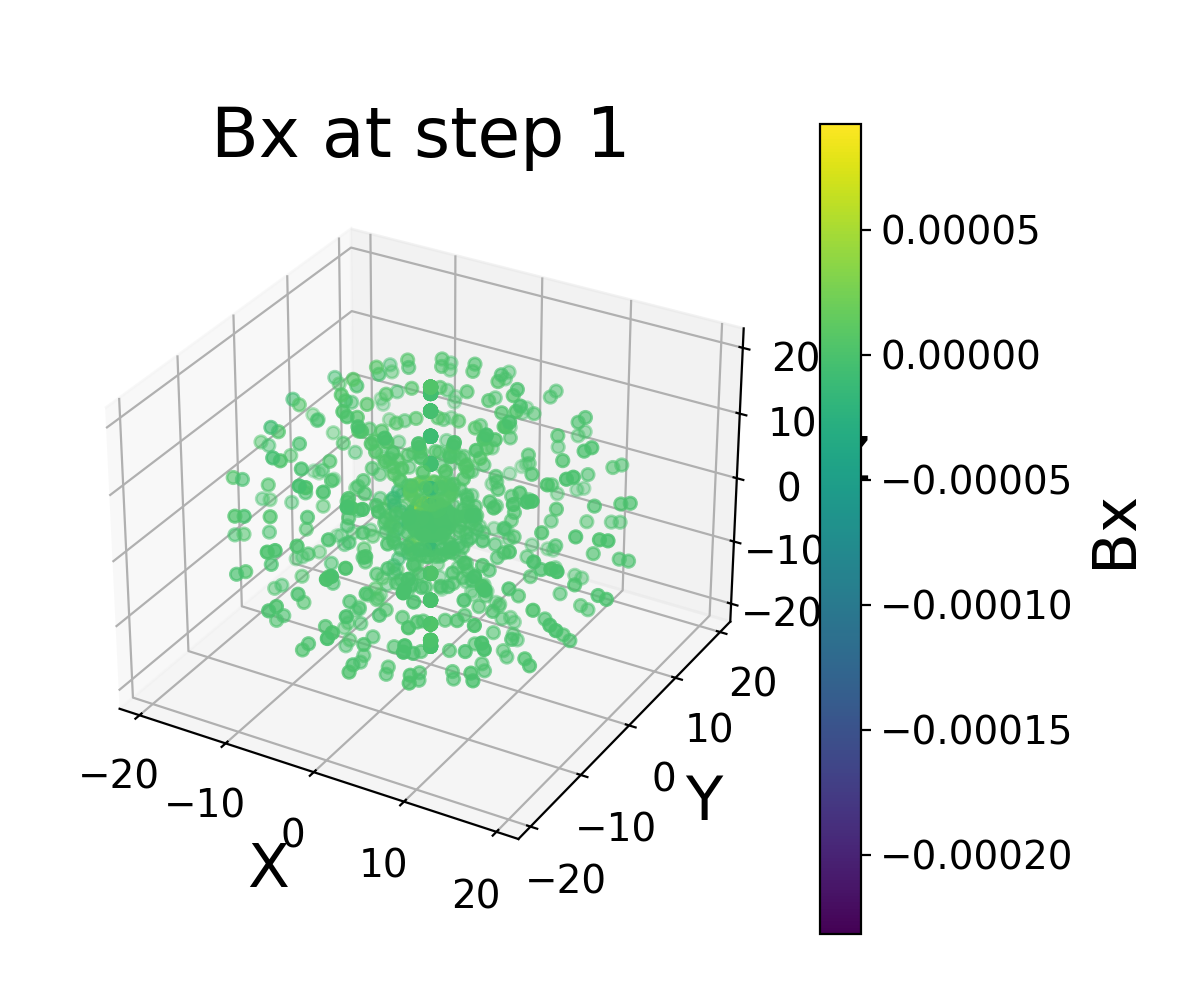}
  \includegraphics[width=0.30\linewidth]{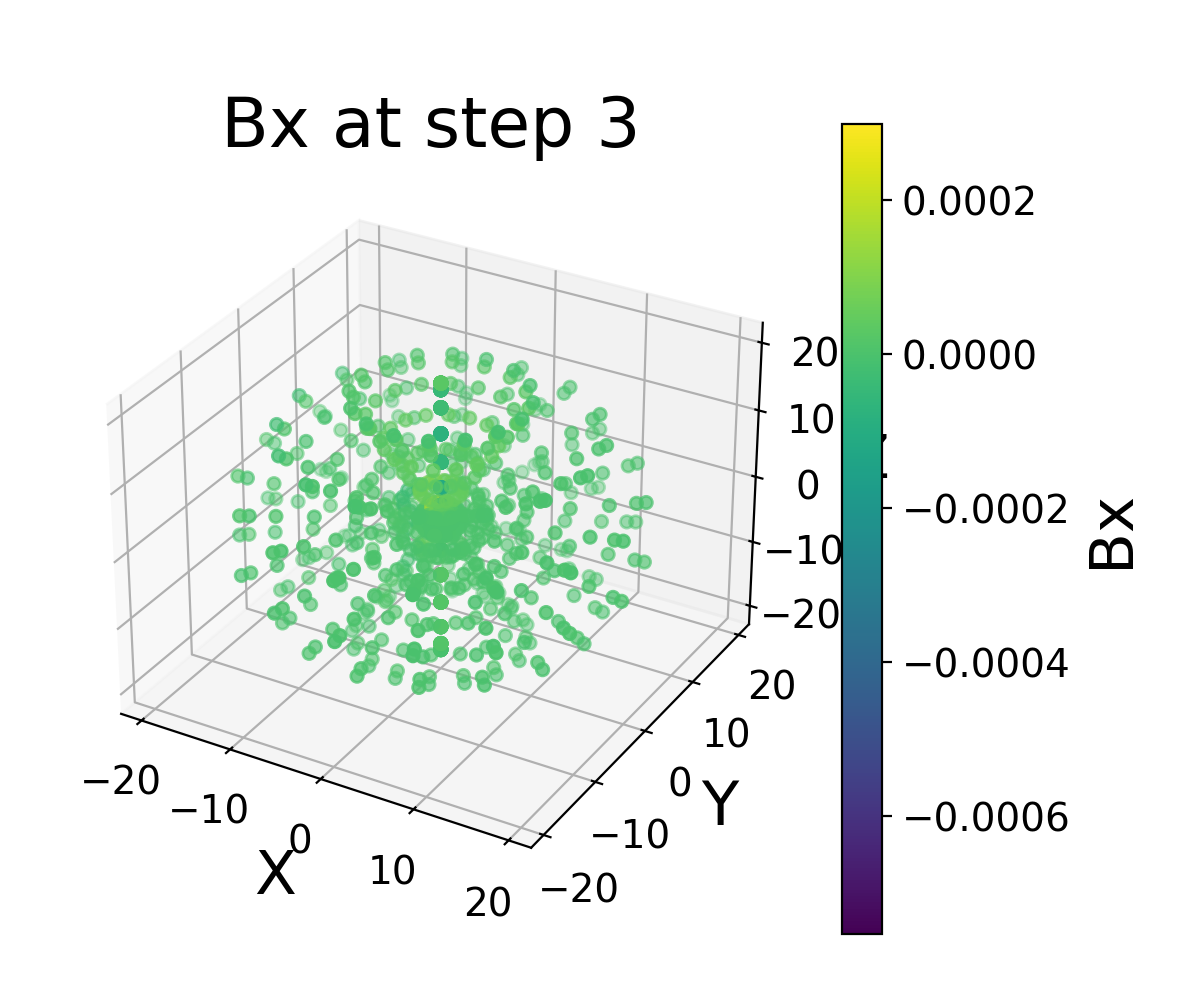}
  \includegraphics[width=0.30\linewidth]{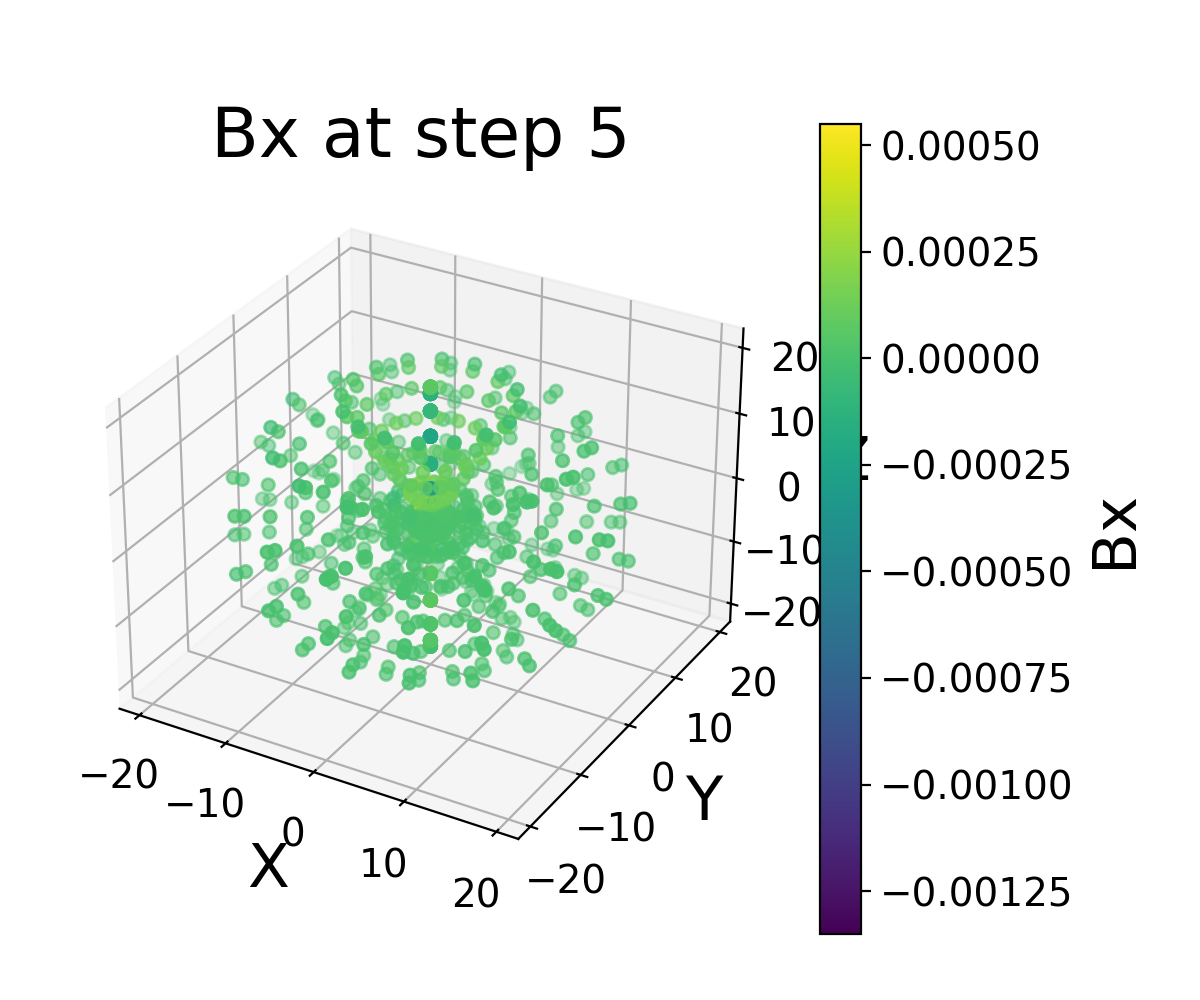}
  \includegraphics[width=0.30\linewidth]{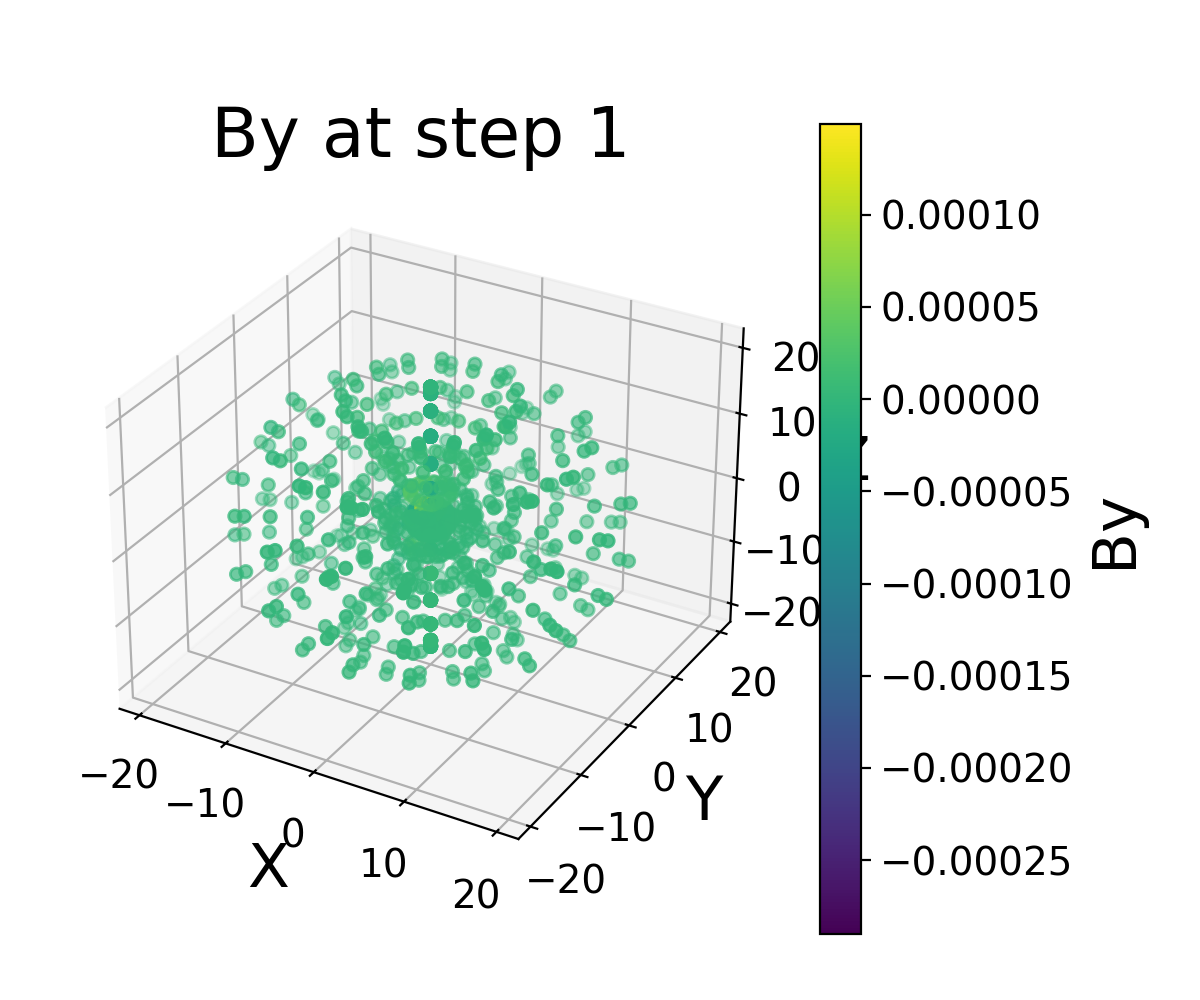}
  \includegraphics[width=0.30\linewidth]{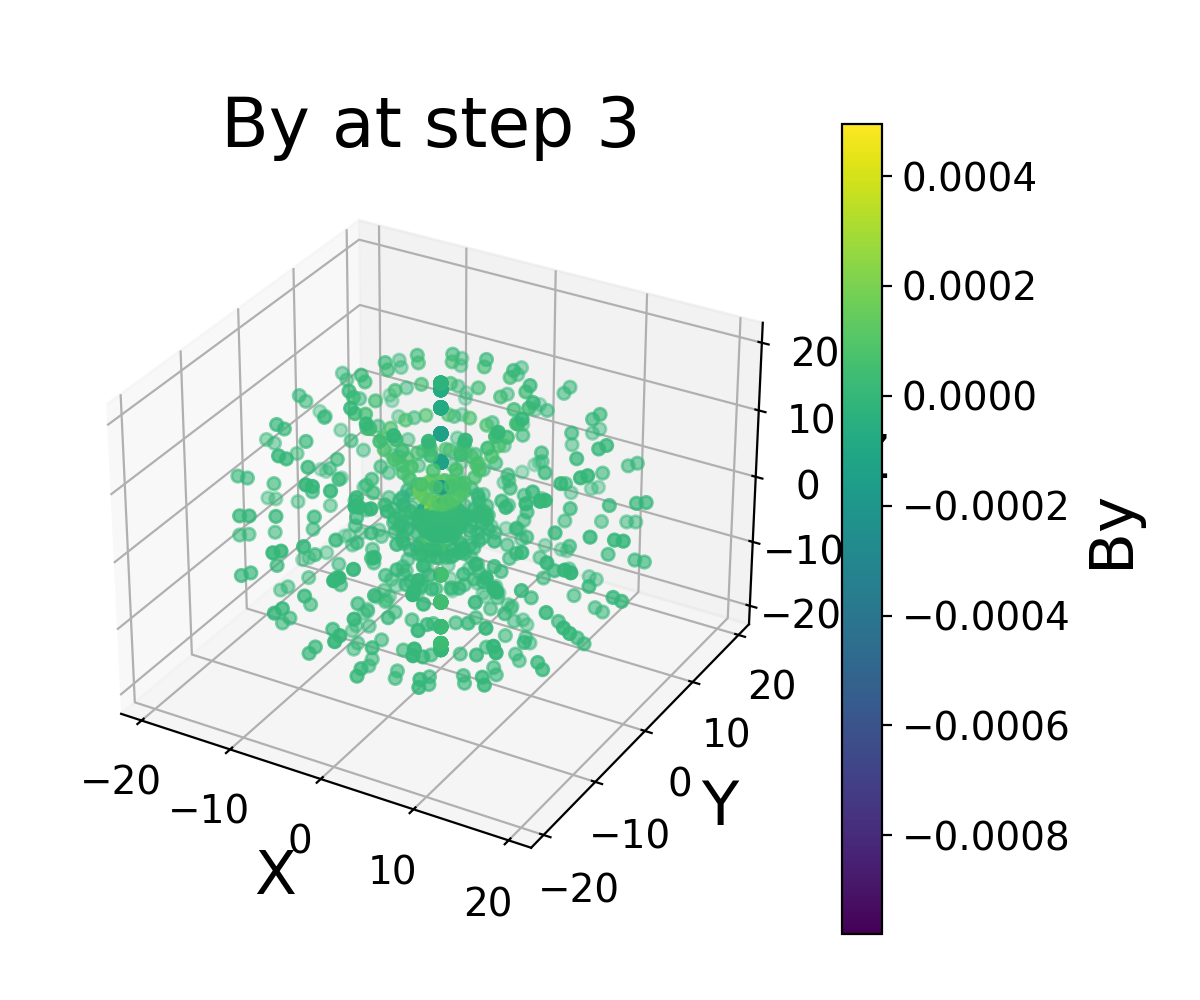}
  \includegraphics[width=0.30\linewidth]{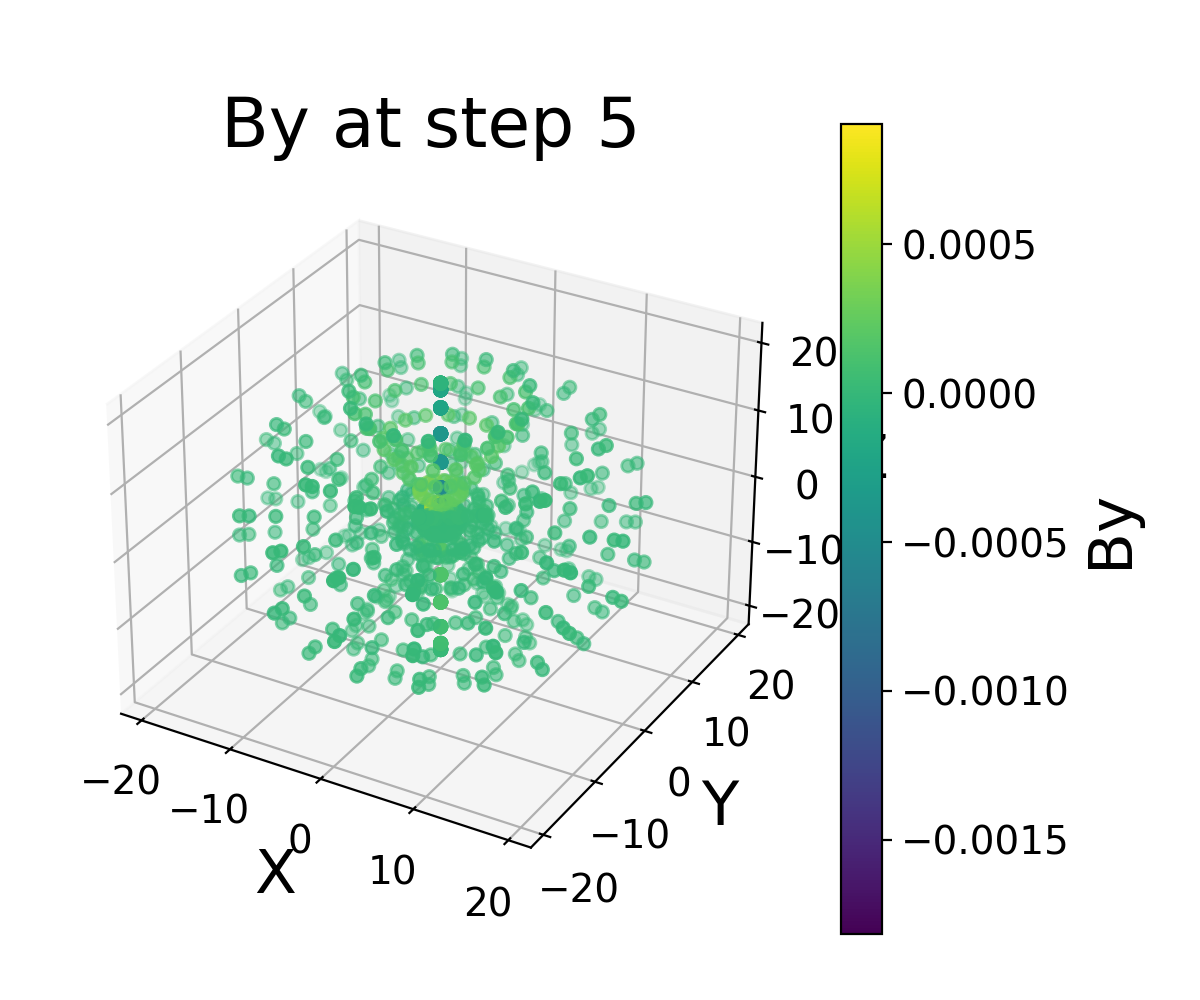}
  \includegraphics[width=0.30\linewidth]{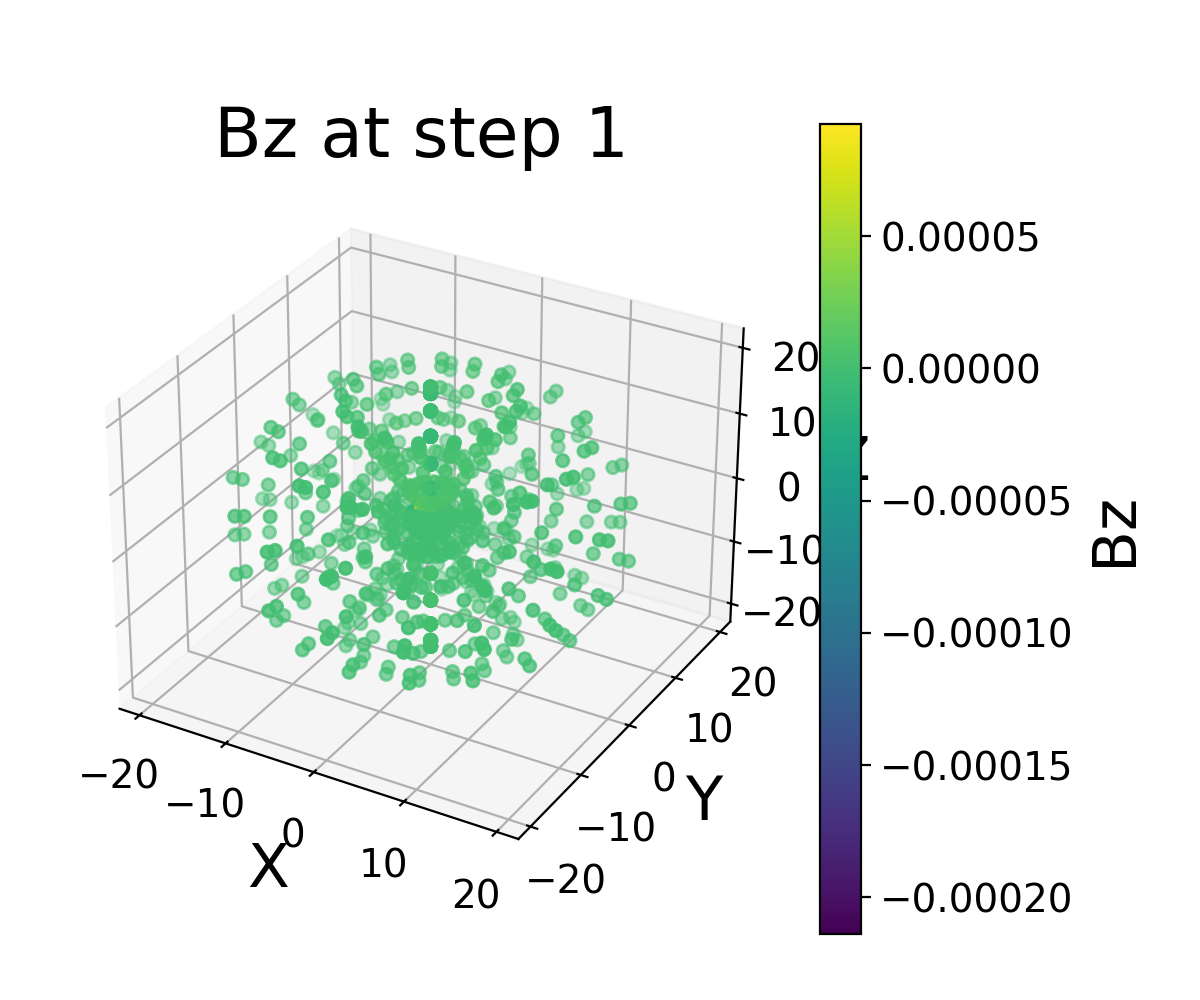}
  \includegraphics[width=0.30\linewidth]{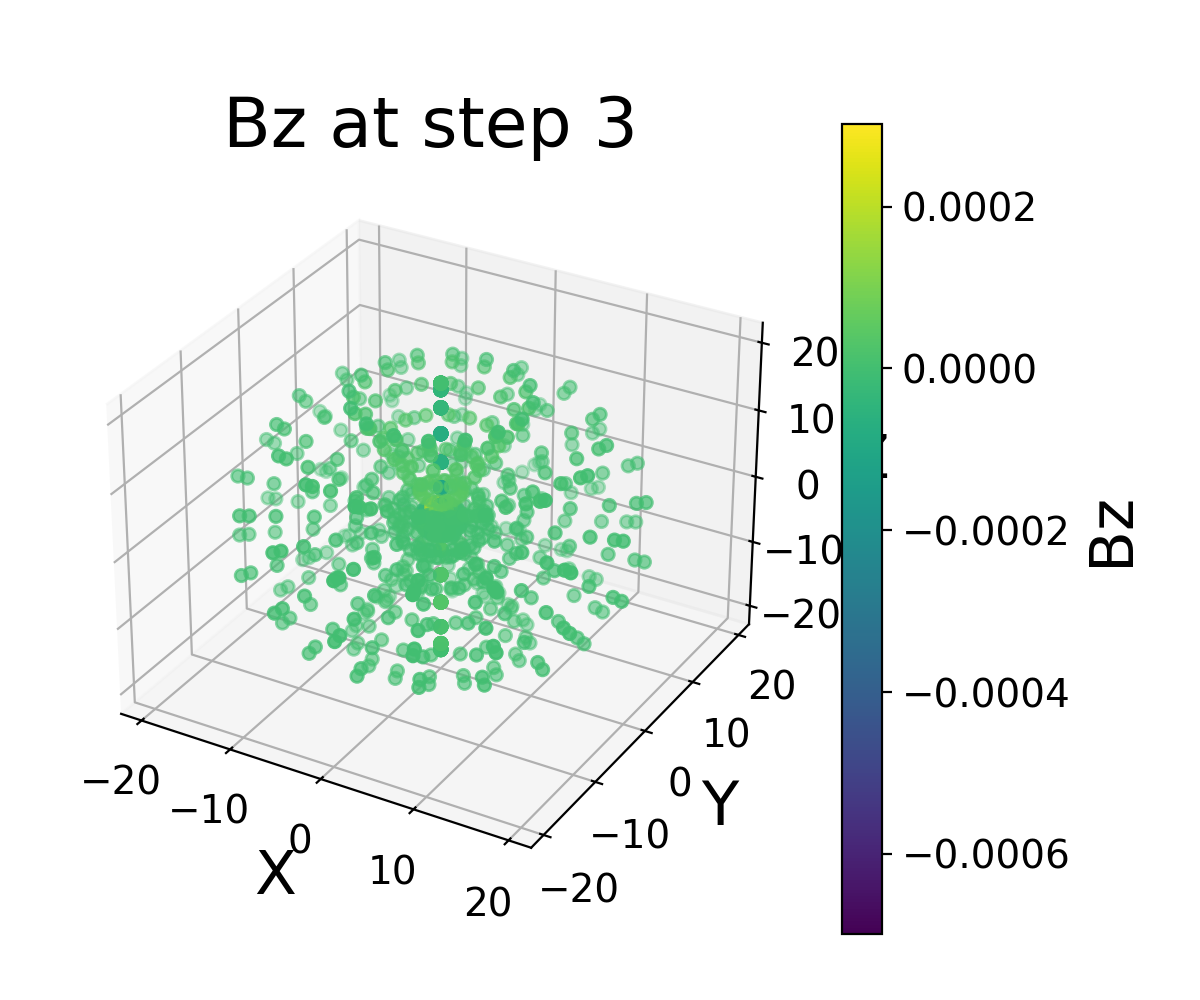}
  \includegraphics[width=0.30\linewidth]{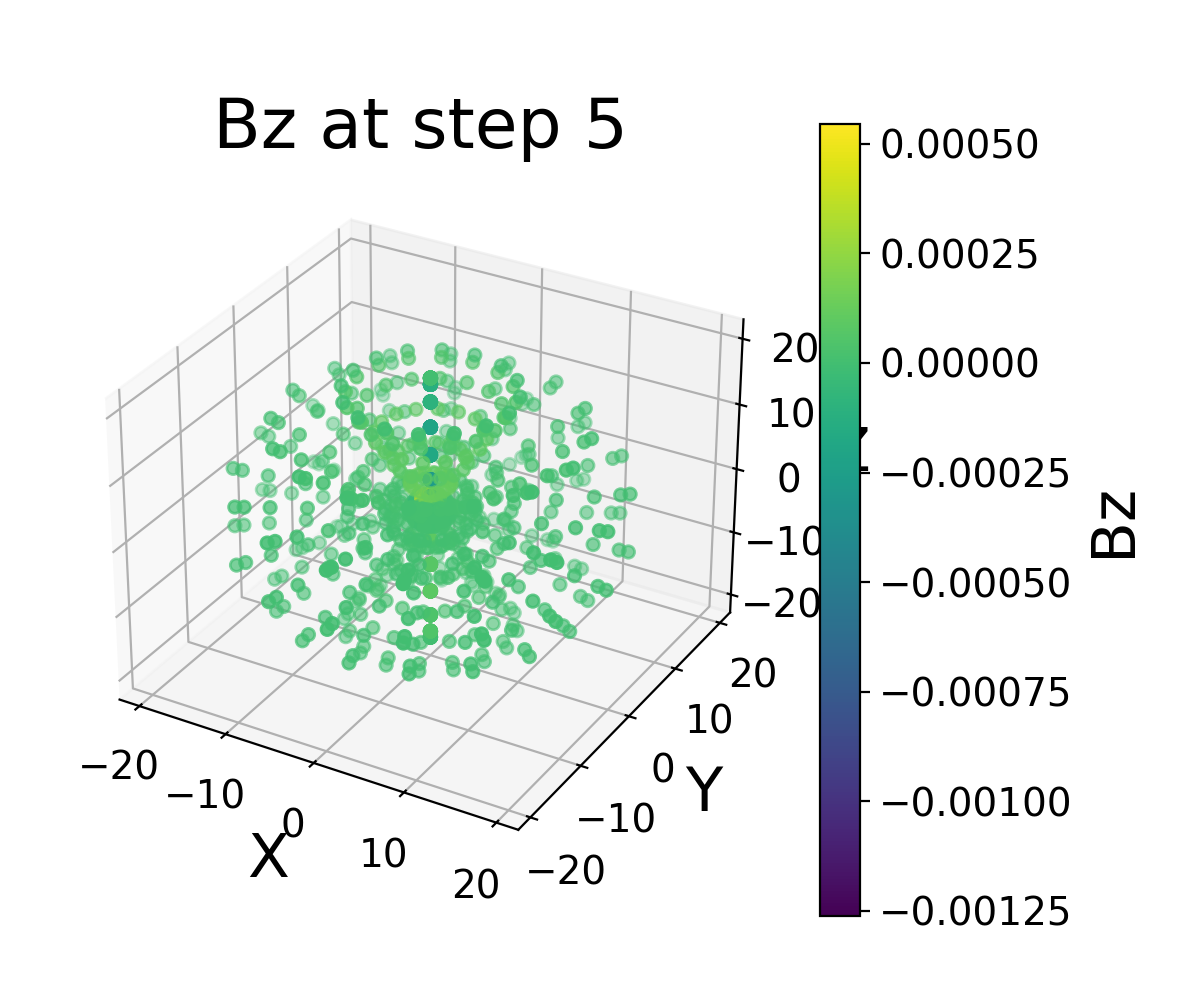}
  \caption{The 3D scatter plots of $B_x$, $B_y$, and $B_z$  at steps 1, 3 and 5 respectively.}
  \label{fig:B}
\end{figure}

\begin{figure}[h]
  \centering
  \includegraphics[width=0.30\linewidth]{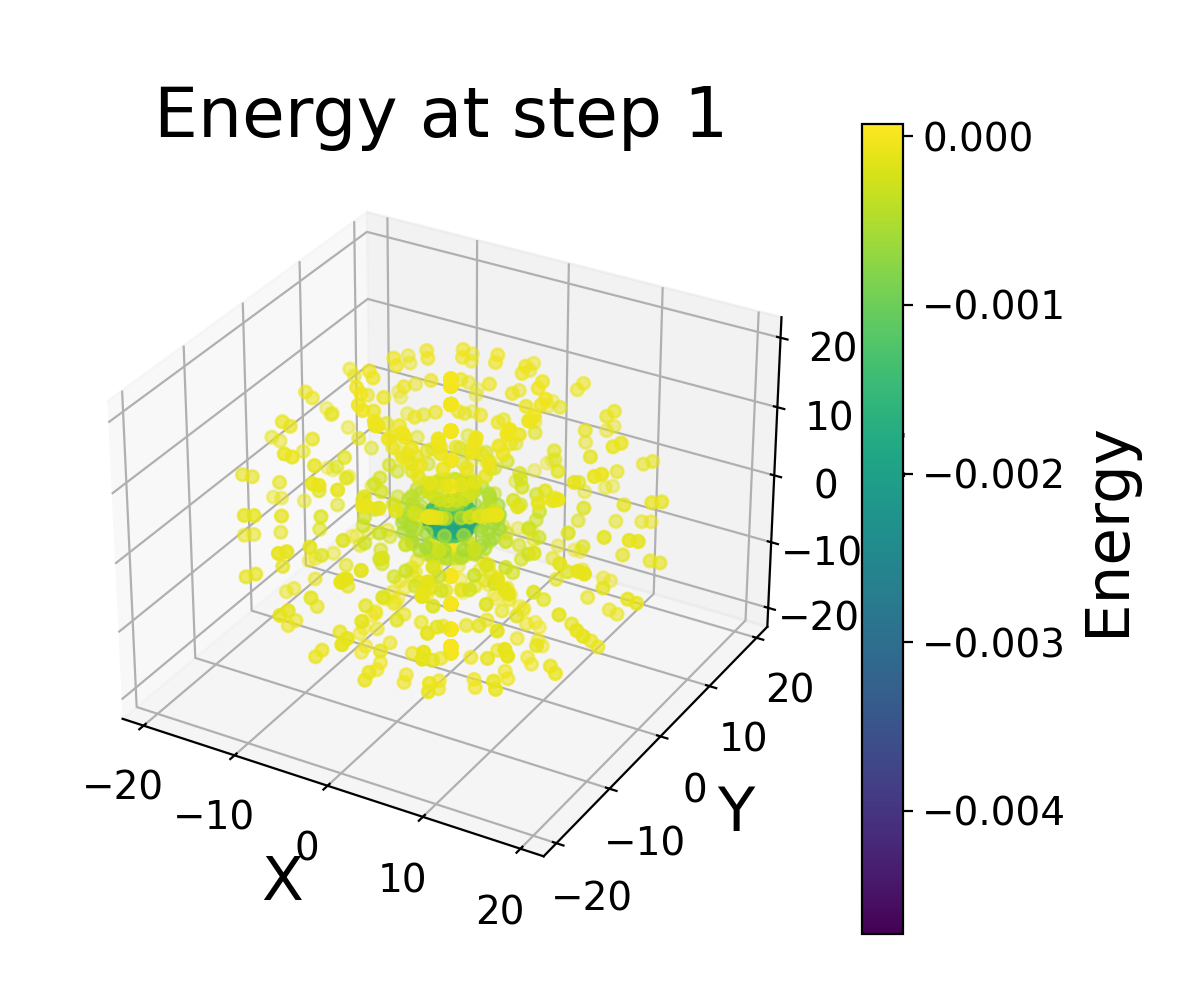}
  \includegraphics[width=0.30\linewidth]{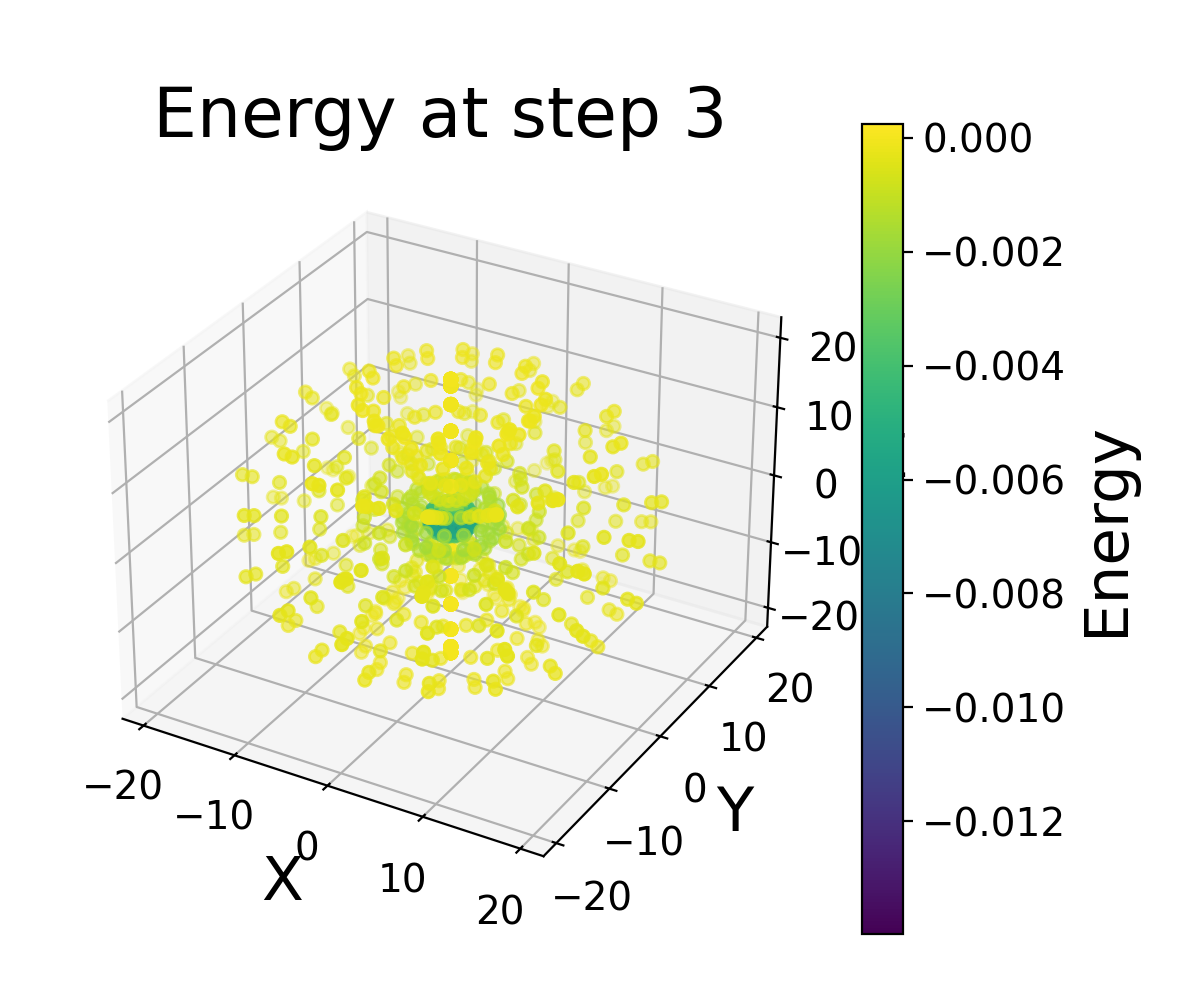}
  \includegraphics[width=0.30\linewidth]{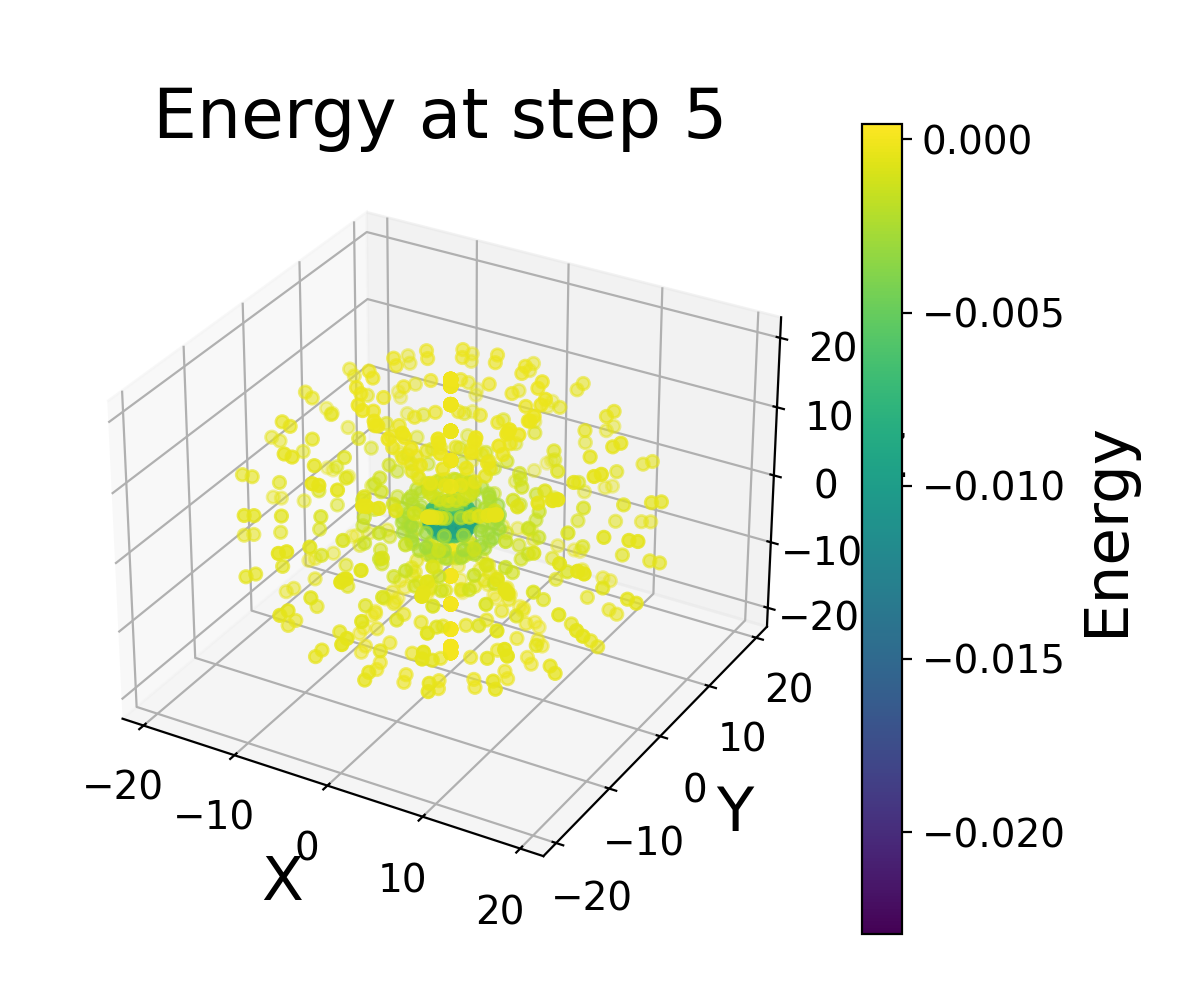}
  \caption{The 3D scatter plots of energy at steps 1, 3 and 5 respectively.}
  \label{fig:E}
\end{figure}

The script scatters all grid points $X,Y,Z$ colored by a chosen conserved variable at selected time steps 
\[
  U = \bigl[D,\,S_x,\,S_y,\,S_z,\,\tau,\,B_x,\,B_y,\,B_z\bigr],
\]
where, for example, $\tau$ is the energy density.  Each subplot in Figure ~\ref{fig:B}, ~\ref{fig:E} and ~\ref{fig:m}  shows one variable’s value at every point in 3D, giving a volumetric sense of how density, momentum, and magnetic‐field components are distributed in the Kerr background.


\subsection{The Results of Numerical Implementation of Coupled GRMHD–Einstein Evolution}
In the coupled version, recovered primitive variables density, velocity, internal energy, magnetic field are assembled into the stress–energy tensor $T^{\mu\nu}$, and an XCTS solver iteratively satisfies the Hamiltonian and momentum constraints to generate initial data for the spatial metric $\gamma_{ij}$ and extrinsic curvature $K_{ij}$. At each time step, the code alternates between a half‐step GRMHD update including divergence cleaning and primitive recovery to compute $T^{\mu\nu}$ and a BSSN update that integrates the conformal metric, extrinsic curvature, conformal factor, trace of $K$, and Gamma variables, with gauge conditions applied after each metric evolution. This strategy yields a fully self‐consistent evolution of both the fluid and the dynamical spacetime.

\begin{figure}[h]
    \centering
    \includegraphics[width=0.3\textwidth]{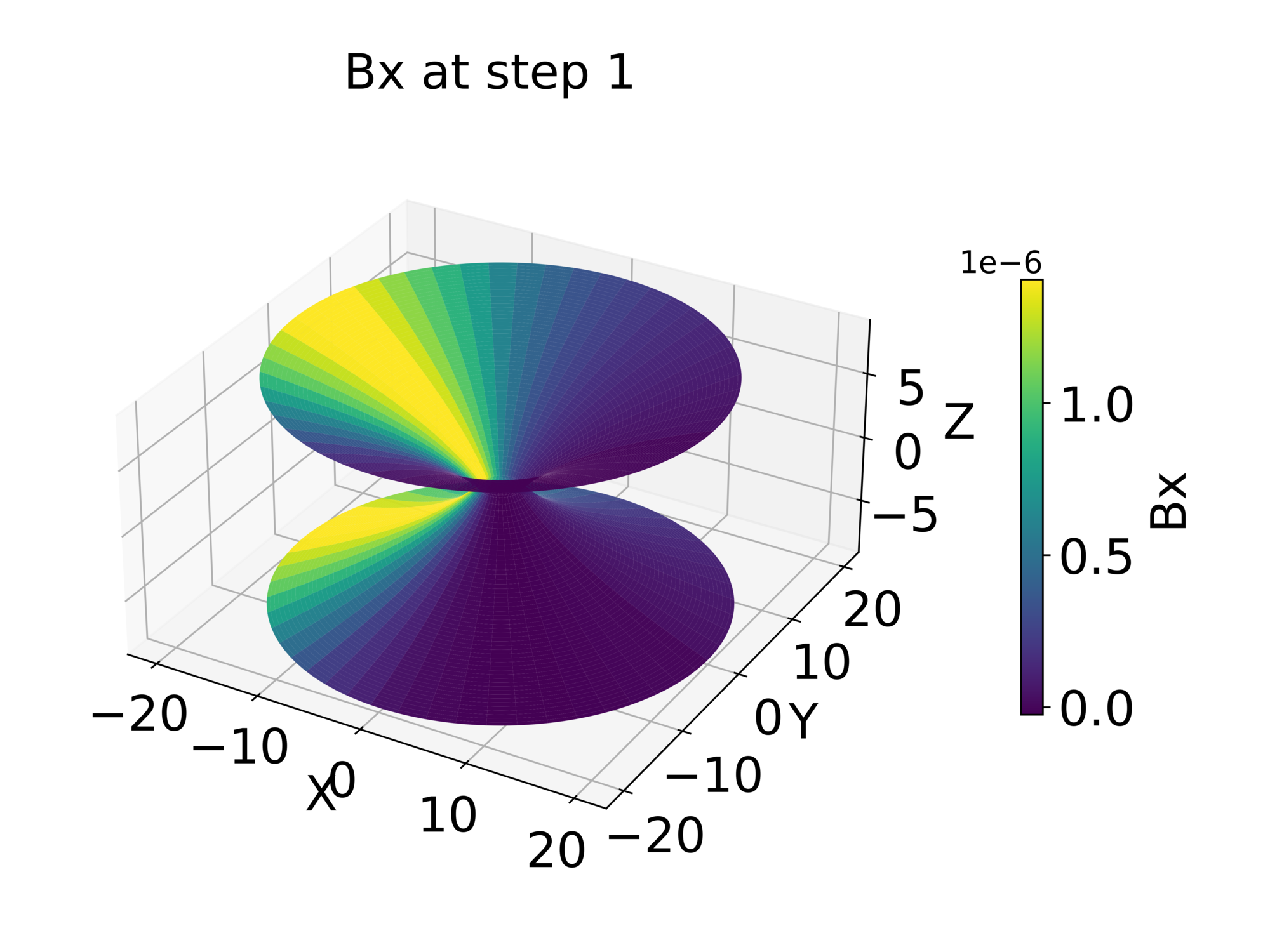}
    \includegraphics[width=0.3\textwidth]{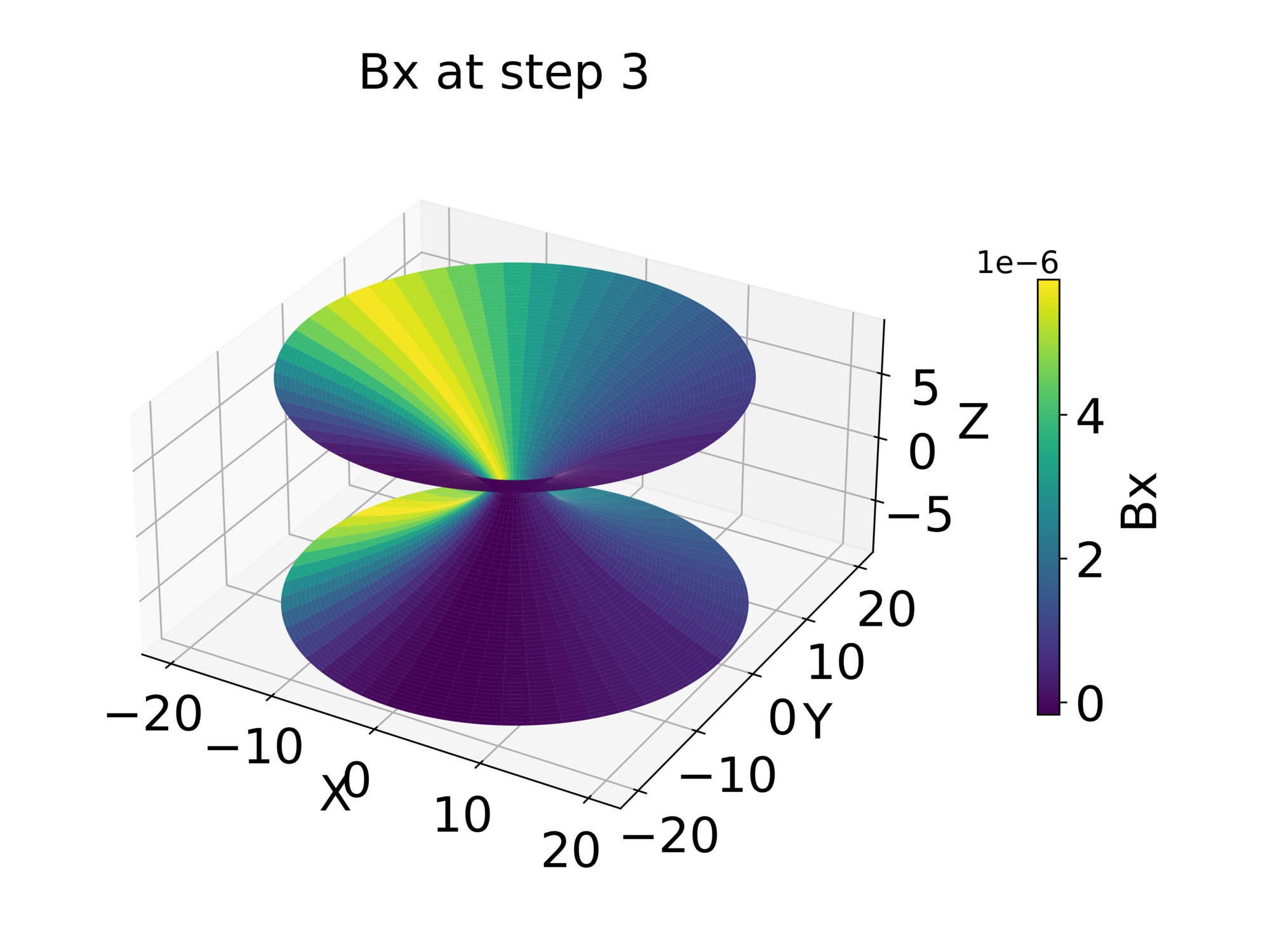}
    \includegraphics[width=0.3\textwidth]{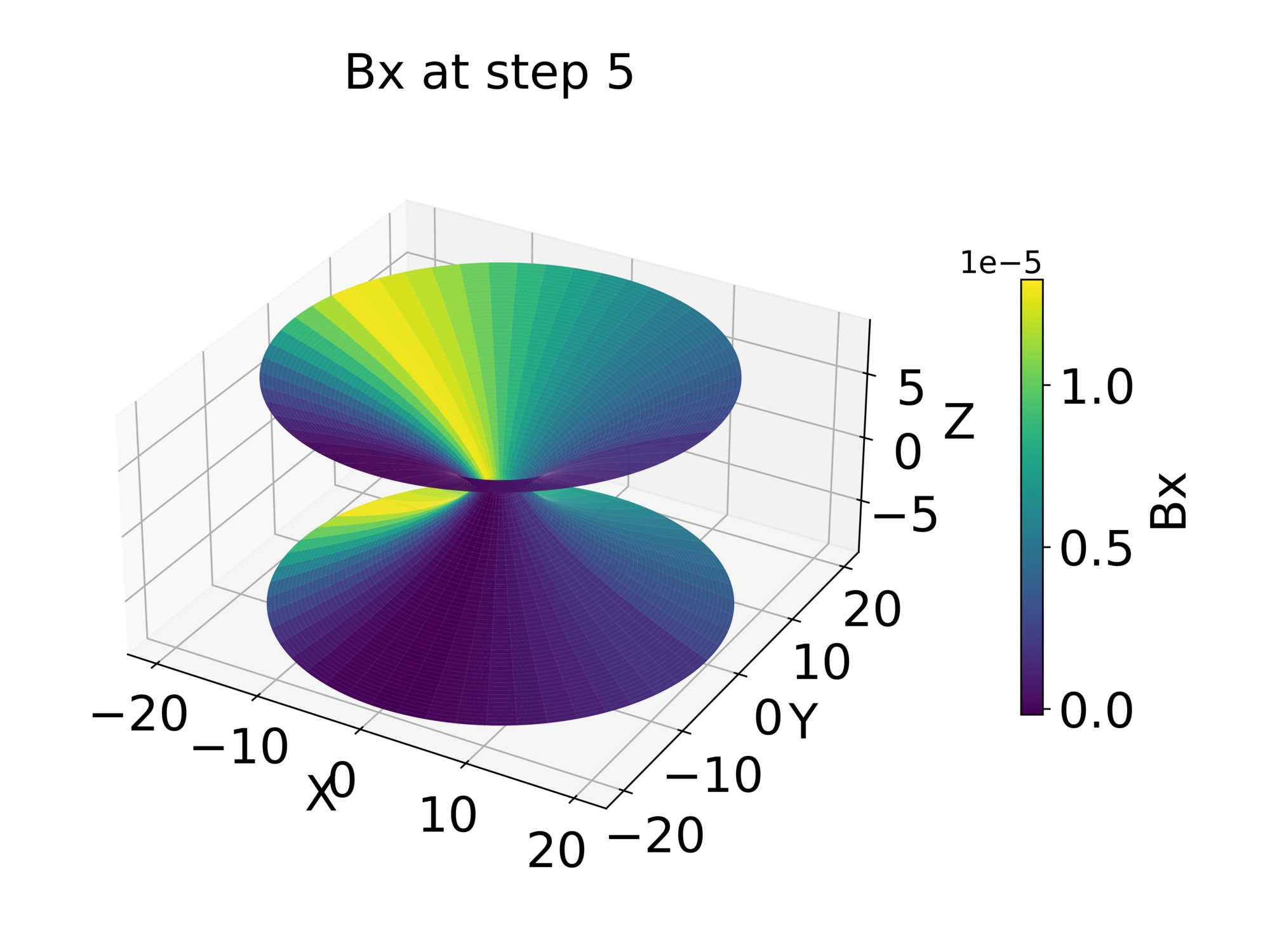}
    \includegraphics[width=0.3\textwidth]{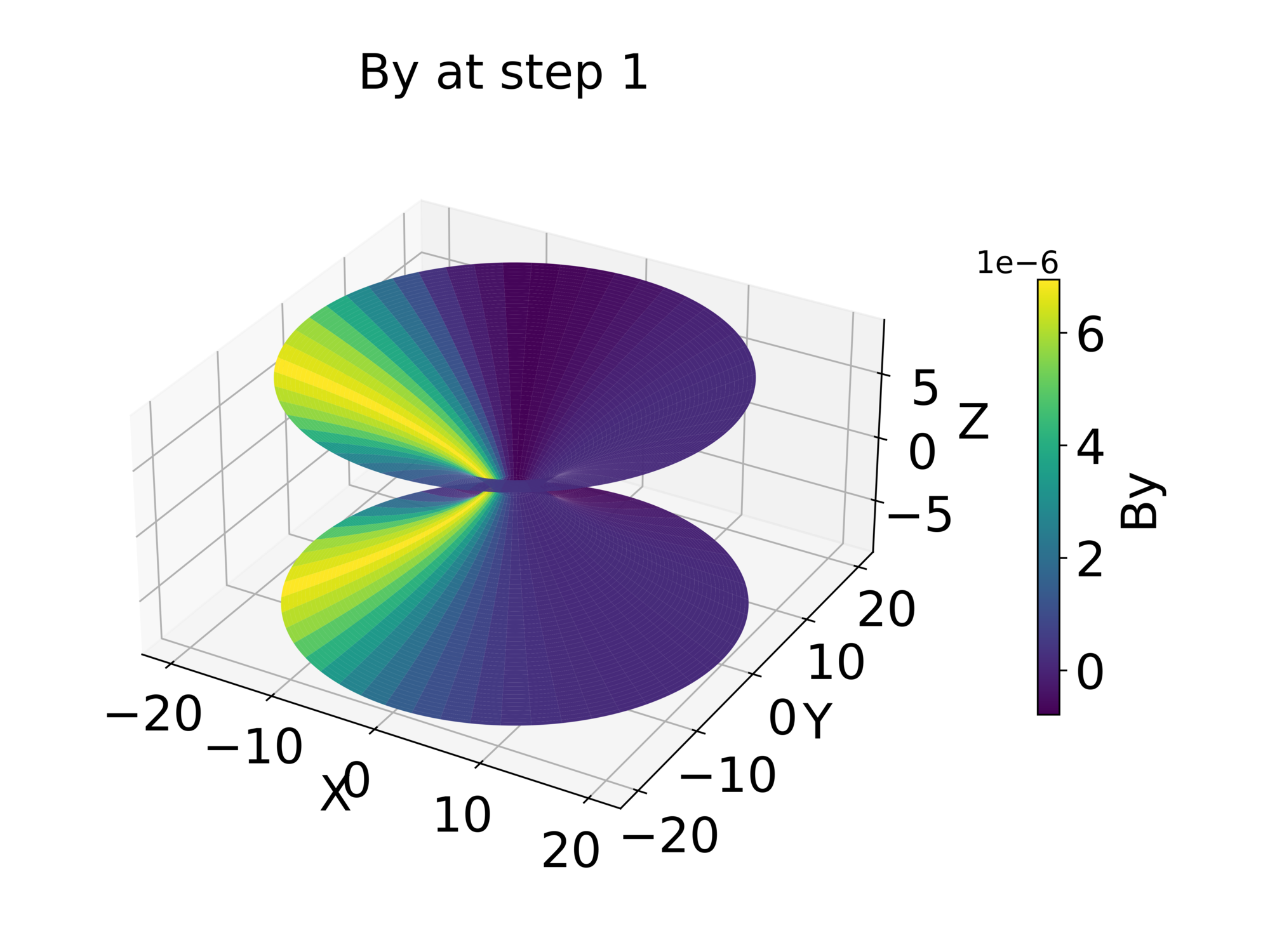}
    \includegraphics[width=0.3\textwidth]{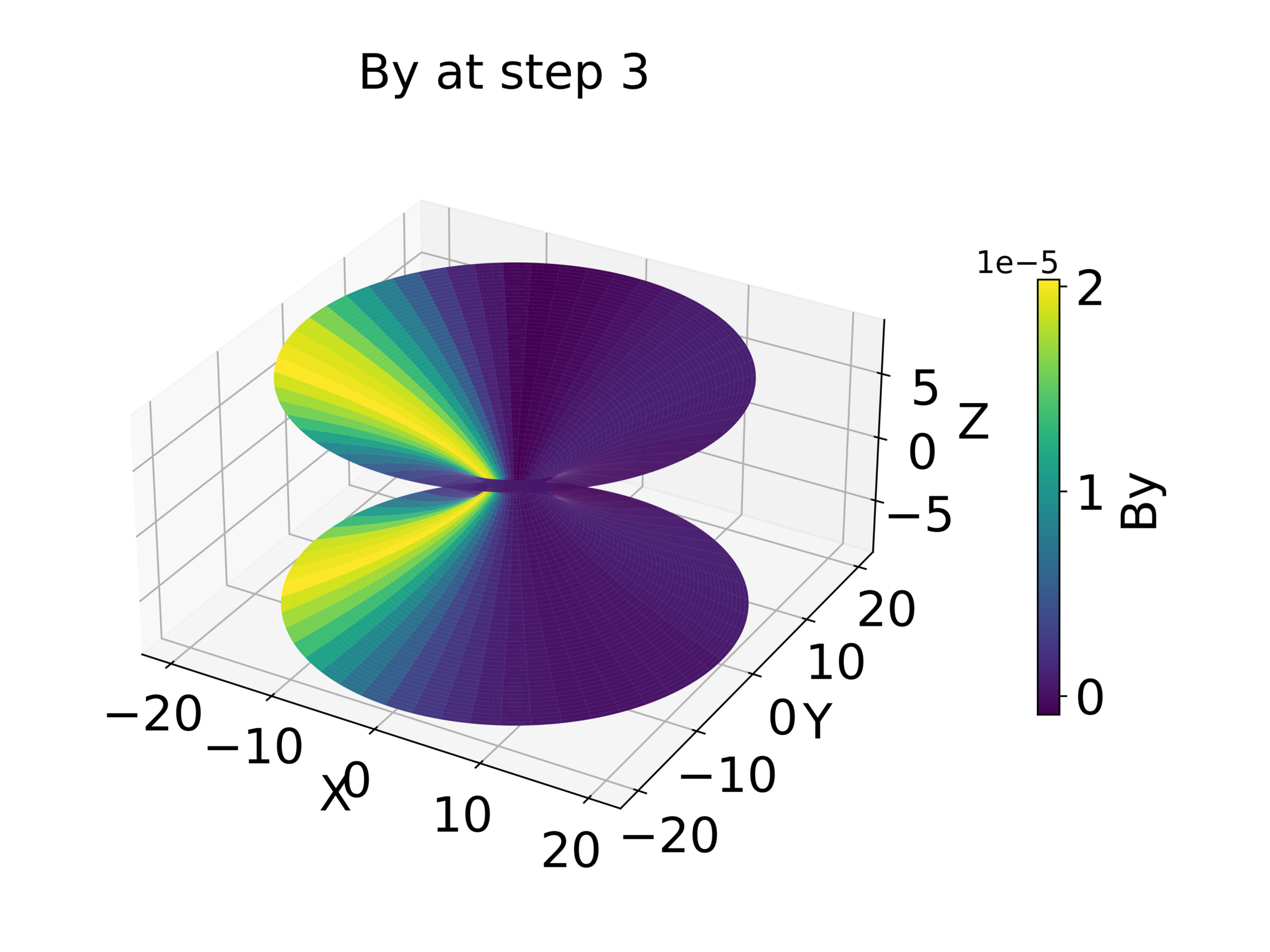}
    \includegraphics[width=0.3\textwidth]{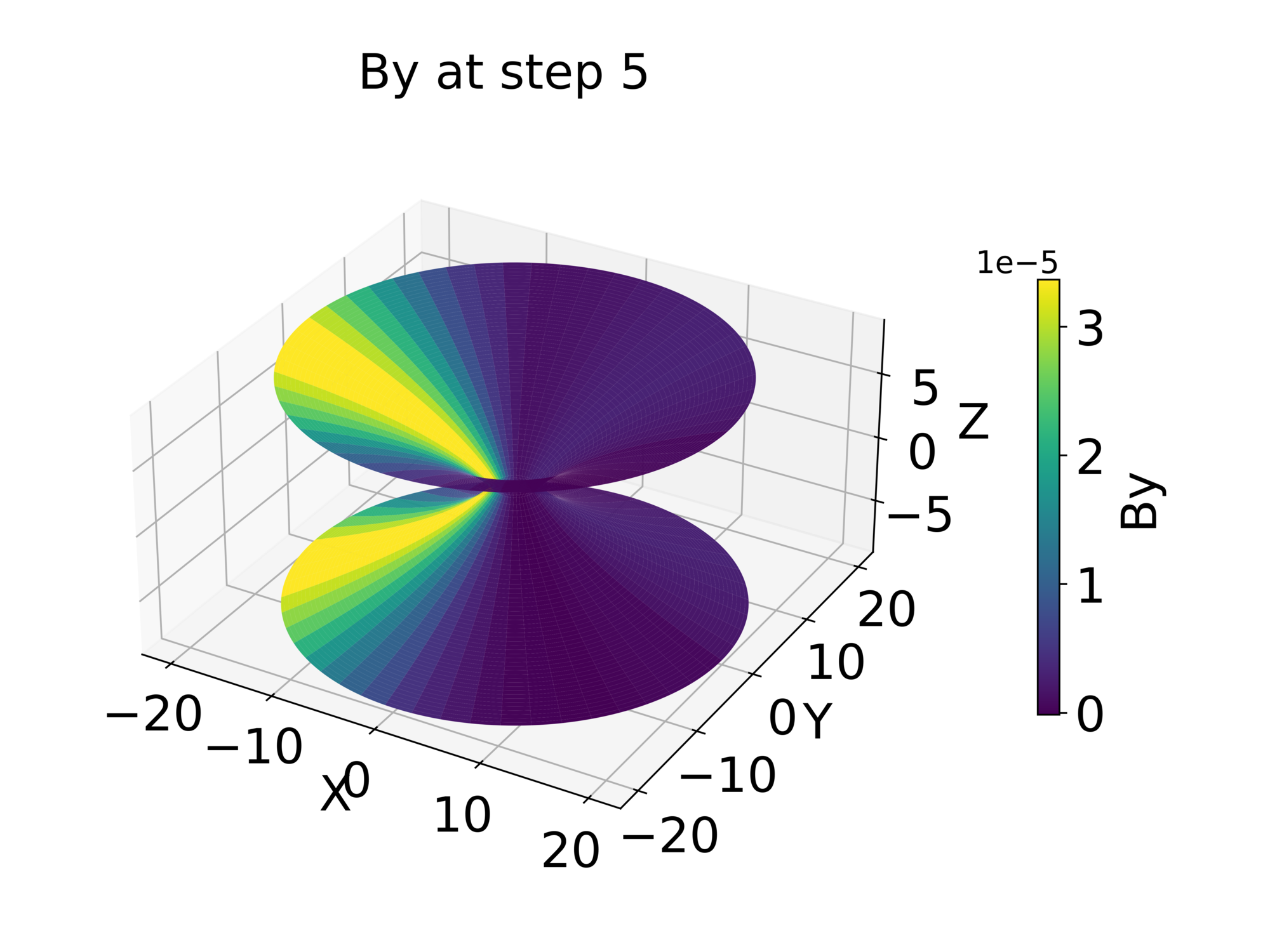}
    \includegraphics[width=0.3\textwidth]{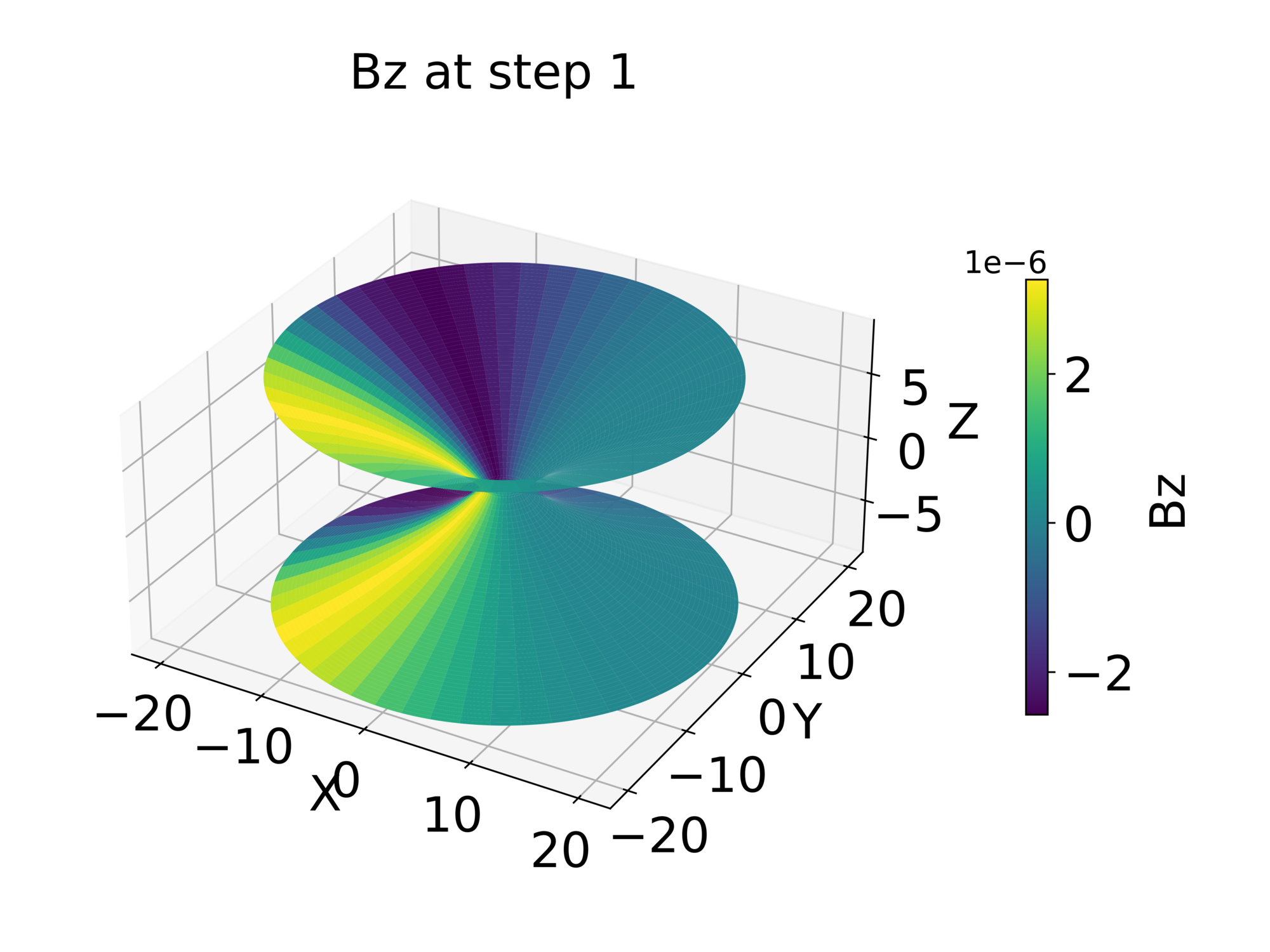}
    \includegraphics[width=0.3\textwidth]{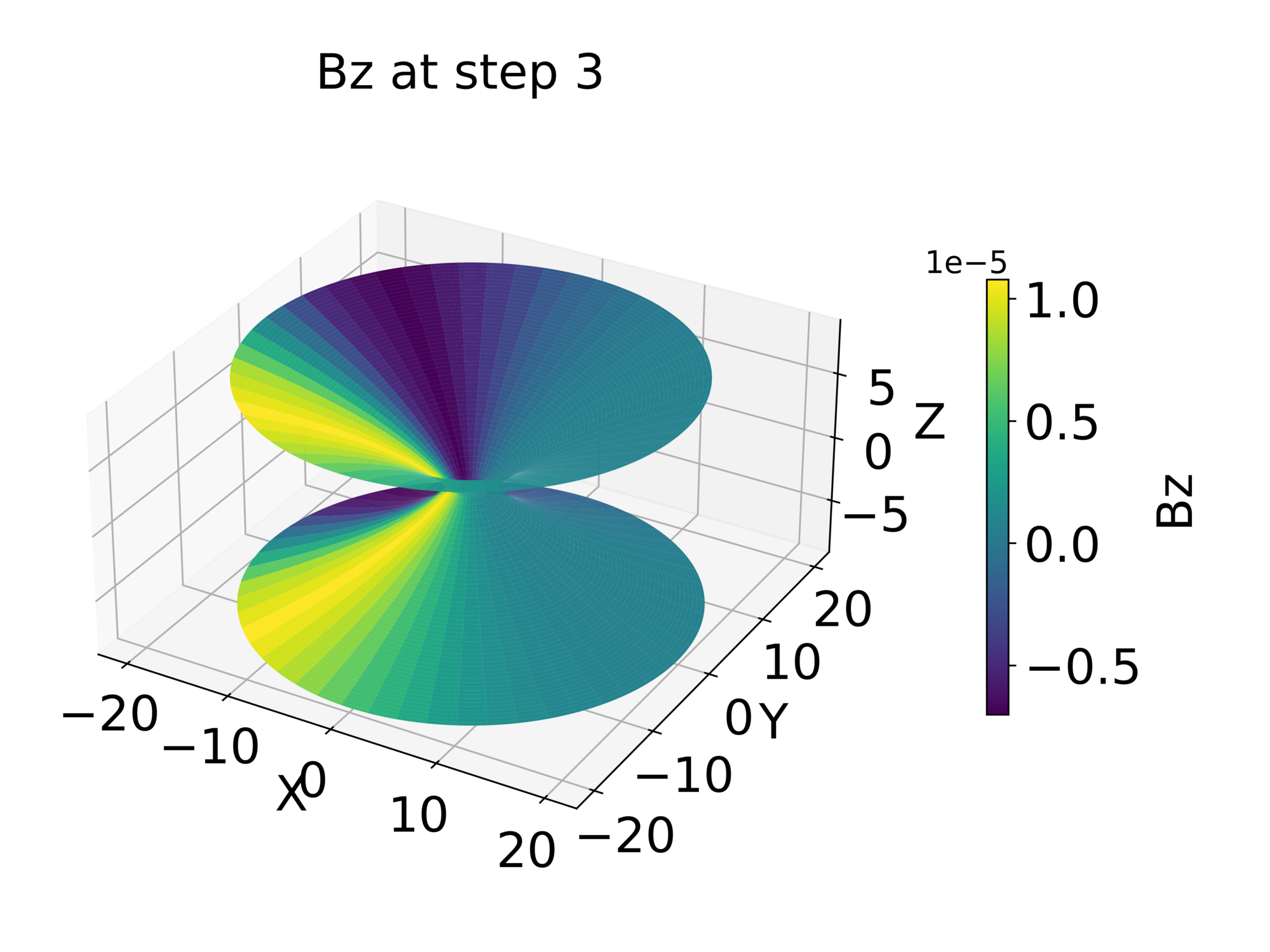}
    \includegraphics[width=0.3\textwidth]{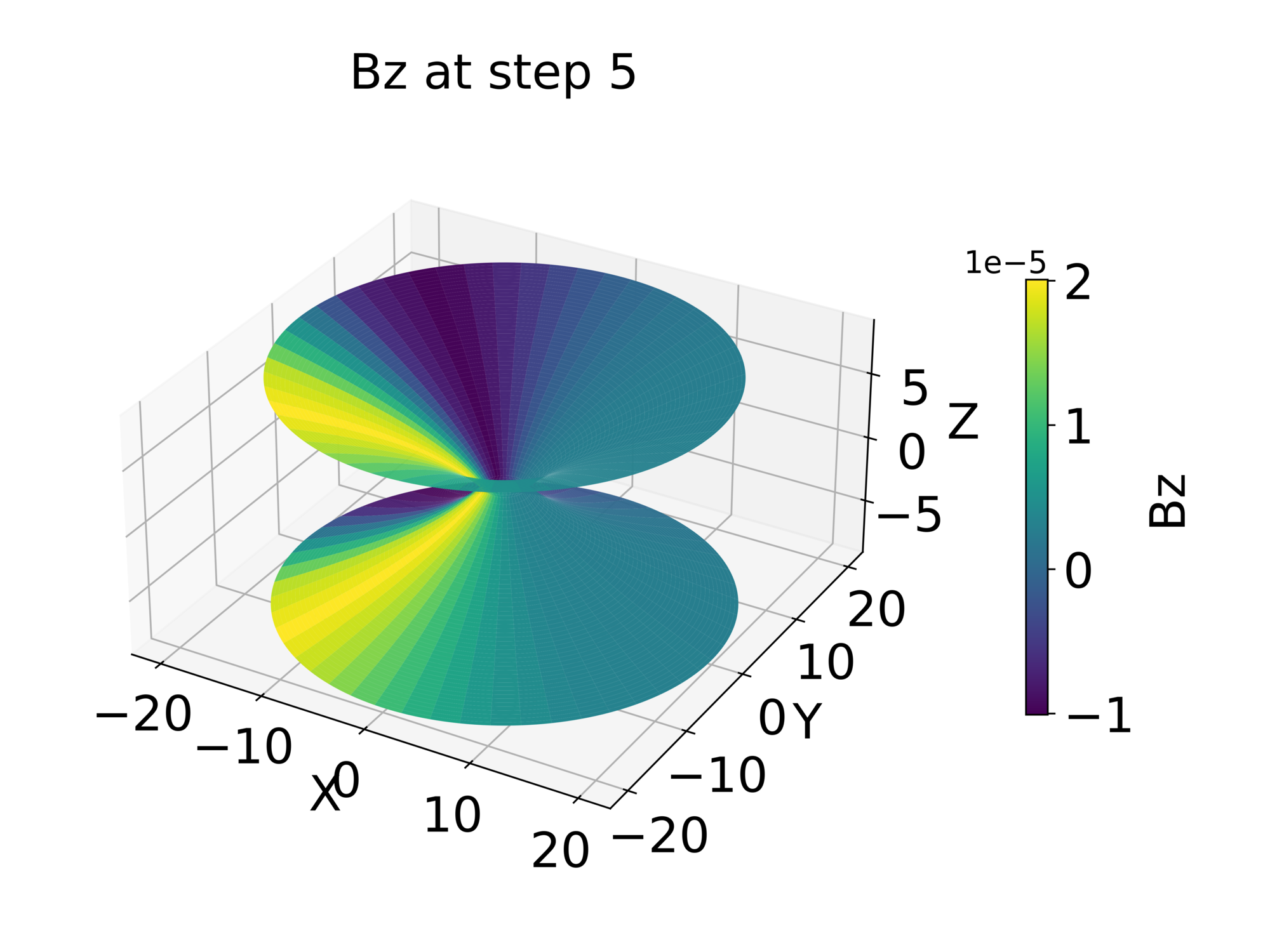}
    \includegraphics[width=0.3\textwidth]{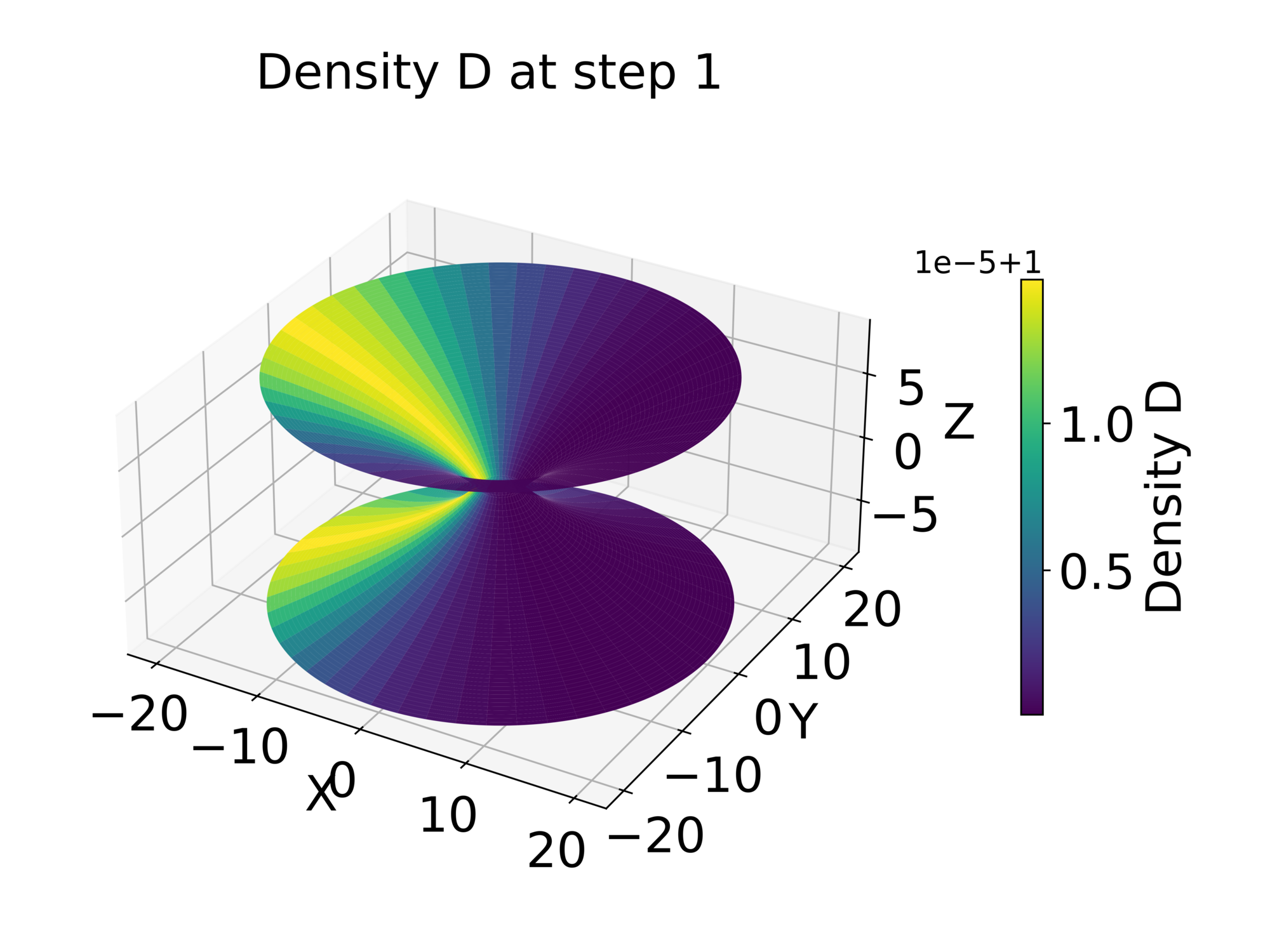}
    \includegraphics[width=0.3\textwidth]{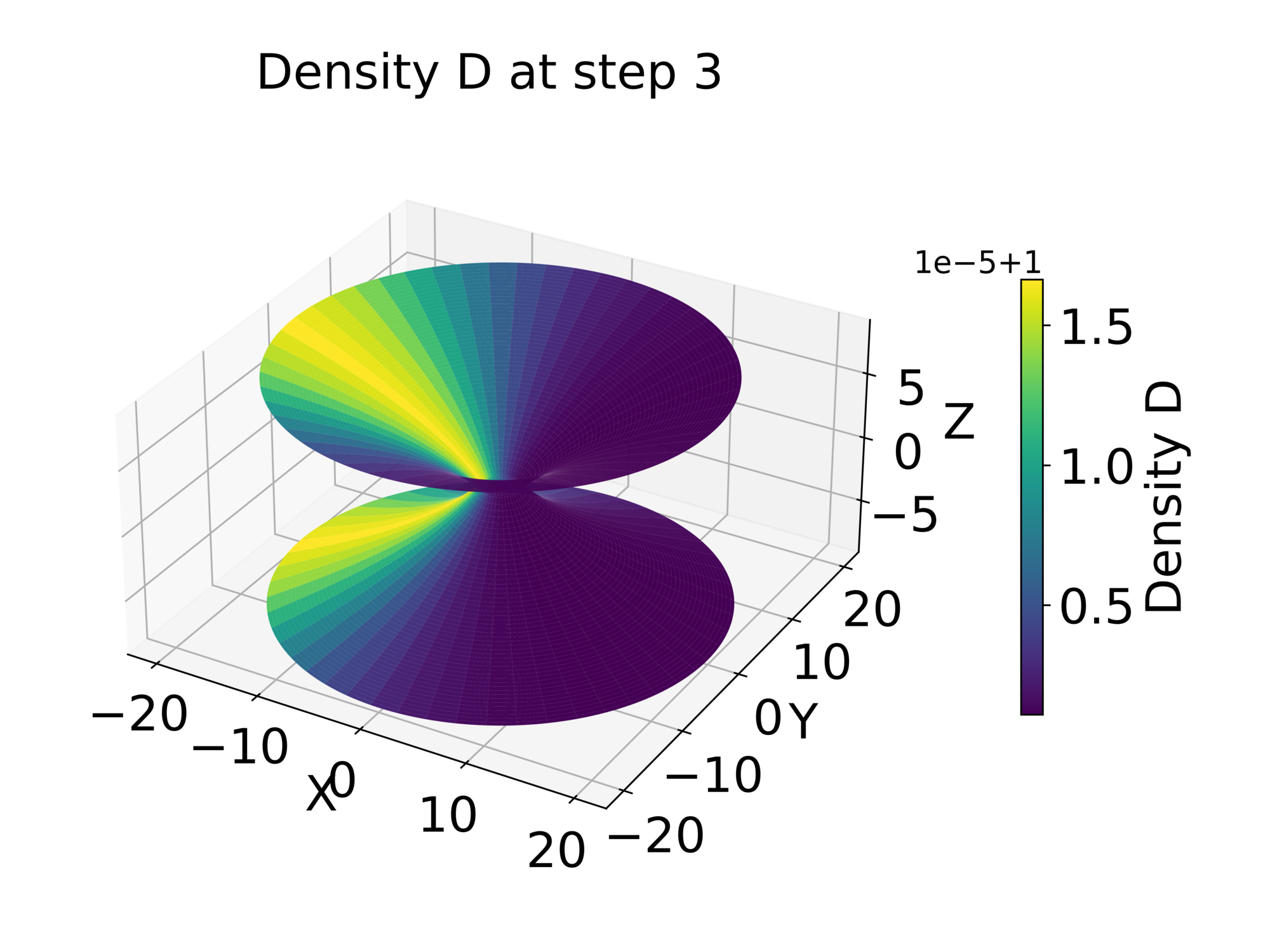}
    \includegraphics[width=0.3\textwidth]{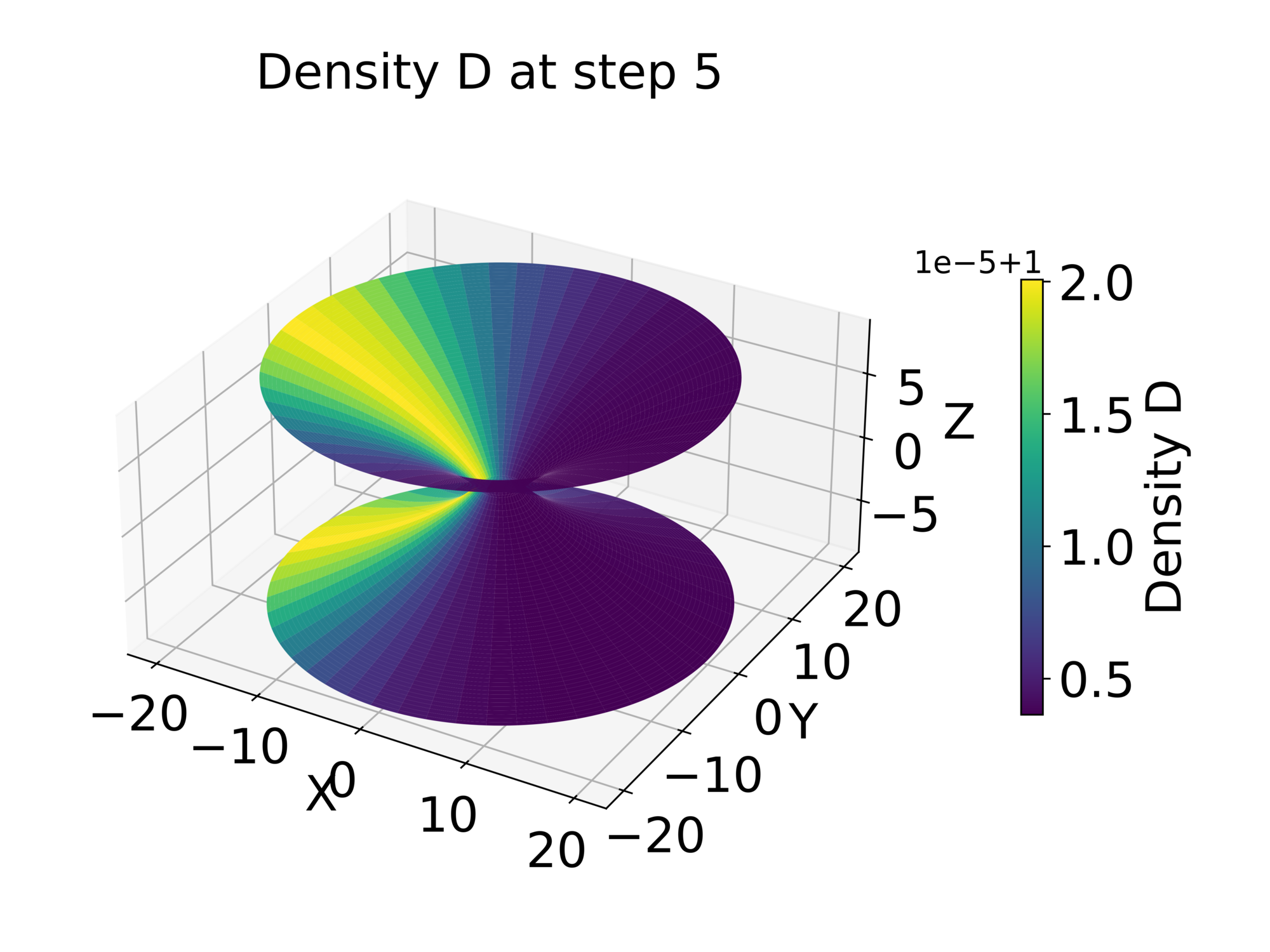}
    \includegraphics[width=0.3\textwidth]{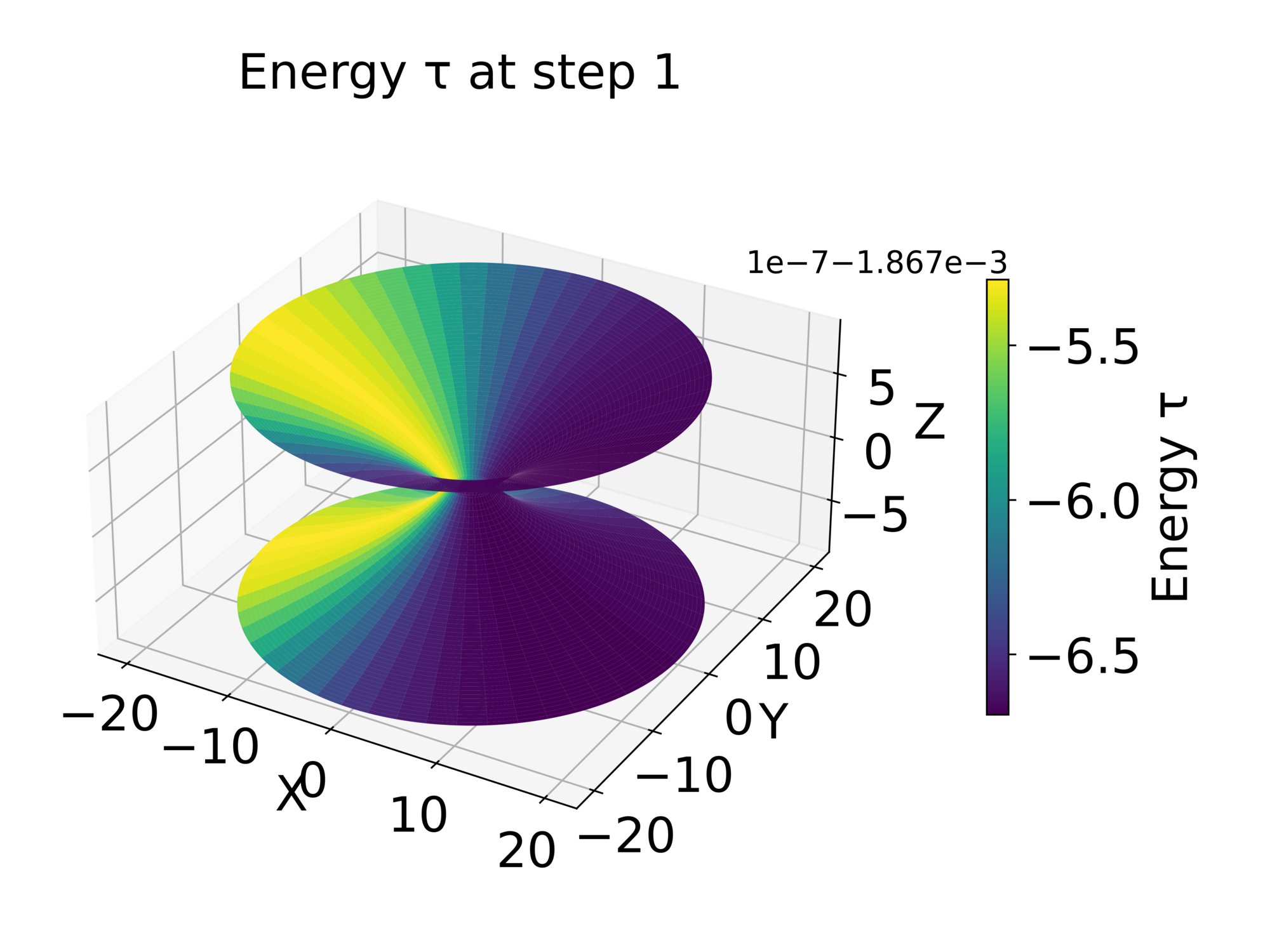}
    \includegraphics[width=0.3\textwidth]{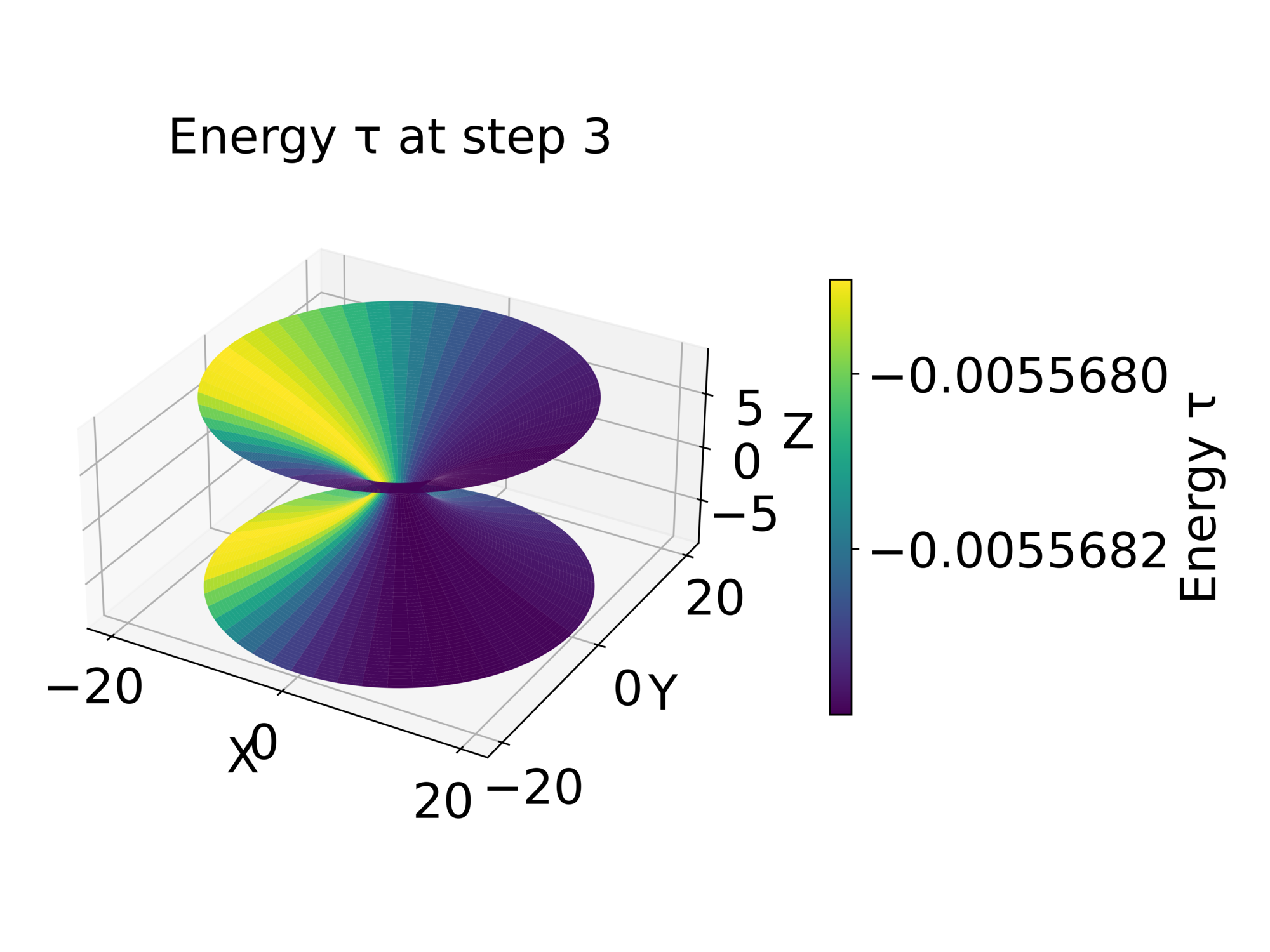}
    \includegraphics[width=0.3\textwidth]{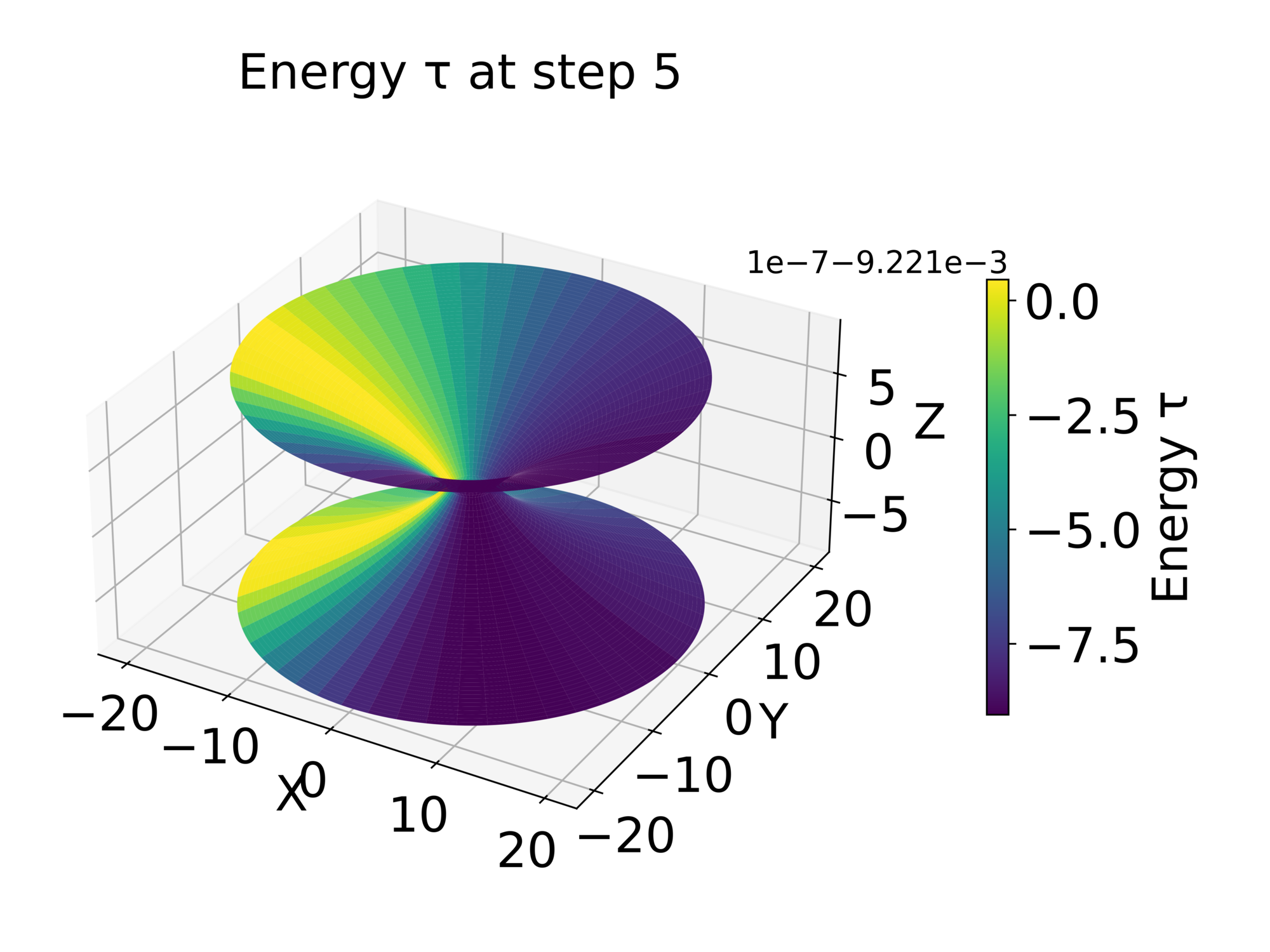}
    \caption{Equatorial embedding diagrams of the Kerr black hole event horizon \( r_h = M + \sqrt{M^2 - a^2} \)), shown as a Flamm paraboloid in fictitious 3D Euclidean space. The surface is parametrized by \( X = r_h \cos\phi \), \( Y = r_h \sin\phi \), and the embedding height \( Z \) computed from the spatial metric. Colors indicate the value of each conserved variable of magnetic field components, density and energy along the equatorial ring \( \theta = \pi/2 \) at the event horizon. Both upper \( +Z \) and lower \( -Z \) surfaces are shown for symmetry.}
    \label{fig:DSSN}
\end{figure}

We extract one of the conserved GRMHD variables in Figure~\ref{fig:DSSN} and ~\ref{fig:TheDSSN} at different time step, 
  \[
    U = \bigl[D,\;S_x,\;S_y,\;S_z,\;\tau,\;B_x,\;B_y,\;B_z\bigr]
    \quad\text{at}\quad r=r_{\rm h},\ \theta=\tfrac\pi2,
  \]
  as a function of $\varphi$.
We interpolate these values onto a fine $\varphi$-grid and use them to color the embedding surface.
Thus each colored Flamm surface shows the angular variation of the chosen variable \emph{on the event horizon}.
Repeating over several time steps illustrates how the horizon‐crossing flow, momentum or magnetic field structure evolves.

\section{Discussion}

We have introduced a fully coupled, hybrid spectral solver for the Einstein–MHD system that combines exponential convergence in smooth regions. By embedding a discontinuous spectral scheme on mapped Chebyshev–Fourier grids within a Valencia‐BSSN framework and invoking conservative updates where needed, the method preserves exact conservation laws, enforces $\nabla \cdot B=0$, and maintains entropy stability. The convergence tests confirm spectral‐order accuracy for smooth flows, while embedding‐diagram and 3D scatter‐visualizations demonstrate its capacity to resolve complex magnetized structures in both fixed Kerr backgrounds and fully dynamical spacetimes.

This approach opens new possibilities for exascale GRMHD simulations of accreting black holes, jet formation, and neutron star mergers. The ongoing work will extend the solver to include resistive effects, radiation–MHD coupling, and matter with more realistic equations of state. By combining high‐order accuracy, low numerical dissipation, and fully dynamical gravity, our hybrid spectral method lays the groundwork for next‐generation studies of strong‐field astrophysics with unprecedented fidelity.

\section{Methods}
\subsection{Numerical Algorithm and Mathematical Formulation}
We discretize the Kerr–Schild domain using a Chebyshev–Gauss–Lobatto grid in radius and uniform grids in polar and azimuthal angles, achieving spectral accuracy in smooth regions.  Radial derivatives are computed via Chebyshev polynomial expansions, while angular derivatives employ Fourier pseudospectral methods; nonlinear aliasing is controlled by 2/3 truncation and exponential filtering.  At each timestep, conserved GRMHD variables are updated with a fourth‐order Runge–Kutta integrator: we reconstruct primitive variables by a damped Newton–Raphson iteration, compute spectral fluxes and geometric sources in Valencia form, enforce $\nabla\!\cdot\!B=0$ via FFT‐based projection or GLM damping, and advance in time subject to a CFL condition.  We prove entropy stability by monitoring a discrete entropy functional and invoke the Lax–Wendroff and Lax equivalence theorems to guarantee convergence of our conservative, consistent, and entropy‐stable scheme to the physically admissible solution.  

\subsubsection*{Grid Discretization}

\begin{enumerate}
  \item \textbf{Radial Chebyshev–Gauss–Lobatto grid:}\\
   The nodes on the interval $[r_{\rm hor}, r_{\rm outer}]$ are given by
  \[
    r_j = \frac{r_{\rm hor} + r_{\rm outer}}{2}
      + \frac{r_{\rm outer} - r_{\rm hor}}{2}
        \cos\left(\frac{j\pi}{N_r - 1}\right), \quad j=0,1,\dots,N_r-1.
  \]
  This clusters nodes near the boundaries, which is favorable for resolving strong gradients near the black hole horizon and at large radii.
  
  \item \textbf{The Uniform grids in Polar $\theta$ and Azimuthal $\phi$:}
  \[
    \theta_k = \frac{k\pi}{N_\theta - 1},\quad k=0,\dots,N_\theta-1;\qquad
    \phi_\ell = \frac{2\pi\,\ell}{N_\phi},\quad \ell=0,\dots,N_\phi-1.
  \]
Uniform grid on \([0,\pi]\) with \(N_\theta\) points.  Uniform grid on \([0,2\pi)\) with \(N_\phi\) points.

\textbf{The Total grid size:}      \(\displaystyle N_r\times N_\theta\times N_\phi\).  

\end{enumerate}

\subsubsection*{The Spectral Derivative Operators}

\paragraph*{Radial Chebyshev Spectral Derivative}

Given function values $U_j = U(r_j)$ at Chebyshev nodes, expand as
\[
  U(r) \approx \sum_{k=0}^{N_r-1} \hat U_k\,T_k(x(r)), \qquad
  x(r) = \frac{2r - (r_{\rm hor} + r_{\rm outer})}{r_{\rm outer} - r_{\rm hor}},
\]
where $T_k$ is the $k$th Chebyshev polynomial. Using \texttt{numpy.polynomial.chebyshev.chebder}, one can directly compute $m$th derivative coefficients $\{ \hat{U}^{(m)}_k \}$ in coefficient space, then interpolate back to physical space:
\[
  \left.\frac{\partial^m U}{\partial r^m}\right|_{r_j}
  \approx \sum_{k=0}^{N_r-1} \hat{U}^{(m)}_k\,T_k(x_j).
\]

\paragraph*{The Fourier Pseudospectral Derivative in $\theta$, $\phi$}

For a periodic direction $x=\theta$ or $\phi$, the DFT of discrete data $U_n$ is
\[
  \hat U_k = \sum_{n=0}^{N-1} U_n\,e^{-2\pi i k n/N}, \quad
  k = -\frac N2 + 1, \dots, \frac N2.
\]
The $d$th derivative in spectral space:
\[
  \widehat{\partial_x^d U}_k = (i k)^d\,\hat U_k, \qquad
  \partial_x^d U_n = \frac{1}{N} \sum_k \widehat{\partial_x^d U}_k\,e^{2\pi i k n / N}.
\]

\subsubsection*{Dealiasing and Spectral Filtering}

\begin{enumerate}
  \item \textbf{2/3 dealiasing:} \\
    All spectral modes with $|k| > N/3$ are zeroed to eliminate aliasing errors in nonlinear convolutions.
  \item \textbf{Exponential filter:} \\
    Each mode is multiplied by
    \[
      \sigma(k) = \exp\left[-\alpha\left(\frac{|k|}{k_{\max}}\right)^p\right],
      \quad \alpha=36,\, p=8,
    \]
    to smoothly damp Gibbs oscillations.
\end{enumerate}

\subsubsection*{The Primitive Variable Recovery}

Given conserved variables
\[
  U = [D,\, S_i,\, \tau,\, B_i]
\]
and metric data, solve for primitive variables $\{ \rho, v^i, \varepsilon \}$ via iterative procedure:

\begin{enumerate}
  \item \textbf{Definitions:}
  \[
    W = (1 - v^2)^{-1/2}, \quad
    p = (\gamma - 1)\rho\varepsilon, \quad
    h = 1 + \varepsilon + \frac{p}{\rho},
  \]
  \[
    b^0 = W\,\vec{v} \cdot \vec{B}, \quad
    b^i = \frac{B^i + b^0 v^i}{W}.
  \]
  Compute predicted values:
  \[
    S_i^{\rm pred} = (\rho h W^2) v_i - b^0 b_i, \quad
    \tau^{\rm pred} = \rho h W^2 - p - \frac{1}{2}b^2 - D.
  \]
  
  \item \textbf{Residuals:} \\
    $\Delta S_i = S_i - S_i^{\rm pred}$, \quad
    $\Delta \tau = \tau - \tau^{\rm pred}$.
  
  \item \textbf{Damped Newton Update:}
  \[
    v^i \gets v^i + \lambda_v \frac{\Delta S_i}{\rho h W^2}, \qquad
    \varepsilon \gets \varepsilon + \lambda_\varepsilon \frac{\Delta\tau}{\rho W^2},
  \]
  where $\lambda_{v, \varepsilon} \approx 0.5$. If the residual increases, halve the damping parameter and retry.
  
  \item \textbf{Update density:} \\
    $\rho \gets D / W$.
  
  \item \textbf{Convergence criterion:} \\
    $\|\Delta S\|_2 + |\Delta\tau| < \mathrm{tol}$.
  
  Perform at most $30$--$40$ iterations. If still not converged, a final Newton–Raphson update for $\varepsilon$ is performed.
\end{enumerate}

\subsubsection*{The Conserved Fluxes and Geometric Source Terms}

\paragraph*{Conserved Fluxes}

For each direction $d \in \{ r, \theta, \phi \}$ at each grid point:
\[
  \mathbf{F}^d =
  \begin{pmatrix}
    D v^d \\
    S_j v^d + p^* \delta^d_j - \frac{b_j B^d}{W} \\
    \tau v^d + p^* v^d - \frac{b^0 B^d}{W} \\
    v^j B^d - v^d B^j
  \end{pmatrix},
\]
where $p^* = p + \frac{1}{2} b^2$, $b^2 = b^\mu b_\mu$.

\paragraph*{The Geometric Source Terms}

Define the stress-energy tensor:
\[
  T^{\mu\nu} = \rho h\,u^\mu u^\nu + p^* g^{\mu\nu} - b^\mu b^\nu,
\]
The source term is:
\[
  S_\nu = \sqrt{-g}\,T^{\kappa\lambda} \Gamma^\lambda_{\nu\kappa}, \qquad
  \sqrt{-g} = \alpha \sqrt{\det \gamma_{ij}},
\]
with
\[
  \Gamma^\lambda_{\mu\nu} = \frac{1}{2} g^{\lambda\sigma}
    \left( \partial_\mu g_{\sigma\nu} + \partial_\nu g_{\sigma\mu} - \partial_\sigma g_{\mu\nu} \right).
\]

\subsubsection*{Divergence Cleaning $\nabla \cdot B = 0$}

On a periodic grid, using FFT:
\begin{enumerate}
  \item $\widehat{\nabla \cdot B}(k) = i\,k \cdot \widehat{\mathbf B}(k)$
  \item Solve Poisson: $\widehat{\phi}(k) = \widehat{\nabla \cdot B}(k) / (-|k|^2)$
  \item Projection: $\widehat{\mathbf B}_{\rm new} = \widehat{\mathbf B} - i k \widehat{\phi}$
  \item Inverse FFT to obtain divergence-free $\mathbf B_{\rm new}$
\end{enumerate}

\subsubsection*{GLM Constraint Damping Source}

Introduce auxiliary scalar $\psi$:
\[
  \partial_t \psi = -c_h^2 (\nabla \cdot B) - \frac{\psi}{\tau}, \qquad
  S_{B^i} = \partial_i \psi.
\]

\subsubsection*{Time Integration: The Fourth-Order Runge–Kutta}

Let $U_{\rm hat}$ denote the spectral coefficients. One RK4 step is:
\[
\begin{aligned}
k_1 &= f(U^n), \\
k_2 &= f(U^n + \frac{\Delta t}{2} k_1), \\
k_3 &= f(U^n + \frac{\Delta t}{2} k_2), \\
k_4 &= f(U^n + \Delta t\,k_3), \\
U^{n+1} &= U^n + \frac{\Delta t}{6} (k_1 + 2k_2 + 2k_3 + k_4).
\end{aligned}
\]

\subsubsection*{The Stability and Complexity}

\begin{itemize}
  \item \textbf{CFL condition:}
  \[
    \Delta t \lesssim
    \frac{ \min ( \Delta r,\, r \Delta\theta,\, r \sin\theta\, \Delta\phi ) }
         { \max(|v| + c_{\rm fast}) }.
  \]
  \item \textbf{Computational cost:}
  \begin{itemize}
    \item Each step requires 4 RHS evaluations, inverse FFT, primitive recovery, flux/source, forward FFT.
        \[
      4\times\bigl(T_{\rm invFFT}+T_{\rm primRec}+T_{\rm flux/Src}+T_{\rm FFT}\bigr).
    \]
    \item FFT: $O(N\log N)$; Chebyshev derivatives: $O(N_r^2)$; primitive recovery: $O(N_{\rm iter})$ per point.
    \item Efficient for $N = N_r N_\theta N_\phi \lesssim 10^3$.
    \item \textbf{Storage:} \(O(N_rN_\theta N_\phi)\) for fields and spectra.
  \end{itemize}
\end{itemize}

\subsubsection*{The Entropy Stability and Convergence}

We define the \emph{discrete total entropy} at time‐level \(n\) by
\[
E^n \;=\; \sum_{i,j,k}
   \underbrace{\eta\bigl(u^n_{i,j,k}\bigr)}_{\rho\,s}
   \;\underbrace{\sqrt{\gamma(r_i,\theta_j)}}_{\text{metric factor}}
   \;\Delta V_{i,j,k},
\]
where
\[
\eta(u)=\rho\,s,\qquad
s=\frac{1}{\gamma-1}\ln\!\Bigl(\frac{p}{\rho^\gamma}\Bigr),
\]
\(\sqrt\gamma\) is the spatial metric determinant factor, and
\(\Delta V_{i,j,k}=(\Delta r_i)(\Delta\theta_j)(\Delta\phi_k)\)
is the cell volume.

\medskip

\noindent\textbf{Discrete entropy inequality.}
The numerical results satisfy
\[
E^{n+1}\;\le\;E^n
\quad\text{for all }n,
\]
i.e.\ the scheme obeys
\[
\sum_{i,j,k}\eta\bigl(u^{n+1}_{i,j,k}\bigr)\,\Delta V_{i,j,k}
\;\le\;
\sum_{i,j,k}\eta\bigl(u^{n}_{i,j,k}\bigr)\,\Delta V_{i,j,k}.
\]
This guarantees \emph{entropy stability} which is no non‐physical entropy growth.

\medskip

\noindent\textbf{Consistency.}
\begin{itemize}
  \item The Chebyshev–Fourier spatial discretization is spectrally accurate (truncation error \(\to0\) exponentially fast for smooth solutions).
  \item The RK4 time integrator has local truncation error \(O(\Delta t^5)\) and hence global error \(O(\Delta t^4)\).
\end{itemize}

\noindent\textbf{Convergence.}
\begin{itemize}
  \item For nonlinear conservation laws, the \emph{Lax–Wendroff–Tadmor theorem} states that any conservative, consistent, and entropy‐stable scheme converges to the unique entropy solution.
  \item For linear problems, the \emph{Lax equivalence theorem} asserts that consistency plus stability implies convergence.
\end{itemize}

\medskip

\noindent\textbf{Conclusion.}
Since the method is both consistent and entropy‐stable as shown by the monotonic decrease of \(E^n\), it follows rigorously that the numerical solution converges to the correct entropy‐satisfying solution.

\subsection{Numerical Implementation of Coupled GRMHD–Einstein Evolution}
In this section we outline an updated solver that evolves the GRMHD fields on a fixed Kerr–Schild background, and a fully coupled version that also advances the spacetime geometry via the BSSN form of Einstein’s equations.  The two versions differ primarily in whether they evolve only the GRMHD fields on a fixed background or also evolve the spacetime geometry itself via the Einstein equations. Specifically:

\begin{enumerate}
  \item \textbf{Stress–Energy Tensor}  
    \begin{itemize}
      \item \emph{Second version:} Adds a function that assembles
      \(\,T^{\mu\nu}\) from the recovered primitives (density, velocity,
      internal energy, magnetic field) to serve as the source term in
      Einstein’s equations.
    \end{itemize}

  \item \textbf{Initial Data Solver XCTS Approximation}  
    \begin{itemize}
      \item \emph{Second version:} Introduces a fixed-point iteration that
      solves the Hamiltonian and momentum constraints to produce an
      initial spatial metric \(\gamma_{ij}\) and extrinsic curvature
      \(K_{ij}\) satisfying the constraints.
    \end{itemize}

  \item \textbf{Gauge Conditions}  
    \begin{itemize}
      \item \emph{Second version:} Implements 1+log slicing and
      Gamma-driver shift via
      \(\texttt{apply\_gauge\_conditions}( \alpha,\beta,\gamma,K,\dots)\),
      updating the lapse \(\alpha\) and shift \(\beta^{i}\) each step.
    \end{itemize}

  \item \textbf{BSSN Evolution Step}  
    \begin{itemize}
      \item \emph{Second version:} Adds
      \(\texttt{bssn\_evolution\_step}(\gamma,A,\phi,K,\Gamma;\alpha,\beta,T^{\mu\nu},\dots)\)
      which performs a first-order Euler/RK1 BSSN update of the
      conformal metric, extrinsic curvature, conformal factor, trace of
      \(K\), and Gamma-variables, coupling in the matter sources
      \(T^{\mu\nu}\).
    \end{itemize}

  \item \textbf{Main Coupled Evolution Loop}  
    \begin{itemize}
      \item \emph{Second version:} 
      \begin{enumerate}
        \item Evolves the fluid half–step on a fixed background.
        \item Recovers primitives and computes \(T^{\mu\nu}\).
        \item Evolves the metric via the BSSN step.
        \item Updates gauge variables.
        \item Prints progress and visualizes the final state.
      \end{enumerate}
    \end{itemize}
\end{enumerate}

\subsubsection*{Background Geometry: ADM Split in Kerr–Schild Coordinates}

\paragraph*{Kerr–Schild metric}
\[
  g_{\mu\nu} = \eta_{\mu\nu} + 2H\,\ell_\mu\ell_\nu,
  \quad
  \eta_{\mu\nu} = \mathrm{diag}(-1,1,1,1),
\]
\[
  H(r,\theta)
    = \frac{M\,r}{r^2 + a^2\cos^2\theta},
  \quad
  \ell_\mu
    = \bigl(1,\,1,\,0,\,\tfrac{a\sin^2\theta}{r^2 + a^2\cos^2\theta}\bigr).
\]

\paragraph*{ADM decomposition}
\[
  ds^2 = -\alpha^2\,dt^2
         + \gamma_{ij}\bigl(dx^i + \beta^i\,dt\bigr)\bigl(dx^j + \beta^j\,dt\bigr),
\]
where
\[
  \alpha = (1+2H)^{-\tfrac12}, 
  \quad
  \beta^i = \text{shift vector},
  \quad
  \gamma_{ij} = \text{spatial 3-metric}.
\]

\subsection*{Spectral-Domain Derivative Operators}

\paragraph*{Chebyshev nodes}
\[
  x_j
    = \frac{a+b}{2}
      + \frac{b-a}{2}\cos\Bigl(\frac{j\pi}{N-1}\Bigr),
  \quad j=0,\dots,N-1.
\]

\subsubsection*{Chebyshev multi-derivative}
Expand data in Chebyshev polynomials, use \texttt{chebder} to differentiate in coefficient space, then transform back.

\paragraph*{Fourier derivative}
\[
  \frac{\partial f}{\partial \phi}
    = \mathcal{F}^{-1}\bigl(i\,k\,\mathcal{F}[f]\bigr),
  \quad
  k=0,1,\dots,\tfrac{n}{2},-\tfrac{n}{2}+1,\dots,-1.
\]

\subsubsection*{Pseudo-spectral \(\theta\) derivative}
Use Fourier on an evenly spaced \(\theta\) grid.

\subsubsection*{GRMHD Conservation Variables, Primitives \& EOS}

\paragraph*{Conserved variables \(U\)}
\[
  U = \bigl[D,\,S_i,\,\tau,\,B^i\bigr],
\]
\[
  D = \rho\,W,
  \quad
  S_i = (\rho h^* + b^2)\,W^2 v_i - b^0 b_i,
  \quad
  \tau = (\rho h^* + b^2)\,W^2 - \bigl(p + \tfrac12 b^2\bigr) - (b^0)^2 - D.
\]

\paragraph*{Primitive variables}
\(\rho,\;v^i,\;\varepsilon\) recovered by a damped Newton iteration.

\paragraph*{Ideal-gas EOS}
\[
  p = (\gamma-1)\,\rho\,\varepsilon,
  \quad
  \gamma = \tfrac{5}{3}.
\]

\subsubsection*{The Fluxes and Source Terms}

\paragraph*{Fluxes \(\mathbf{F}^i(U)\)}
Computed in the standard GRMHD form for each conserved variable.

\paragraph*{Geometric source}
\[
  S_\nu = \sqrt{-g}\;T^{\kappa\lambda}\,\Gamma^\mu_{\lambda\nu},
  \quad
  \sqrt{-g} = \alpha\,\sqrt{\det\gamma}.
\]

\subsubsection*{Stress–Energy Tensor \(T^{\mu\nu}\)}

\[
  T^{\mu\nu}
    = \rho\,h^*\,u^\mu u^\nu
      + p^*\,g^{\mu\nu}
      - b^\mu b^\nu.
\]

\subsubsection*{Initial Data: Approximate XCTS Solution}

Solve
\[
  \tilde\nabla^2\psi = -2\pi\,\rho\,\psi^5,
  \quad
  \tilde\nabla_j A^{ij} = 8\pi\,S^i\,\psi^{10},
\]
with fixed-point relaxation on \(\psi\) and \(A_{ij}=0\).  
Then
\[
  \gamma_{ij} = \psi^4\,\tilde\gamma_{ij},
  \quad
  K = 0.
\]

\subsubsection*{Gauge Conditions}

\paragraph*{1+log slicing}
\[
  \partial_t\alpha = -2\,\alpha\,K + \beta^i\,\partial_i\alpha.
\]

\subsubsection*{Gamma-driver shift}
\[
  \partial_t\beta^i = \tfrac34\,B^i,
  \quad
  \partial_t B^i = \partial_t\tilde\Gamma^i - \eta\,B^i.
\]

\subsubsection*{BSSN Evolution Equations}

Define
\(\tilde\gamma_{ij},\;\phi=\tfrac14\ln\det\gamma_{ij},\;\tilde A_{ij},\;K,\;\tilde\Gamma^i.\)

\[
\begin{cases}
\partial_t\tilde\gamma_{ij}
  = -2\alpha\,\tilde A_{ij}
    + \mathcal{L}_\beta\tilde\gamma_{ij}
    - \tfrac23\,\tilde\gamma_{ij}\,\partial_k\beta^k,\\[6pt]
\partial_t\phi
  = -\tfrac16\alpha\,K
    + \mathcal{L}_\beta\phi
    + \tfrac16\,\partial_k\beta^k,\\[6pt]
\partial_t\tilde A_{ij}
  = e^{-4\phi}\bigl[-D_iD_j\alpha + \alpha\,(R_{ij}-8\pi S_{ij})\bigr]^{\!TF}
    + \alpha\,(K\,\tilde A_{ij}-2\,\tilde A_{ik}\tilde A^k{}_j)\\[3pt]
  \quad + \mathcal{L}_\beta\tilde A_{ij}
    - \tfrac23\tilde A_{ij}\,\partial_k\beta^k,\\[6pt]
\partial_t K
  = -\gamma^{ij}D_iD_j\alpha
    + \alpha\bigl(\tilde A_{ij}\tilde A^{ij} + \tfrac13K^2\bigr)
    + 4\pi\alpha\,(\rho+S)
    + \mathcal{L}_\beta K,\\[6pt]
\partial_t\tilde\Gamma^i
  = 2\alpha\bigl(\tilde\Gamma^i_{jk}\tilde A^{jk}
    - \tfrac23\,\tilde\gamma^{ij}\partial_jK - 8\pi\,j^i\bigr)
    -2\,\tilde A^{ij}\partial_j\alpha\\[3pt]
  \quad + \tilde\gamma^{jk}\partial_j\partial_k\beta^i
    + \tfrac13\,\tilde\gamma^{ij}\partial_j\partial_k\beta^k
    + \mathcal{L}_\beta\tilde\Gamma^i
    + \tfrac23\,\tilde\Gamma^i\,\partial_j\beta^j.
\end{cases}
\]

\subsubsection*{Coupled Main Loop}

\begin{enumerate}
  \item \textbf{Fluid evolution GRMHD:}
    \begin{enumerate}
      \item Half-step RK4 update \(U\to U_{\rm phys}\).
      \item Divergence cleaning of \(B\).
      \item Primitive recovery \(\to T^{\mu\nu}\).
    \end{enumerate}
  \item \textbf{Spacetime evolution BSSN + gauges:}
    \begin{enumerate}
      \item Evolve metric with source \(T^{\mu\nu}\).
      \item Update lapse \(\alpha\) and shift \(\beta^i\).
    \end{enumerate}
\end{enumerate}

\begin{appendices}

\section{Extended Figures}\label{secA1}
\begin{figure}[h]
    \centering
    \includegraphics[width=0.45\textwidth]{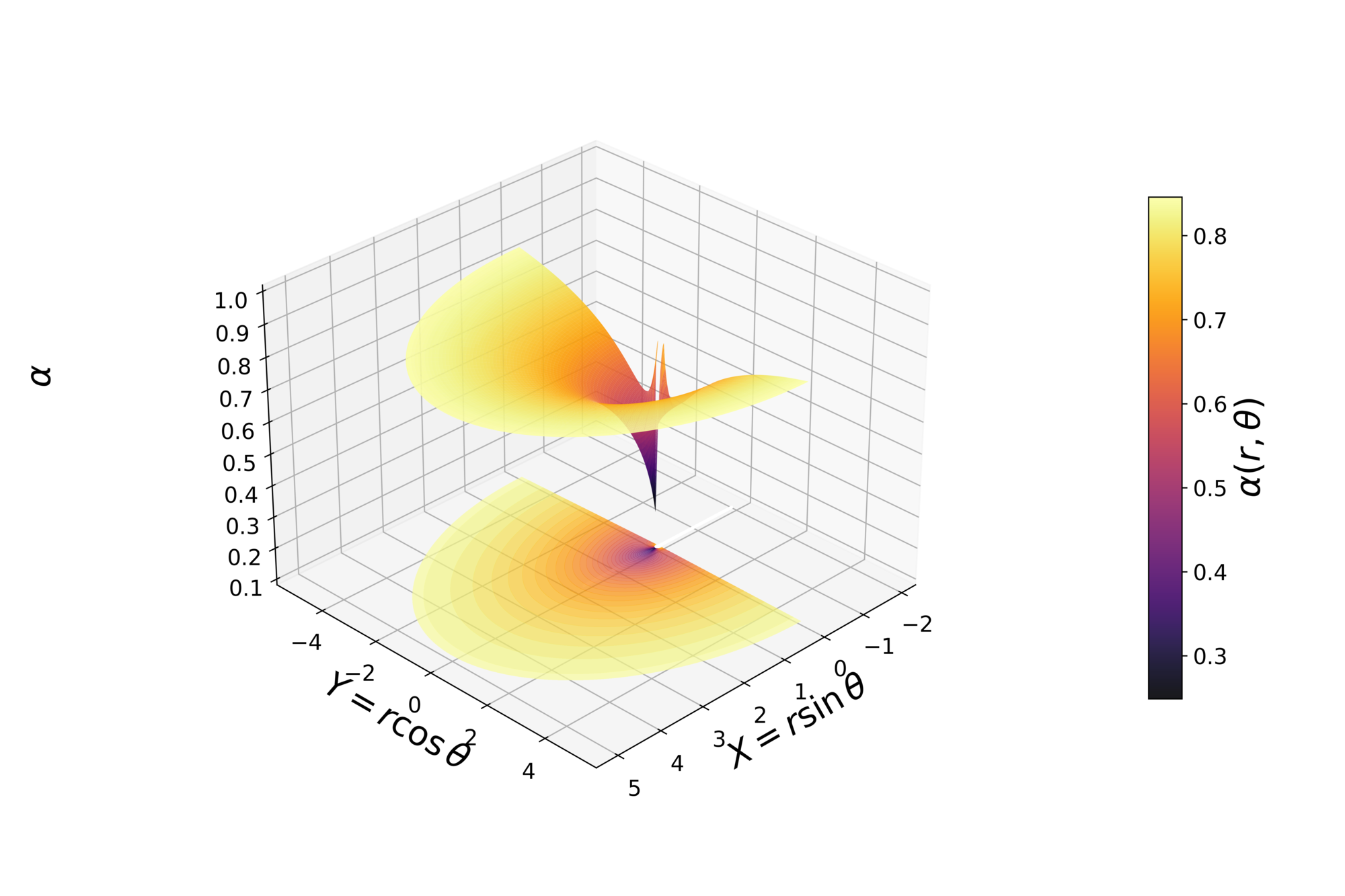}
    \includegraphics[width=0.45\textwidth]{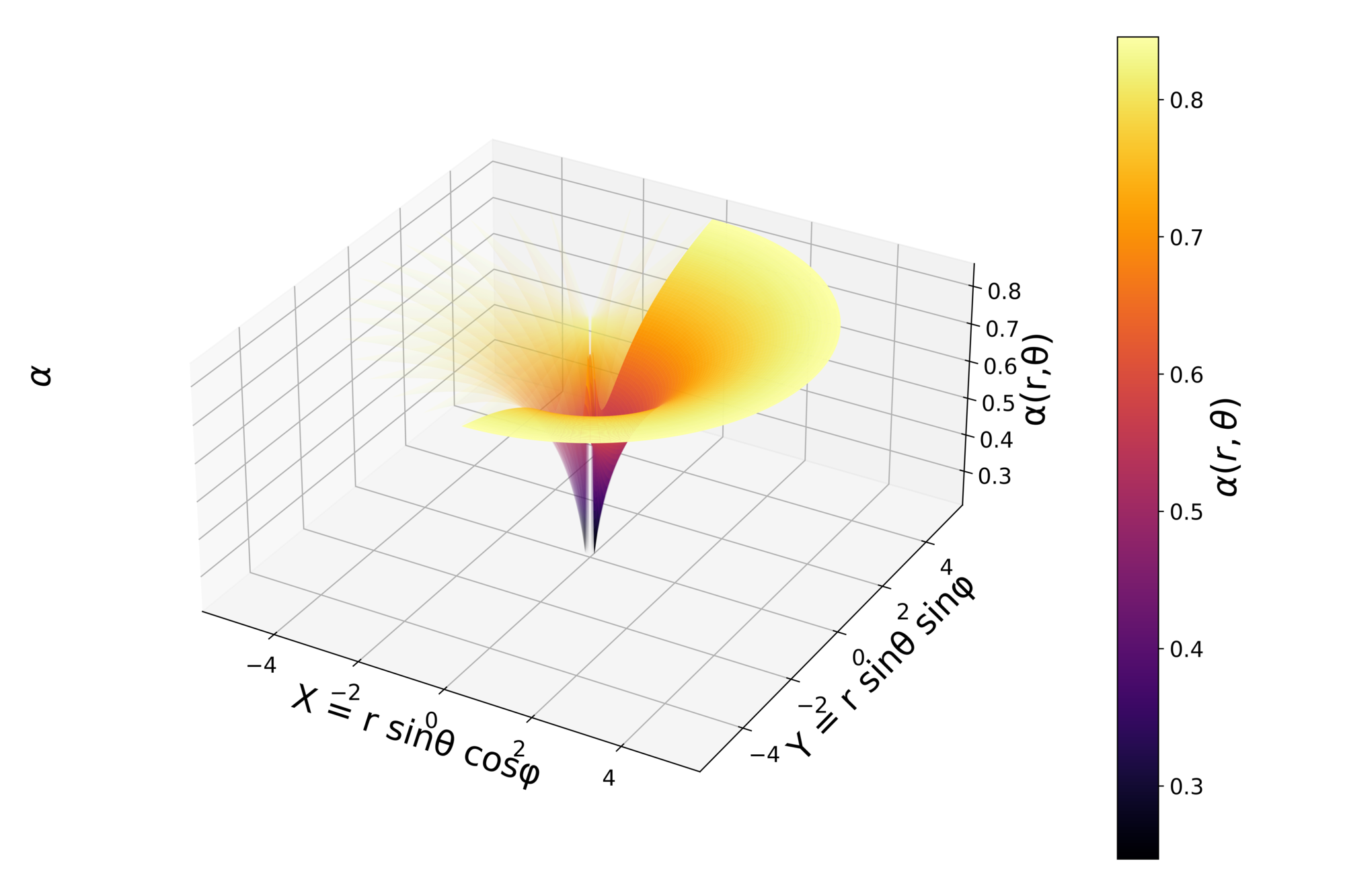}
    \caption{The 3D surface of the Kerr--Schild lapse $\alpha(r,\theta)$, plotted as $Z = \alpha$ over $X, Y$ with color showing the lapse value.}
    \label{fig:K}
\end{figure}

\begin{table}[h]
\centering
\begin{tabular}{lll}
\hline
Field               & Physical Meaning                         & Formula \\
$\alpha$            & Lapse time dilation factor             & $\sqrt{\Sigma / (\Sigma + 2r)}$ \\
$\beta^r$           & Radial component of shift vector         & $2 r^2 / \Sigma$ \\
$\beta^\theta$      & Poloidal component of shift vector       & $0$ \\
$g_{x_1x_1}$        & Spatial metric in $x_1$ ($r$) direction  & $r^2 \left(1 + \frac{2r}{\Sigma}\right)$ \\
$g_{\theta\theta}$  & Spatial metric in $\theta$ direction     & $\Sigma$ \\
$g_{\phi\phi}$      & Spatial metric in $\phi$ direction       & $\left(r^2 + a^2 + \frac{2a^2 r \sin^2\theta}{\Sigma}\right)\sin^2\theta$ \\
\hline
\end{tabular}
\caption{Summary of the Kerr–Schild ADM split geometric fields and their expressions. Here, $\Sigma = r^2 + a^2 \cos^2\theta$.}
\label{tab:kerr-adm-fields}
\end{table}
\begin{figure}[h]
    \centering
    \includegraphics[width=0.7\textwidth]{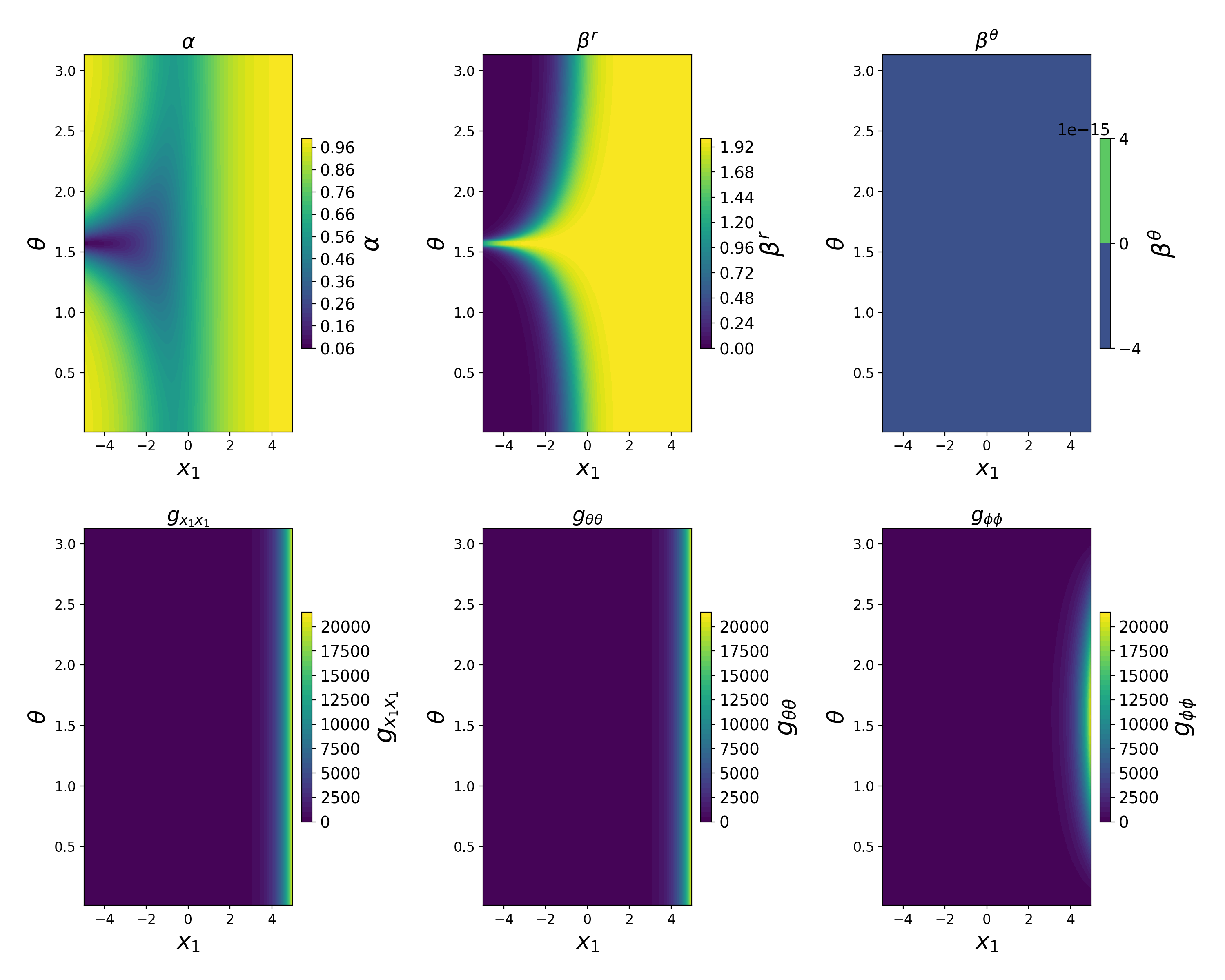}
    \caption{
        Contour plots of the Kerr–Schild ADM split geometric quantities in the $(x_1,\,\theta)$ plane, where $x_1 = \ln r$. Each panel displays a different field: the lapse $\alpha$, shift vector components $\beta^r$ and $\beta^\theta$, and spatial metric components $g_{x_1x_1}$, $g_{\theta\theta}$, and $g_{\phi\phi}$ for a Kerr black hole with spin parameter $a = 0.5$. The color indicates the value of each field at each point.
    }
    \label{fig:kerr-adm-contours}
\end{figure}

\begin{figure}[h]
  \centering
  \includegraphics[width=0.30\linewidth]{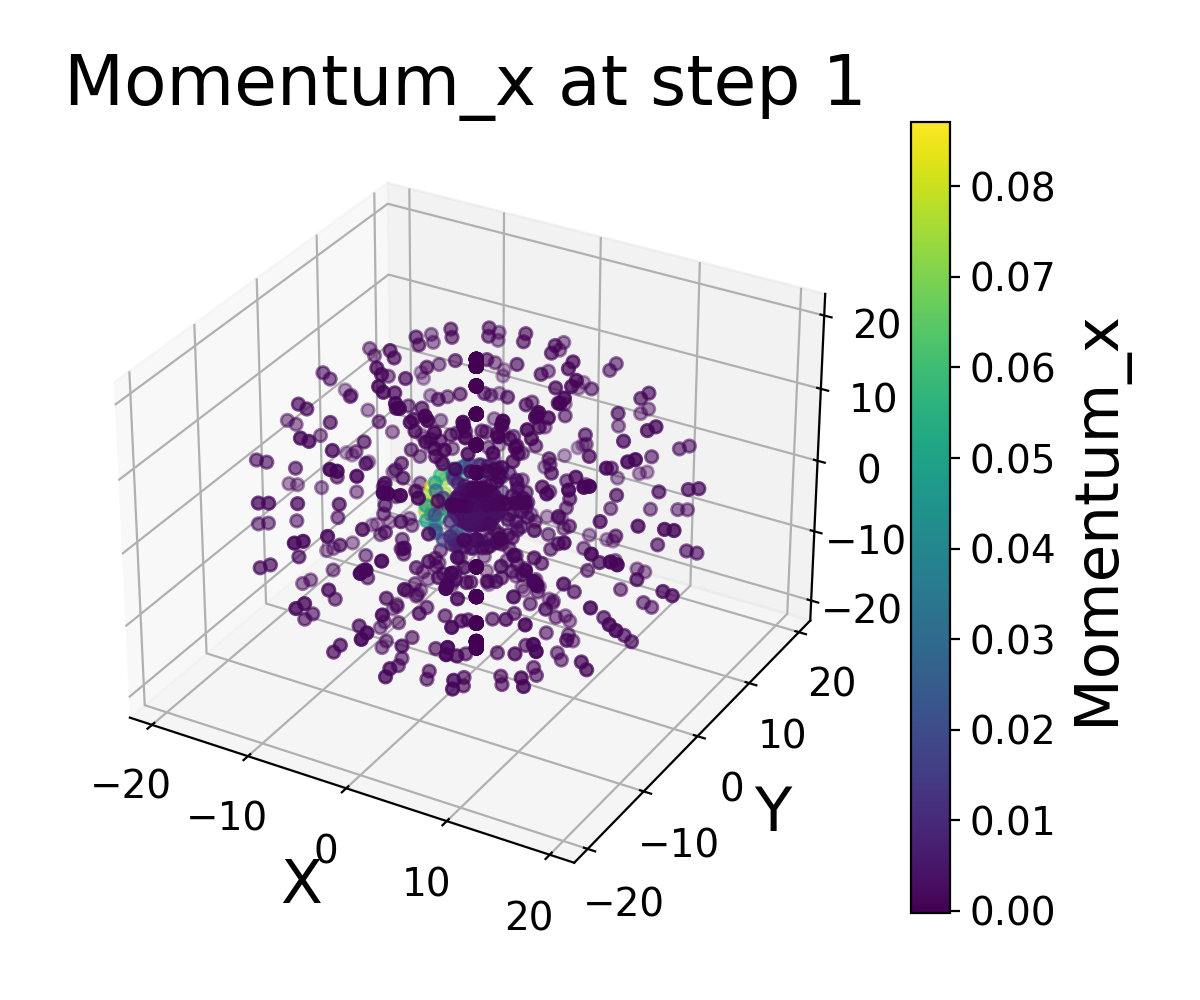}
  \includegraphics[width=0.30\linewidth]{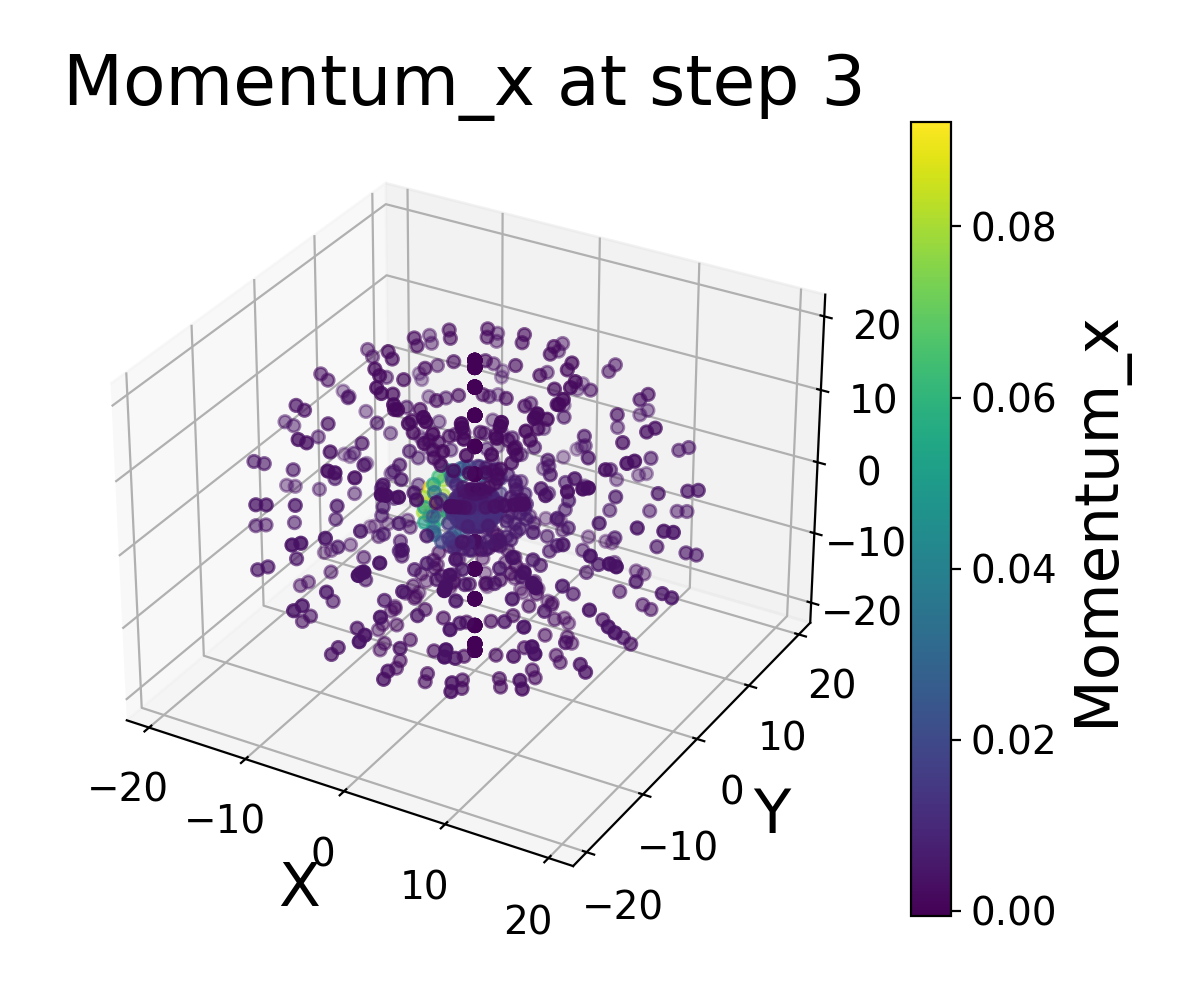}
  \includegraphics[width=0.30\linewidth]{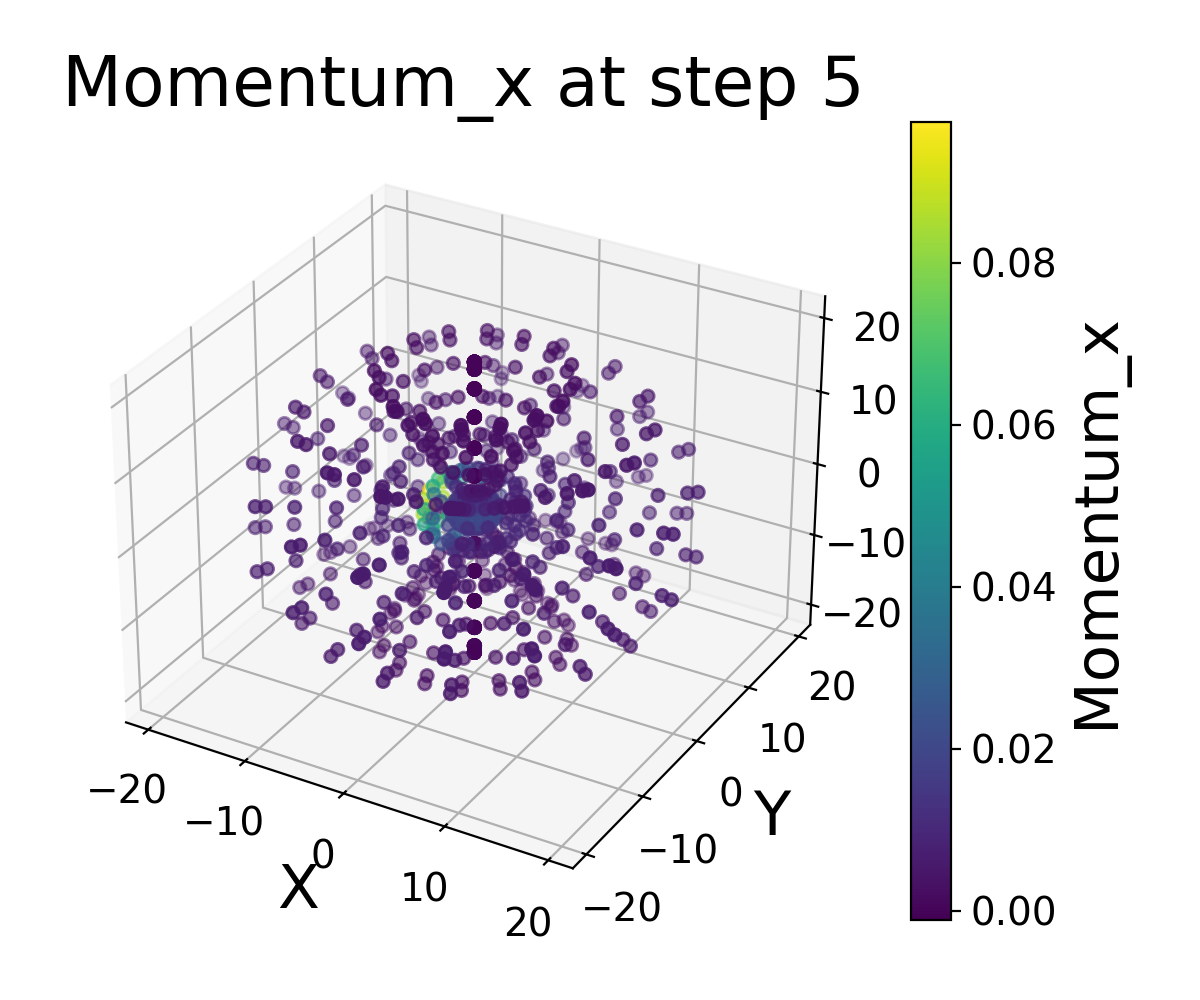}
  \includegraphics[width=0.30\linewidth]{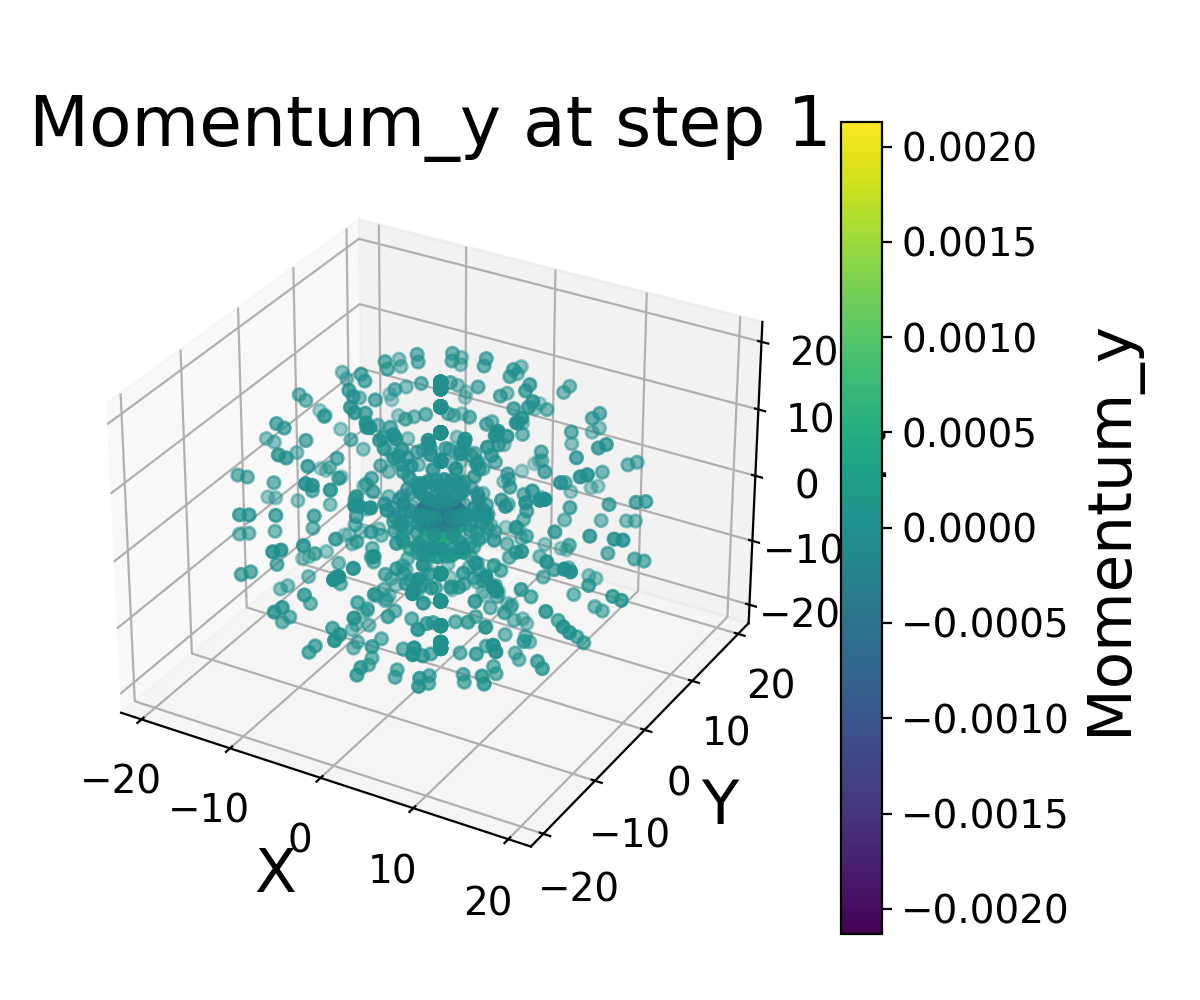}
  \includegraphics[width=0.30\linewidth]{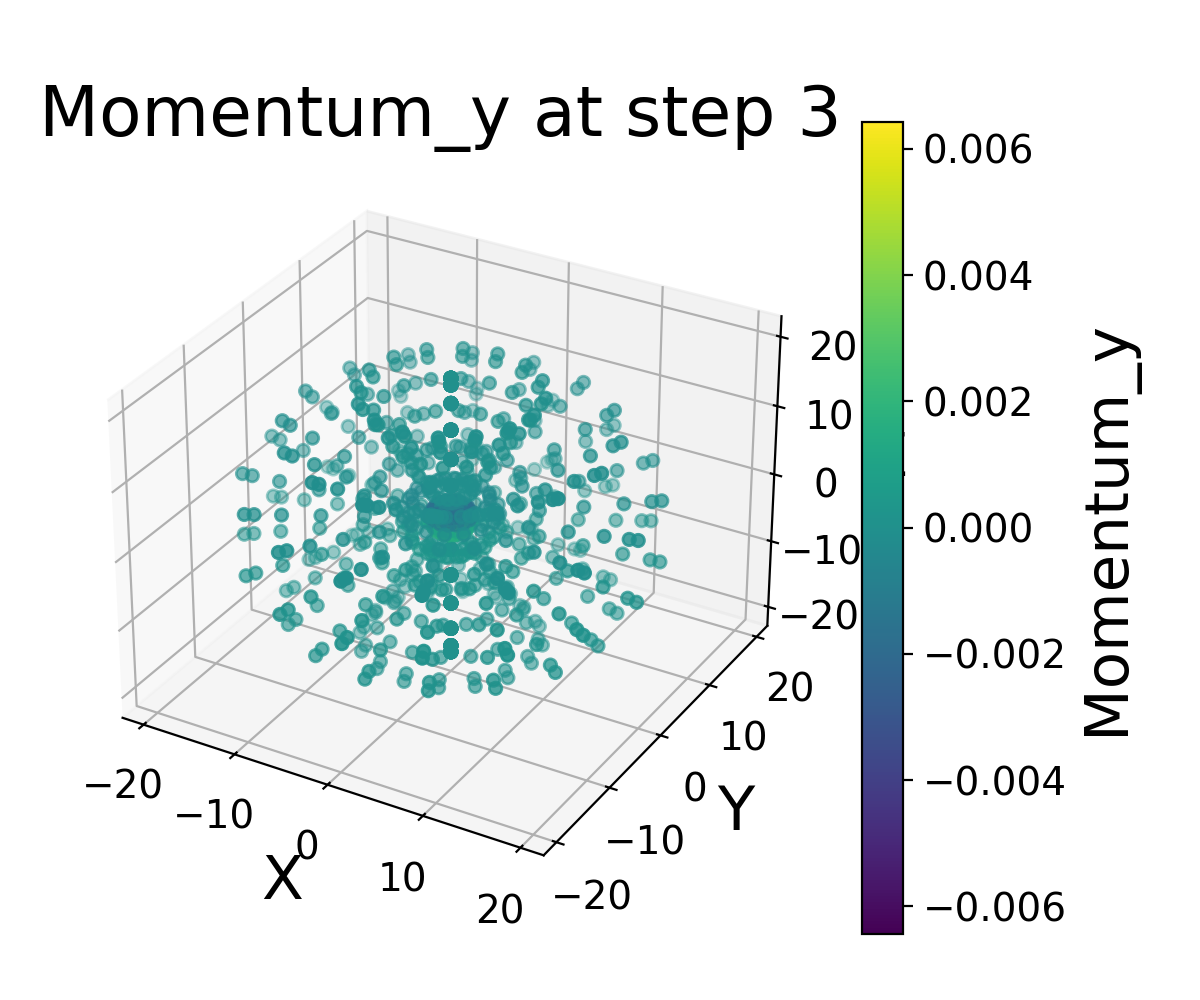}
  \includegraphics[width=0.30\linewidth]{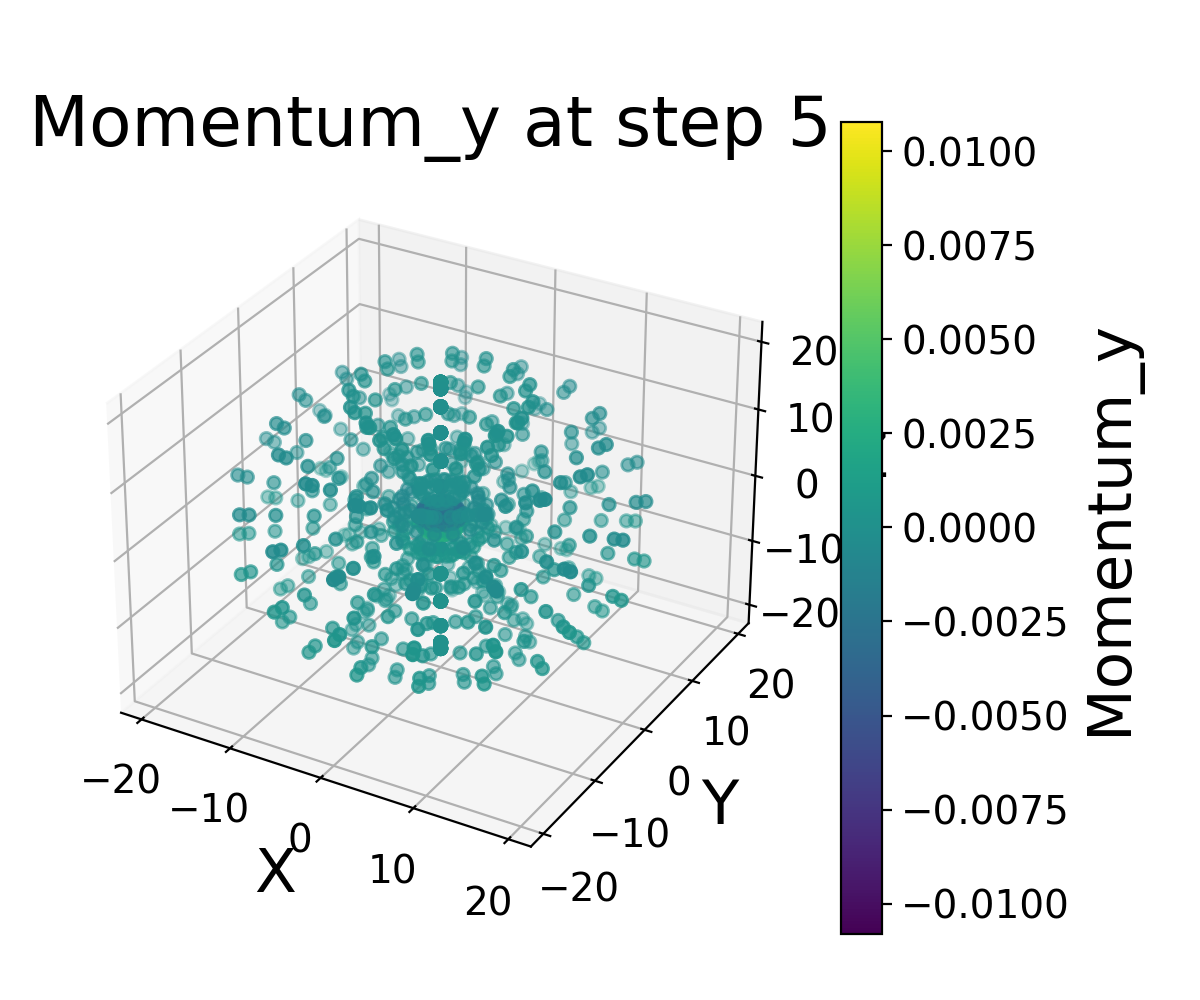}
  \includegraphics[width=0.30\linewidth]{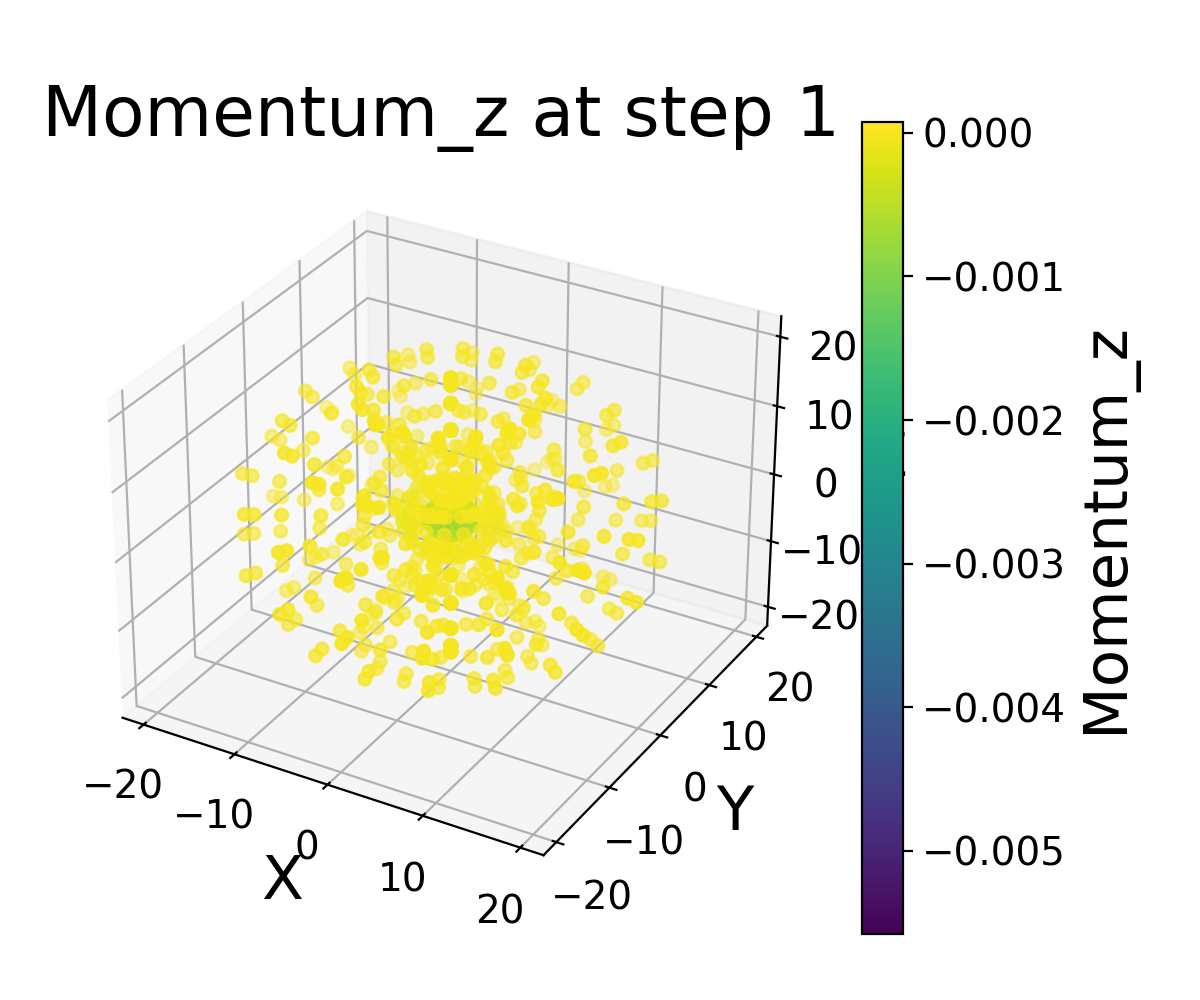}
  \includegraphics[width=0.30\linewidth]{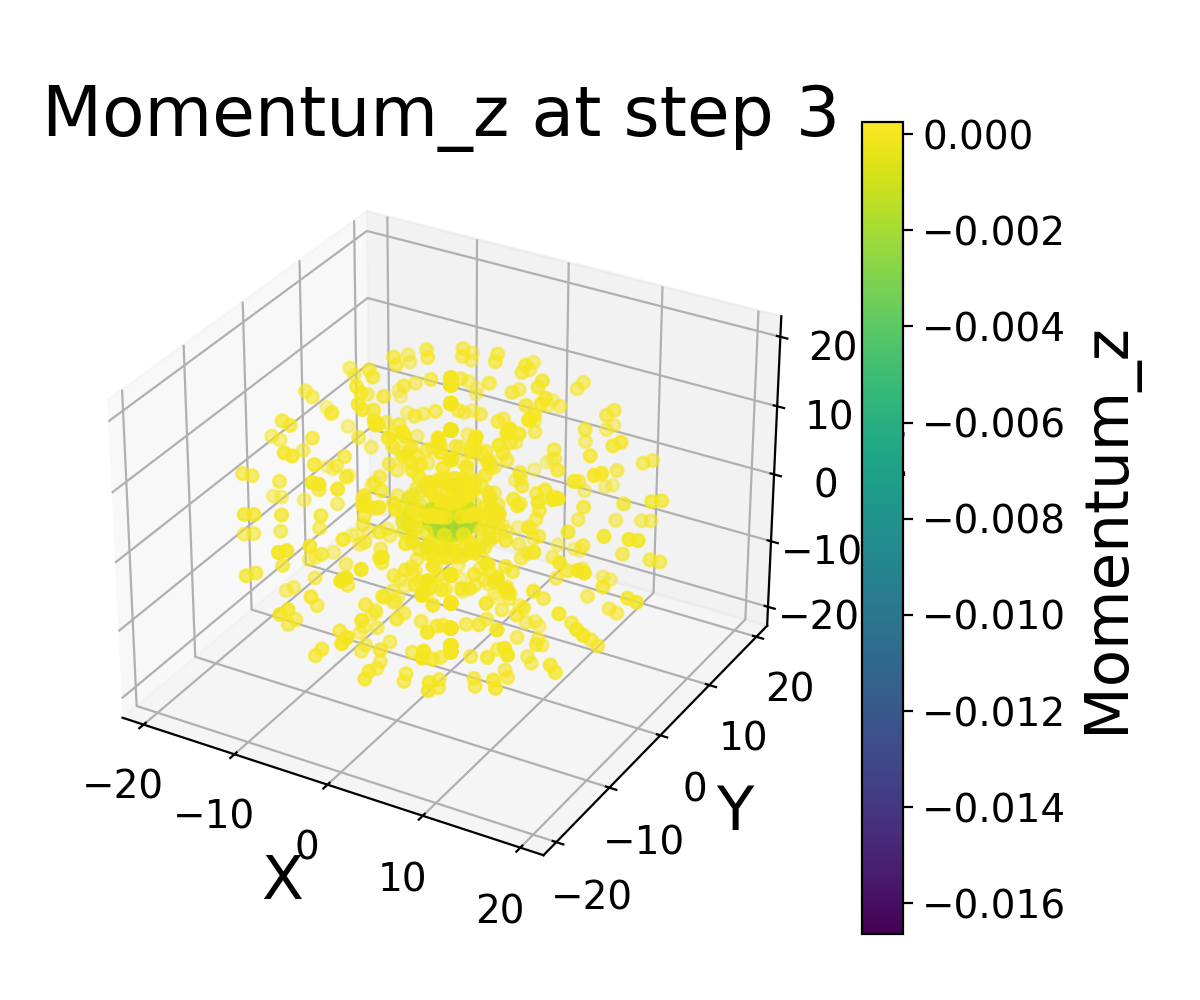}
  \includegraphics[width=0.30\linewidth]{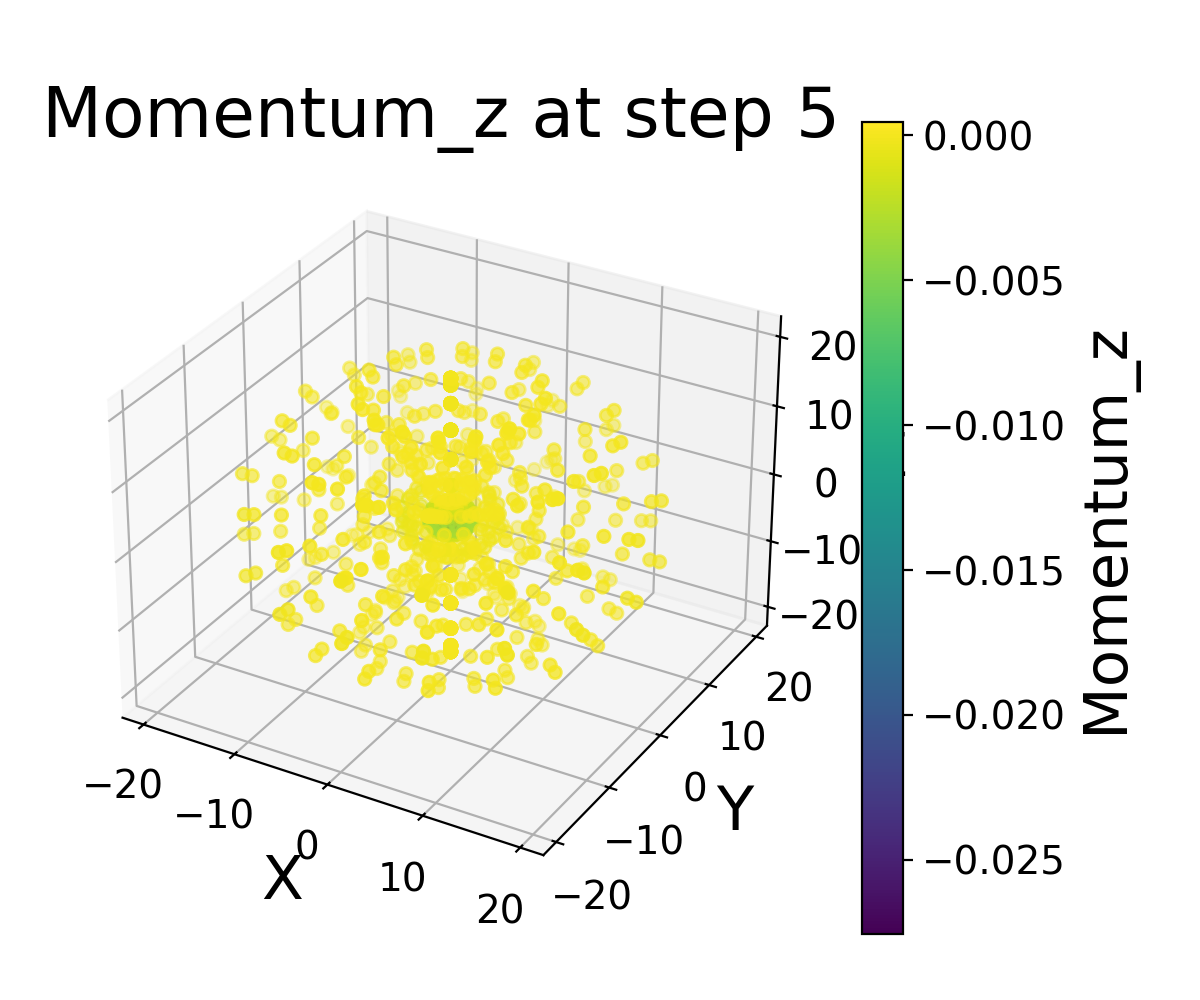}
  \caption{The 3D scatter plots of $Momentum_x, Momentum_y, Momentum_z$ at steps 1, 3 and 5 respectively.}
  \label{fig:m}
\end{figure}

\begin{figure}[h]
    \centering
    \includegraphics[width=0.3\textwidth]{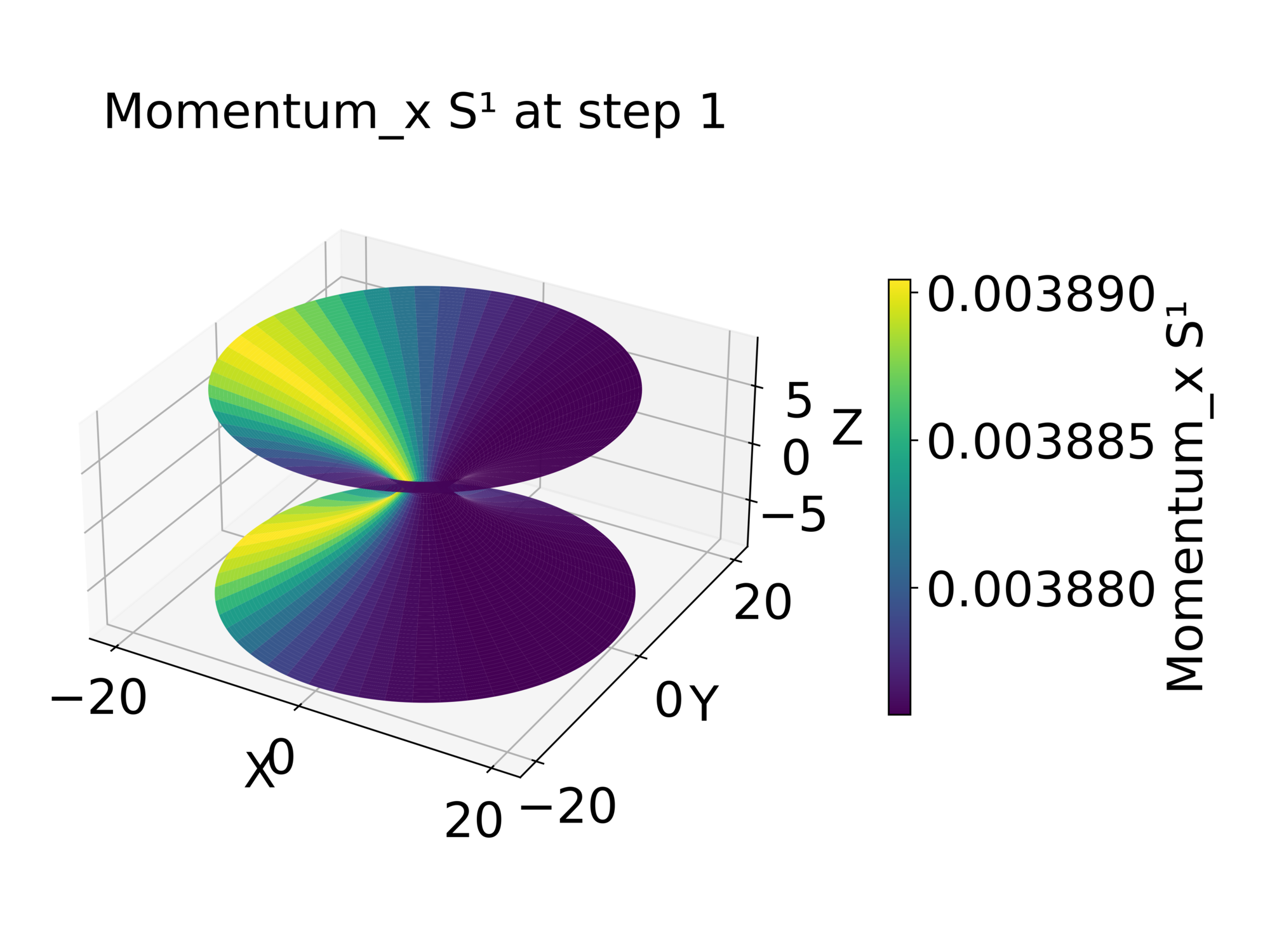}
    \includegraphics[width=0.3\textwidth]{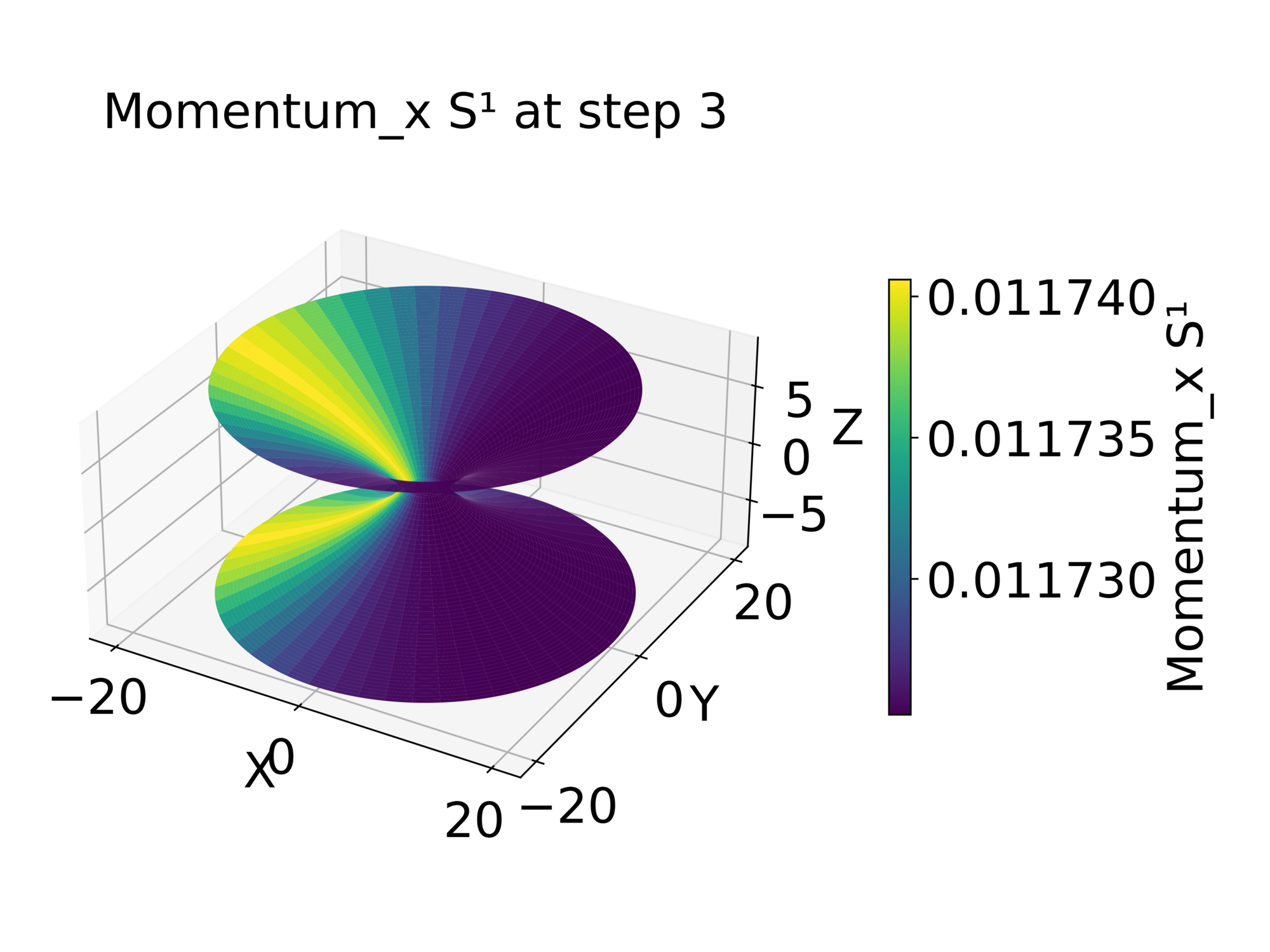}
    \includegraphics[width=0.3\textwidth]{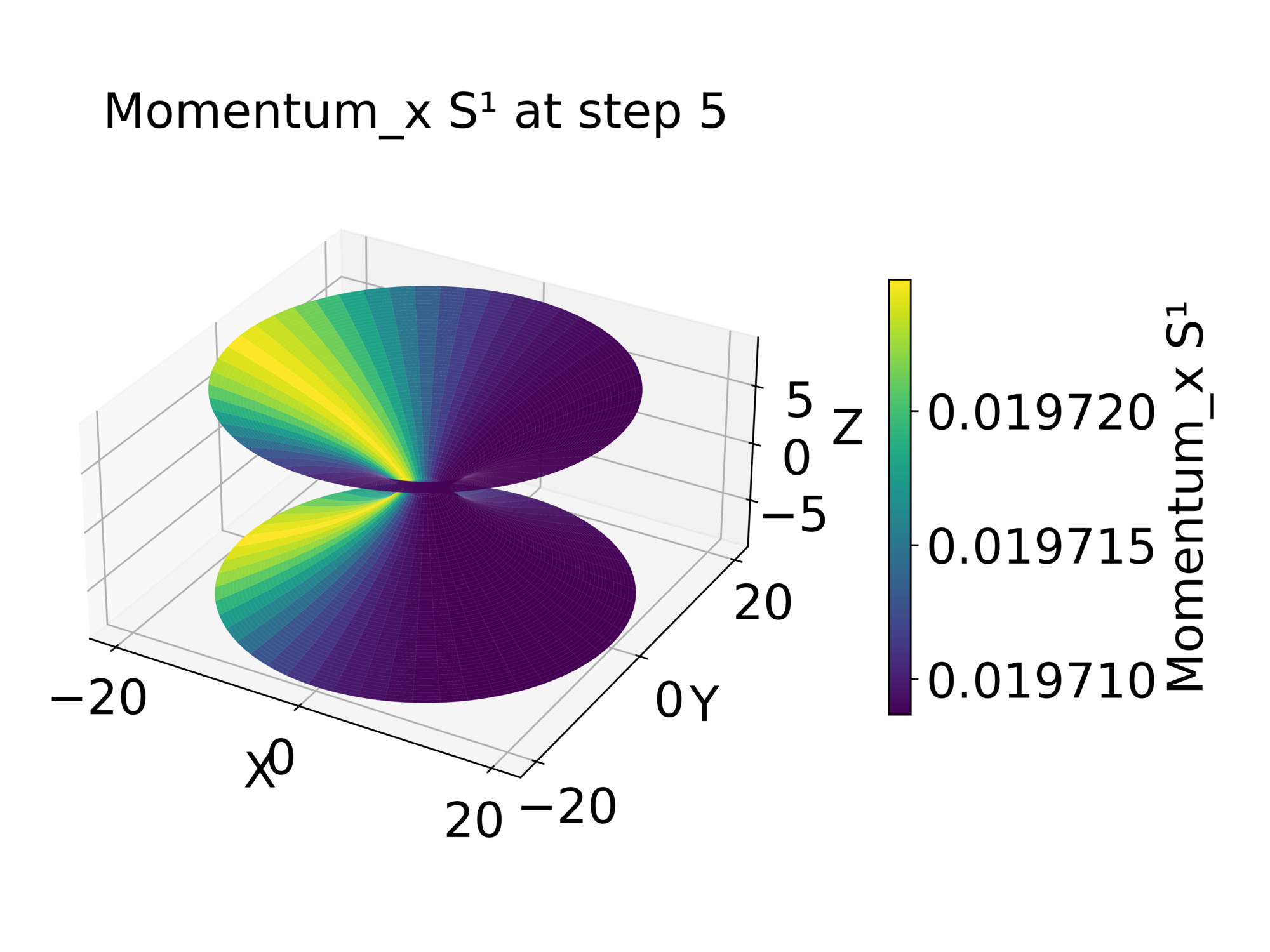}
    \includegraphics[width=0.3\textwidth]{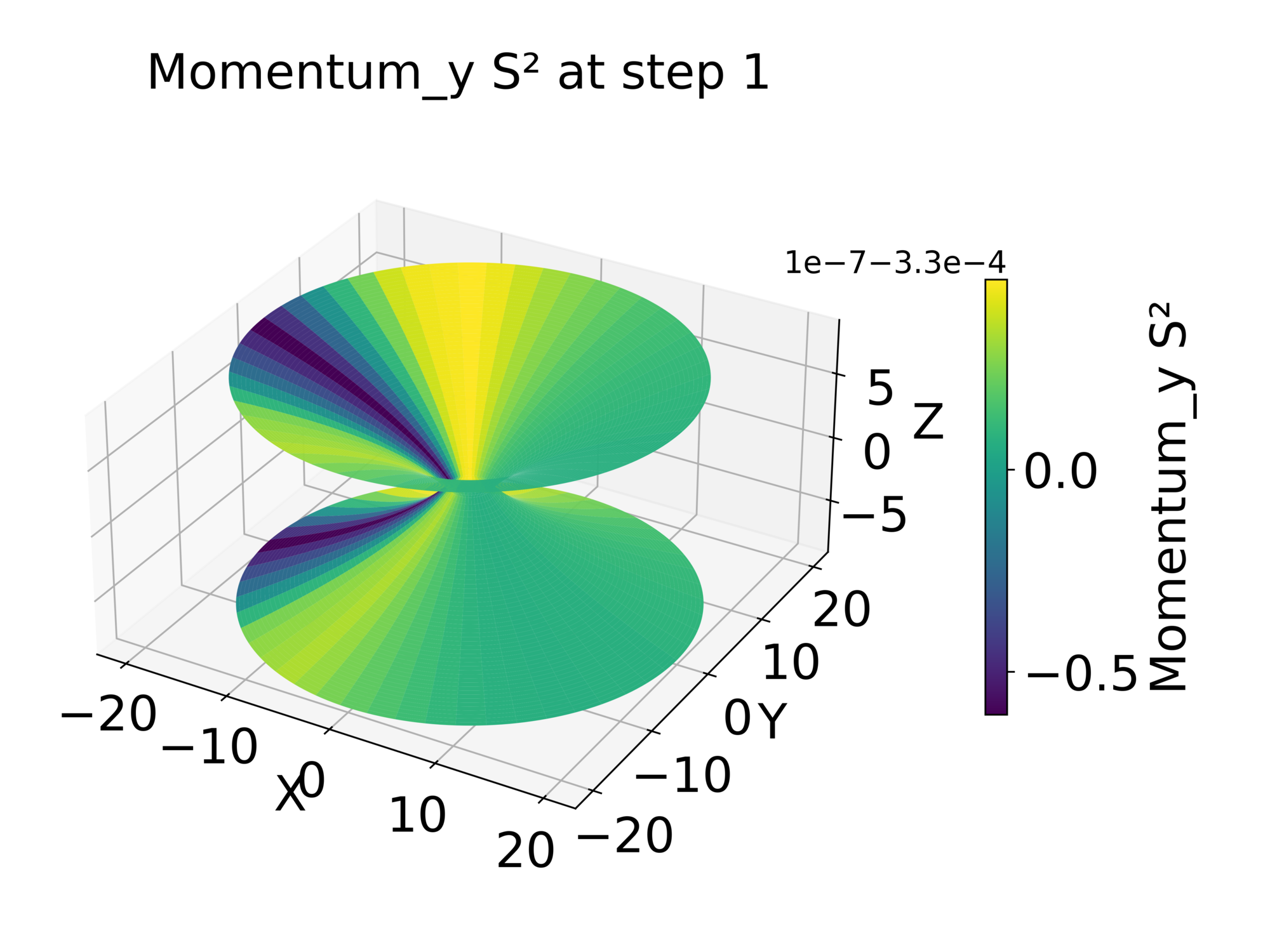}
    \includegraphics[width=0.3\textwidth]{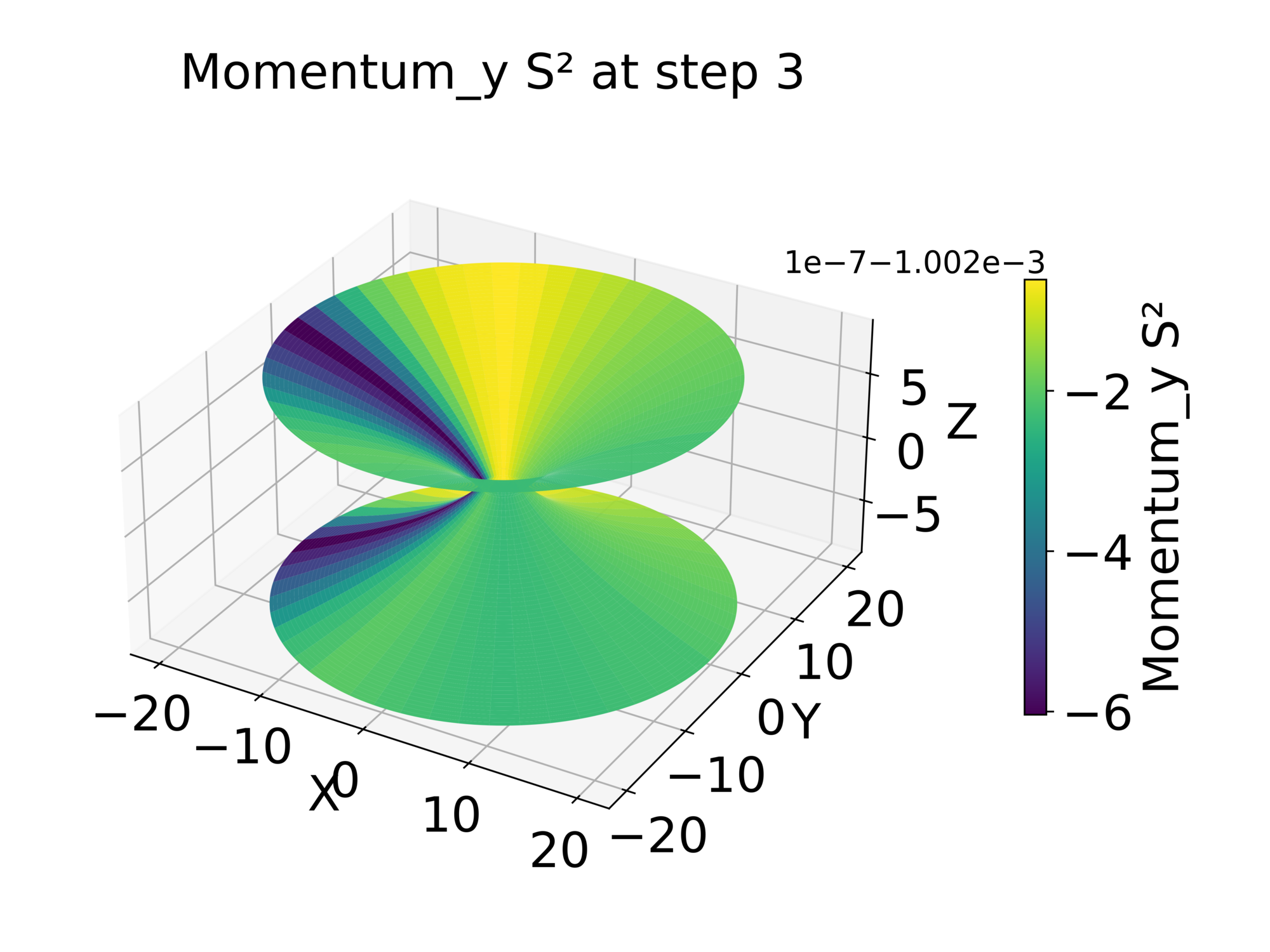}
    \includegraphics[width=0.3\textwidth]{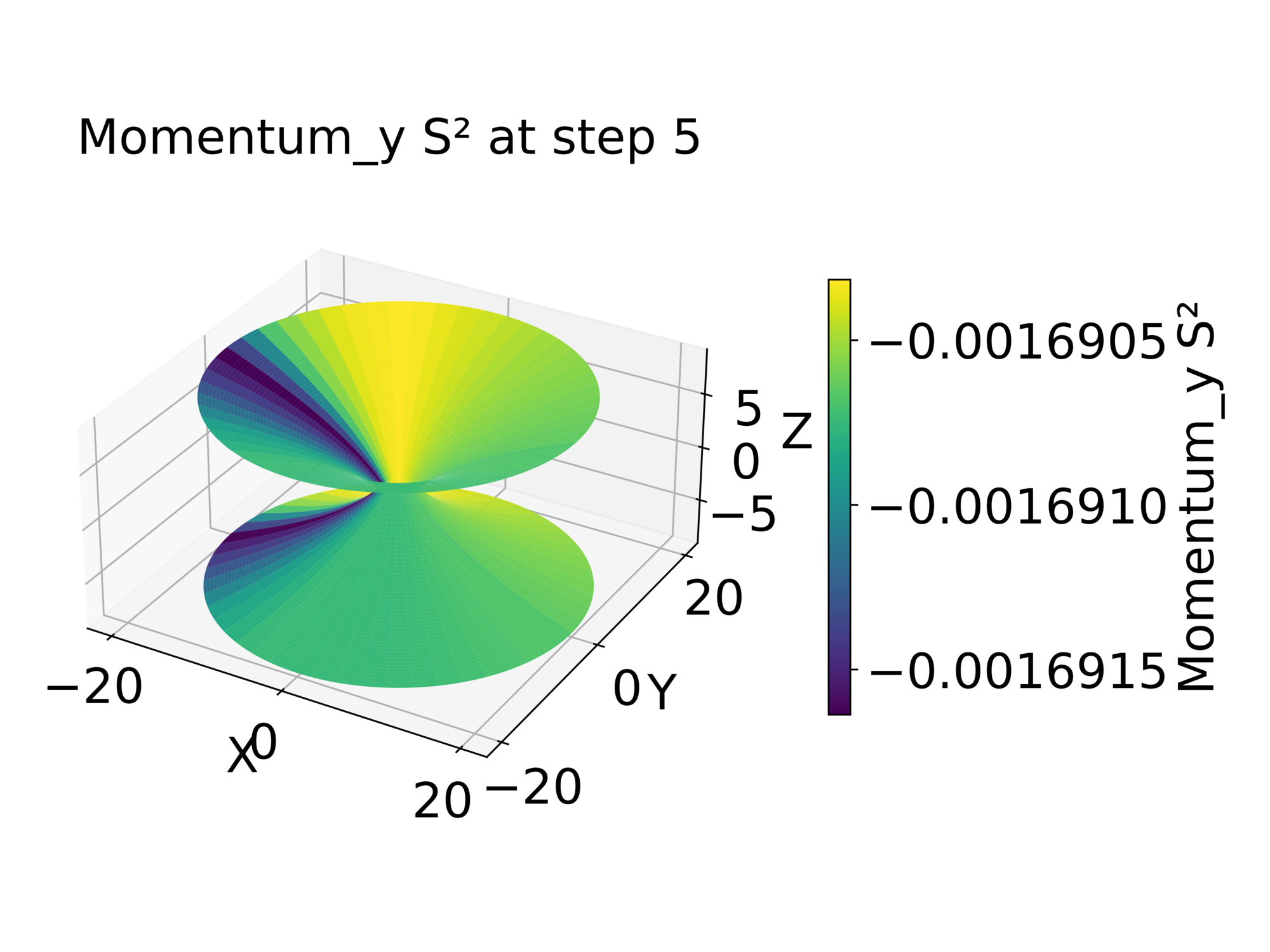}
    \includegraphics[width=0.3\textwidth]{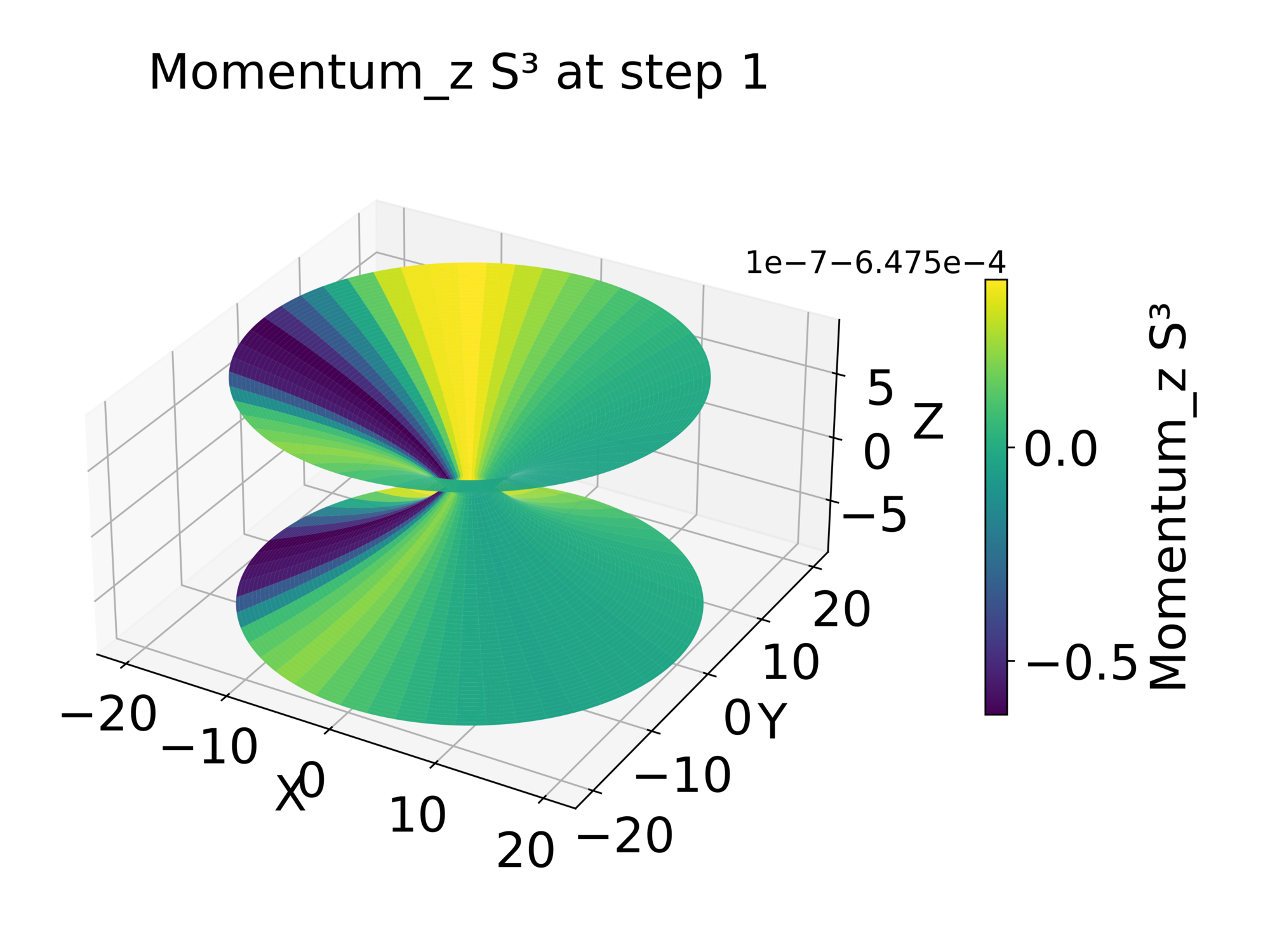}
    \includegraphics[width=0.3\textwidth]{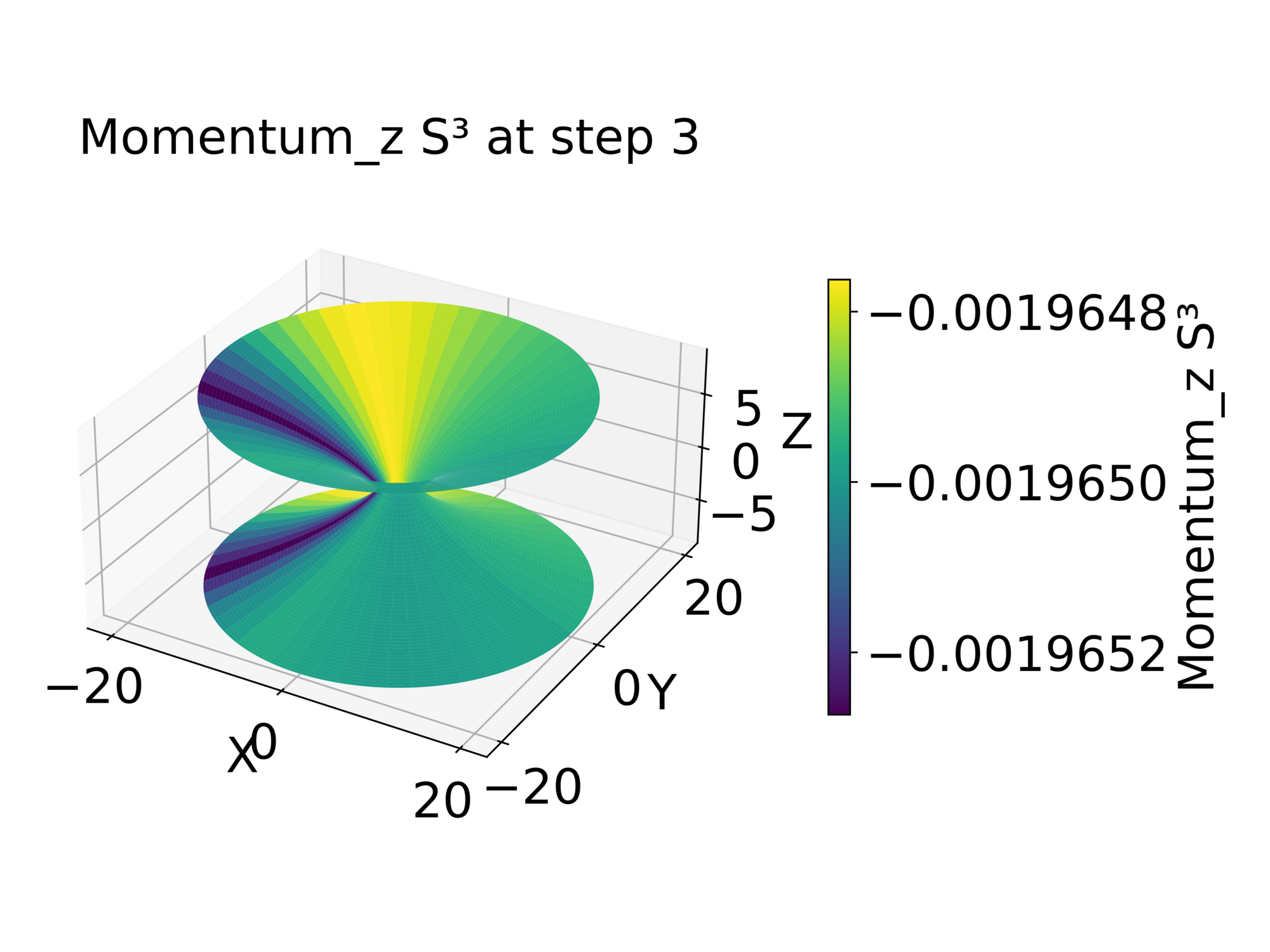}
    \includegraphics[width=0.3\textwidth]{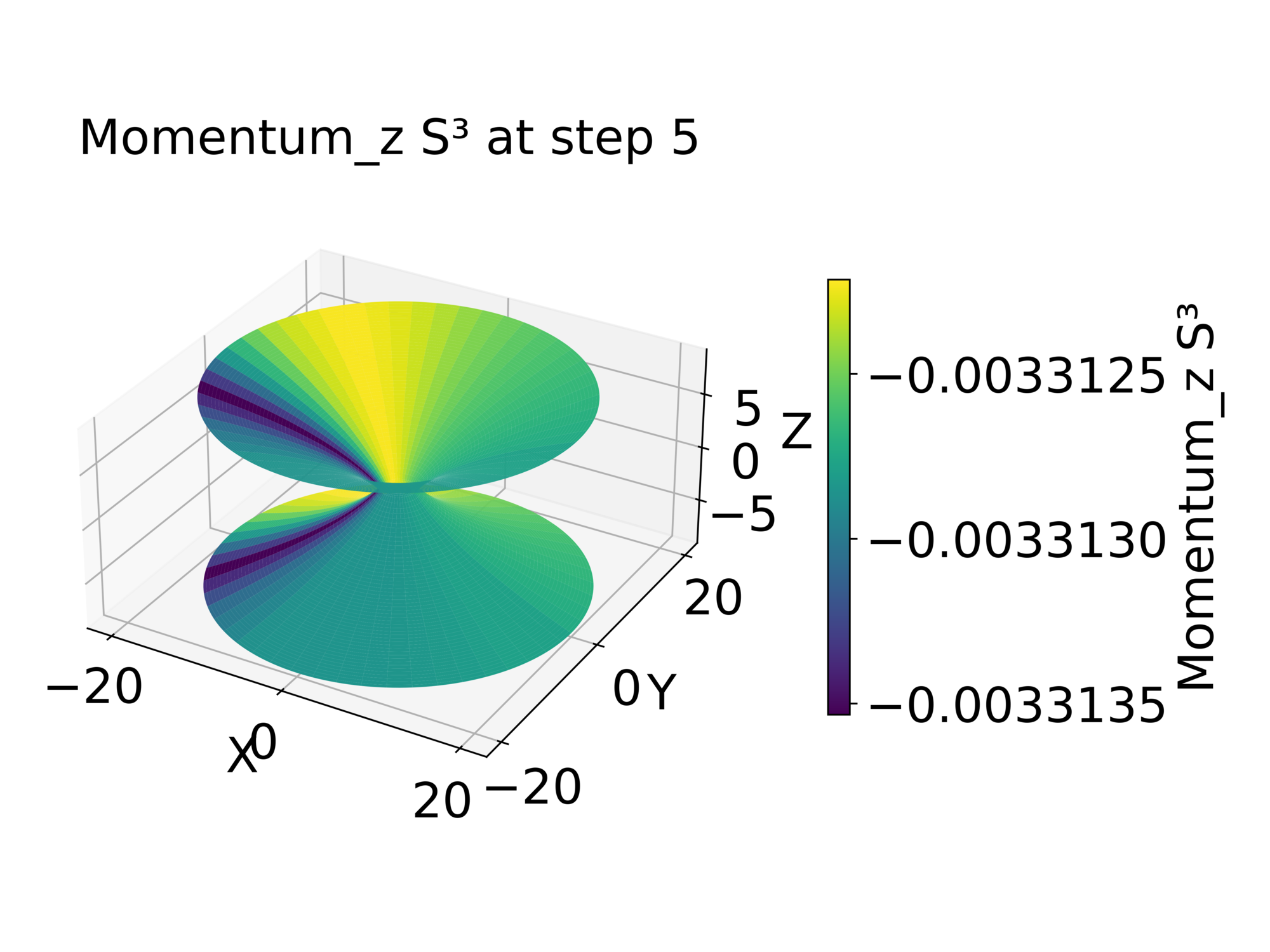}
    \caption{Equatorial embedding diagrams of the Kerr black hole event horizon \( r_h = M + \sqrt{M^2 - a^2} \), shown as a Flamm paraboloid in fictitious 3D Euclidean space. The surface is parametrized by \( X = r_h \cos\phi \), \( Y = r_h \sin\phi \), and the embedding height \( Z \) computed from the spatial metric. Colors indicate the value of each conserved variable of momentum along the equatorial ring \( \theta = \pi/2 \) at the event horizon. Both upper \( +Z \) and lower \( -Z \) surfaces are shown for symmetry.}
    \label{fig:TheDSSN}
\end{figure}

\end{appendices}

\bibliography{grmhd}

\end{document}